
%
%
%
%
 \let\SELECTOR=P      
%
%
%
\expandafter\ifx\csname phyzzx\endcsname\relax \input phyzzx \fi

\newif\iffigureexists
\newif\ifepsfloaded
\openin 1 epsf
\ifeof 1 \epsfloadedfalse \else \epsfloadedtrue \fi
\closein 1
\ifepsfloaded
    \input epsf
\else
    \immediate\write20{Warning:
         No EPSF file --- cannot imbed diagrams!!}
\fi
\def\checkex#1 {\relax
    \ifepsfloaded \openin 1 #1
	\ifeof 1 \figureexistsfalse
	\else \figureexiststrue
	\fi \closein 1
    \else \figureexistsfalse
    \fi }
\def\diagram#1 {\vcenter{\def\epsfsize##1##2{0.4##1} \epsfbox{#1}}}
\def\missbox#1{\vbox{\hrule\hbox{\vrule width 0pt height 14pt depth 4pt
	\vrule \kern 4pt #1\kern 4pt \vrule}\hrule }}
\ifepsfloaded
\checkex Triangle.eps
    \iffigureexists \else
    \immediate\write20{Diagrams are packaged in a separate file}
    \immediate\write20{You should unpack them and TeX again!!}
    \fi
\fi

\newdimen\doublewidth
\doublewidth=73.5pc
\newinsert\LeftPage
\count\LeftPage=0
\dimen\LeftPage=\maxdimen
\def\PageBox{\vbox{\makeheadline \pagebody \makefootline }}
\hsize=35pc
\vsize=48.5pc
\pagebottomfiller=0pt
\skip\footins=\bigskipamount
\normalspace

\if R\SELECTOR
    \mag=833
    \voffset=-2.5truepc
    \hoffset=-3.5truepc
    \output={\ifvoid\LeftPage \insert\LeftPage{\floatingpenalty 20000 \PageBox}
        \else \shipout\hbox to\doublewidth{%
            \box\LeftPage \hfil \PageBox }\fi
        \advancepageno
        \ifnum\outputpenalty>-20000 \else \dosupereject \fi }
    \immediate\write20{Warning:
        Some DVI drivers cannot handle reduced output!!!}
\else
    \mag=1000
    \voffset=0pt
    \hoffset=0pt
\fi
\interdisplaylinepenalty=10000

\def\mpl{M_{\rm Pl}}

\def\modul{\Phi}
\def\modulb{\smash{\overline\Phi}\vphantom{\Phi}}
\def\Kmod{K_{\rm mod}}
\def\matter{Q}
\def\matterb{\smash{\overline\matter}\vphantom{\matter}}
\def\QQ{{\cal Q}}
\def\ccf{\varphi}
\def\ccfb{\smash{\overline\ccf}\vphantom{\ccf}}
\def\uu{{\cal U}}
\def\uuh{\smash{\widehat\uu}\vphantom{\uu}}
\def\Wtree{W_{\rm tree}}
\def\Weff{W_{\rm eff}}
\def\Wmod{W_{\rm mod}}
\def\Vmod{{\cal V}_{\rm mod}}
\def\schi{\chi}
\def\symmat{\Upsilon}
\def\ib{{\bar\imath}}
\def\Ib{{\bar I}}
\def\ind{a}
\def\ww{w}
\def\nmax{{n_{\rm max}}}
\def\cano{{\bf c}}
\def\cren{{\bf b}}
\def\secondb{\cren^{(2)}}
 
\def\Re{\mathop{\rm Re}\nolimits}
\def\Im{\mathop{\rm Im}\nolimits}
\def\Sdet{\mathop{\rm Sdet}\nolimits}
\def\StackThemUp#1{\vbox{\baselineskip=0.45\normalbaselineskip
            \ialign{\the\scriptscriptfont0 \hfil ##\hfil\cr #1\crcr }}}
\def\NHMM{{\StackThemUp{non-Higgs\cr massive} \matter^I}}
\def\MV{{\StackThemUp{massive\cr vectors}}}
\def\ol{{\scriptscriptstyle\rm 1-loop}}
\def\dalamthick#1{{\setbox0=\hbox{$Q$} \dimen0=\ht0
    \advance\dimen0 -#1 \advance\dimen0 -#1
    \hbox{\vrule width #1 \vbox to \ht0{
	    \hrule width \dimen0 height 0pt depth #1 \vss
            \hrule height #1 depth 0pt}%
        \vrule width #1 \vrule width #1 height 0pt }}}
\def\dalamord{\mathop{\dalamthick{0.4pt}}\nolimits}
\def\dalamcov{\mathop{\dalamthick{0.8pt}}\nolimits}
\def\dalamFC{\mathop{\triangle}\nolimits}
\def\K{K\"ahler}
\def\ldf{\REF}
\def\JJjournal#1#2{\unskip\space{\sfcode`\.=1000\sl #1 \bf #2}\space }
\def\nup#1 {\JJjournal{Nucl. Phys.}{B#1}}
\def\plt#1 {\JJjournal{Phys. Lett.}{#1}}
\def\cmp#1 {\JJjournal{Comm. Math. Phys.}{#1}}
\def\prp#1 {\JJjournal{Phys. Rep.}{#1}}
\def\prl#1 {\JJjournal{Phys. Rev. Lett.}{#1}}
\def\prv#1 {\JJjournal{Phys. Rev.}{#1}}
\def\mplt#1 {\JJjournal{Mod. Phys. Lett.}{#1}}

\Pubnum{UTTG--94--1\cr LMU--TPW--94--1}
\date{January 1994}

\titlepage
\title{Field Dependent Gauge Couplings \penalty-1000
	in Locally Supersymmetric Effective Quantum Field Theories
\foot{Research supported in part by:
	the NSF, under grant PHY--90--09850 (V.~K.);
	the Robert A.~Welch Foundation (V.~K.);
	the Heisenberg Fellowship of the DFG (J.~L.);
	the NATO, under grant CRG~931380 (both authors).}}
\author{Vadim Kaplunovsky
	\foot{Email: \tt vadim@bolvan.ph.utexas.edu}}
\address{Theory Group, Dept. of Physics, University of Texas\break
	Austin, TX 78712, USA}
\andauthor{Jan Louis
	\foot{Email: \tt jlouis@lswes8.ls-wess.physik.uni-muenchen.de}}
\address{Sektion Physik, Universit\"at M\"unchen\break
	Theresienstrasse 37, D-80333 M\"unchen, Germany}
\par\nobreak\vfil\nobreak\medskip
\centerline{Dedicated to the Memory of \fourteenrm Brian Warr}
\endpage
%
\abstract
We investigate the field dependence of the gauge couplings of locally
supersymmetric effective quantum field theories.
We find that the Weyl rescaling of supergravity gives rise to Wess-Zumino
terms that affect the gauge couplings at the one-loop level.
These Wess-Zumino terms are crucial in assuring supersymmetric consistency
of both perturbative and non-perturbative gauge interactions.
At the perturbative level, we distinguish between the holomorphic
Wilsonian gauge couplings and the physically-measurable
momentum-dependent effective gauge couplings;
the latter are affected by  the Konishi and the
super-Weyl anomalies and their field-dependence is non-holomorphic.
At the non-perturbative level,
we show how consistency of the scalar potential generated
by infrared-strong gauge interactions with the local supersymmetry
requires a very specific form of the effective superpotential.
We use this superpotential to determine the dependence of the
supersymmetric condensates of a strongly interacting gauge theory
on its (field-dependent) Wilsonian gauge coupling and the
Yukawa couplings of the matter fields.
The article concludes with the discussion of the field-dependent
non-perturbative phenomena in the context of string unification.
\endpage

\chapter{Introduction and Summary}
A unified theory of all particle interactions must ultimately account for the
gravitational force, and within the context of General Relativity a
supersymmetric theory must necessarily be {\it locally} supersymmetric.
Locally supersymmetric quantum field theories are non-renormalizable
but nevertheless serve as effective quantum field theories (EQFTs)
for energies well below the Planck scale.
Physics of ultra-high energies of the order $O(\mpl)$ and beyond
has to be governed by some theory which is outside the conventional framework
of local quantum fields in four-dimensional spacetime,
perhaps by a string theory.
The locally supersymmetric EQFT acts as a bridge over the energy chasm
between the string theory at the Planck scale and the non-supersymmetric
Standard Model at the weak scale.
At the high-energy end of this bridge, the low-energy limit of the
string theory determines the high-energy couplings of the EQFT;
couplings at lower energies are then controlled by the field-theoretical
renormalization group equations.
The intermediate-energy phenomena --- which presumably trigger spontaneous
breakdown of the local supersymmetry --- are also governed by the EQFT.

A distinct feature of the string unification is that
all couplings of the  EQFT  depend
on massless scalar fields $\modul^i$  called `moduli'.
{}From the string point of view, vacuum expectation values (VEVs) of the moduli
parametrize  continuously degenerate families of  string vacua.
{}From the EQFT point of view, the moduli are massless neutral scalars that
have
exactly flat potentials to all orders of the perturbation theory;
consequently, conventionally normalized moduli VEVs can be as big as $\mpl$.
This article is about the moduli dependence of the gauge couplings of the EQFT
and its consequences for the nonperturbative phenomena such as
gaugino condensation.
Our analysis is entirely field-theoretical in that
we make no assumptions about the nature of the
unified theory at ultra-high energies and only insist
that its low-energy limit
is a four-dimensional locally supersymmetric EQFT with moduli.
In particular, we do not assume any properties of the moduli that do not
follow from their origin in the Planck-scale vacuum degeneracy and from
the local supersymmetry of the EQFT.
Thus, although our assumptions are inspired by the superstring, our results
apply equally to other unified theories with similar low-energy features.

\ldf\SVa{M.A.~Shifman and A.I.~Vainshtein,
   \nup277 (1986) 456.}
Any discussion of the couplings of an effective {\it quantum} field theory must
distinguish between two very different kinds
of renormalized couplings\refmark{\SVa}:
First, there are coefficients of the quantum operators in the
Lagrangian of the effective theory from which the ultra-high momentum modes
of the quantum fields are integrated out while the remaining modes are
fully quantized.
These coefficients depend on the scale $\Lambda$ of the ultraviolet cutoff
and we call them the {\it Wilsonian couplings} since they renormalize according
to
the Wilsonian renormalization group.
\Ref\WKP{K.~G.~Wilson and J.~G.~Kogut \journal Physics Report &12 (74) 75;\brk
	see also J.~Polchinski \journal Nucl. Phys. &B231 (84) 269.}
Second, there are effective couplings associated with physical processes;
for example, Coulomb-like scattering of charged particles defines
a momentum-dependent gauge coupling $\{g(p)\}$.
We refer to these couplings as the {\it effective couplings}
and denote them with curly braces `$\{\)\}$'.
Unlike the Wilsonian couplings, the effective couplings depend
not on the ultraviolet cutoff scale
but on the momenta of the particles involved;
thus, they do not correspond to any {\it local} effective Lagrangian.
The dependence of the effective couplings on the overall momentum scale
is governed by the Callan-Symanzik equations, which resemble the Wilsonian
renormalization group equations but have different physical meaning and
different $\beta$-functions beyond one loop.

\ldf\SVb{M.A.~Shifman and A.I.~Vainshtein,
    \nup359 (1991) 571.}
\ldf\CFGP{E.~Cremmer, S.~Ferrara, L.~Girardello and A.~Van Proeyen,
    \nup212 (1983) 413.}
\ldf\WB{J.~Wess and J.~Bagger, {\it Supersymmetry and
    Supergravity} (Princeton Unversity Press, 1983).}
\ldf\GGRS{S.~J.~Gates, M.~Grisaru, M.~Ro\v cek and W.~Siegel,
    {\it Superspace},\brk Benjamin/Cummings, 1983.}
\ldf\nilles{H.~P.~Nilles, \plt  180B (1986) 240.}
The analytic properties of the moduli dependence of the Wilsonian and the
effective gauge couplings are very different from each other. \refmark{\SVb}
We find (in section 2) that a properly defined
Wilsonian gauge coupling $1/g^2_W$ of a locally supersymmetric
EQFT must be given by the real part of a holomorphic function $f^W(\modul)$,
plus a Wess-Zumino term originating in the
Weyl rescaling\refmark{\CFGP-\GGRS}
(which is needed to make the $N=1$ supergravity consistent with
 the ordinary Einsteinian gravity).
The Wilsonian gauge couplings renormalize only at
one-loop\refmark{\SVa, \nilles}  and thus
both the holomorphic functions $f^W(\modul)$ and the Wess-Zumino terms
are completely determined at the one-loop level of the EQFT.
This fact leads to exact (\ie, correct to all orders of the
perturbation theory) transformation rules for the $f^W$
under global symmetries of the EQFT,
including anomalous and K\"ahler symmetries (section 3.2).
Similarly, for EQFTs with thresholds at energies
smaller than $\mpl$ but bigger than the supersymmetry breaking scale,
we derive {\sl exact} threshold corrections to the Wilsonian gauge couplings
(section 3.3).

\ldf\DKLb{L.~Dixon, V.~Kaplunovsky and J.~Louis, \nup355 (1991) 649.}
\ldf\JLpascos{A preliminary version of this article was
     presented by J. Louis, in the Proceedings of the
    {\it 2nd International Symposium on
    Particles, Strings and Cosmology}, Boston, MA, March 25-30, 1991,
    ed.~P.~Nath und S.~Reucroft.}
\ldf\DFKZa{J. P. Derendinger, S. Ferrara, C. Kounnas and F. Zwirner,
    \nup372 (1992) 145 and \plt B271 (1991) 307.}
\ldf\CLO{G.~L.~Cardoso and B.~Ovrut, \nup369 (1992) 351;
    \nup392 (1993) 315 and CERN preprint CERN--TH.6961/93.}
\ldf\konishi{T.~E.~Clark, O.~Piguet and K.~Sibold, \nup 159 (79) 1;\brk
    K.~Konishi, \plt  135B (1984) 439.}
The analytic properties of the effective gauge couplings are more complicated.
While the Wilsonian gauge couplings  are related by supersymmetry (SUSY)
to the $\theta$-angles of the EQFT, the effective gauge couplings are related
to the effective axionic couplings,
which (in a theory with massless charged fermions) cannot
be summarized in any kind of a local effective Lagrangian;
consequently, the moduli dependence of the effective gauge couplings is
non-holomorphic.\refmark{\DKLb}
{}From a different point of view, the non-holomorphicity follows from the two
anomalies of locally supersymmetric
EQFTs\refmark{\JLpascos-\CLO}:
The Konishi anomaly\refmark{\konishi} associated with
non-canonically normalized  charged fields
and the super-Weyl anomaly peculiar to local SUSY.
Shifman and Vainshtein\refmark{\SVa} used the Konishi anomaly to integrate
the Callan-Symanzik equations for the effective gauge couplings of a
rigidly supersymmetric gauge theory.
In this article, we extend their results to locally supersymmetric EQFTs
(section 3.1) and use them to determine the moduli dependence of the
confinement scales of the asymptotically-free gauge interactions
(section 4).

\ldf\dindrsw{J.P.~Derendinger, L.E.~Ib\'a\~nez and
	H.P.~Nilles, \plt  155B (1985) 65;\brk
    M.~Dine, R.~Rohm, N.~Seiberg and E.~Witten, \plt  156B (1985) 55.}
\ldf\gauginorev{for a review see H.~P.~Nilles \journal
	Int. J. Mod. Phys. &A5 (1990) 4199,\brk
    J. Louis, in  Proceedings of the 1991 DPF meeting, World
    Scientific, 1992, and references therein.}
\ldf\FGN{S. Ferrara, L. Girardello and H. P. Nilles, \plt  125B (1983) 457.}
String unification often gives rise to `hidden' asymptotically-free and thus
infrared-strong gauge interactions that do not affect the known particles, and
the moduli dependence of their confinement scales leads to
a non-perturbative effective potential for the
moduli fields.\refmark{\dindrsw, \gauginorev}
In section 4 we calculate this effective potential by integrating out all
the strongly interacting degrees of freedom of a locally
supersymmetric gauge theory,\refmark{\FGN}
including the charged (non-singlet) matter multiplets, if present.
The Wess-Zumino term in the Wilsonian gauge coupling plays a crucial
role in the consistency of the resulting potential with the local
supersymmetry of the EQFT.
This consistency also imposes severe constraints on the moduli dependence of
the non-perturbative condensates of the theory:
All the supersymmetric VEVs of the
theory (including the gaugino condensate \penalty 1000 $\vev{\lambda\lambda}$)
are controlled by a holomorphic effective superpotential.

\ldf\TVY{G.~Veneziano and S.~Yankielowicz, \plt 113B (1982) 231;\brk
    T.R.~Taylor, G.~Veneziano and S.~Yankielowicz, \nup 218 (1983) 493;\brk
    T.R.~Taylor, \plt  164B (1985) 43.}
\ldf\ADS{I.~Affleck, M.~Dine and N.~Seiberg, \prl 51 (1984) 1026,
    \nup241 (1984) 493 and \nup256 (1985) 557.}
\ldf\SVc{V.~Novikov, M.A.~Shifman, A.I.~Vainshtein and V.I.~Zakharov,
	\nup 260 (1985) 157;\brk
    M.A.~Shifman and A.I.~Vainshtein,  \nup 296 (1988) 445.}
\ldf\Amati{D. Amati, K. Konishi, Y. Meurice, G. Rossi and
	G. Veneziano, \prp  162 (1988) 169.}
\ldf\FILQ{A.~Font, L.E.~Ib\'a\~nez, D.~L\"ust and F.~Quevedo,
	\plt  B245 (1990) 401.}
\ldf\FMTV{S.~Ferrara, N.~Magnoli, T.~Taylor and G.~Veneziano,
        \plt B245 (1990) 409.}
\ldf\NO{H.~P.~Nilles and M.~Olechowski, \plt B248 (1990) 268.}
\ldf\BG{P.~Bin\'etruy and M. K. Gaillard, \plt B253 (1991) 119
	and \nup358 (1991) 121.}
\ldf\hiddenmatter{D.~L\"ust and T.~Taylor, \plt B253 (1991) 335;\brk
    B.~de Carlos, J.~A.~Casas and C.~Mu\~noz, \plt B263 (1991) 248
	and \nup399 (1993) 621.}
\ldf\MPRV{A.~Masiero, R.~Pettorino, M.~Roncadelli and G.~Veneziano,
	\nup 261 (1985) 633.}
The effective superpotential is
comprised of the tree-level superpotential of the EQFT,
of its Wilsonian gauge couplings
and of the non-perturbative terms,
which depend only on the gauge quantum numbers of the
matter fields but are totally blind to all coupling parameters of the EQFT.
For many gauge theories, the exact form of the non-perturbative terms
is completely determined by the
supersymmetric consistency conditions
of the moduli's potential.\refmark{\TVY-\hiddenmatter, \JLpascos}
For example, SQCD-like theories with more colors than flavors
give rise to the Taylor-Veneziano-Yankielowicz superpotential\refmark{\TVY}
without any further perturbative or non-perturbative corrections
(that are not supressed by $\mpl$),
while for $N_{\rm flavor}=N_{\rm color}$
the effective superpotential is exactly as conjectured in ref.~[\MPRV].
These, and a few more examples are discussed in section 4.3.

In the absence of non-gauge couplings, supersymmetric gauge theories with
massless matter multiplets often suffer from non-perturbative `run-away'
vacuum instabilities.\refmark{\ADS}
In the context of string unification, the charged (non-singlet) matter fields
are
massless, but the EQFT may have Yukawa couplings that prevent the run-away
and keep all the non-perturbative VEVs near the confinement scale of the
strong gauge interactions.
Alternatively, the run-away may be stopped by the {\sl non-renormalizable}
(quartic or higher-order) couplings in the superpotential of the EQFT.
In this case, some scalar VEVs dynamically produce an intermediate scale
between the confinement scale and the Planck scale.
In section 4.4 we give examples of both behaviors.

\ldf\KLa{V.~Kaplunovsky and J.~Louis, \plt B306 (1993) 269.}
Situations where the run-away is not stopped by the tree-level superpotential
of the EQFT are discussed in section 4.5.
There we also discuss the stability or instability of the moduli VEVs once
a non-perturbative effective potential is produced by strong gauge
interactions.
Our discussion of these issues is general rather than specific;
our goal is to survey various scenarios that may happen in the context
of the string unification (or of any other kind of unification with similar
sub-Planck energy properties).
Very briefly, we discuss how the spontaneous breakdown of supersymmetry fits
into those scenarios;
the phenomenology of the supersymmetry breaking among the observable particles
is presented in a separate article.\refmark{\KLa}

\chapter{Locally Supersymmetric Quantum Effective Field Theories}
\section{The Classical Action.}
As a prelude to locally supersymmetric EQFTs,
consider the classical
action of a generic locally supersymmetric effective field theory.
In the standard superfield formalism,\refmark{\CFGP-\GGRS} one has
$$
{\cal S}\
=\ {-3\over\kappa^2} \int\!\! d^8z\,{\bf E}
    \exp\left(-\coeff13\kappa^2 K\right)\
+ \int\!\!d^8z\,{\cal E}\left[W\,
+\,\coeff14 f_{(a)(b)}{\cal W}^{(a)\alpha}{\cal W}^{(b)}_\alpha \right]\
+\ \hbox{h.~c.}
\eqn\SUGRACT
$$
plus higher-derivative terms.
Superfields of the effective field theory correspond to light particles
of the string theory (or some other unified theory).
We distinguish between two kinds of chiral superfields,
namely the moduli $\modul^i$ and the `matter' superfields $\matter^I$.
The VEVs of the scalar components
of the moduli parametrize the degeneracy of the
classical vacua of the unified theory and can be as big as $\mpl$
($\kappa^2 = {8\pi\over \mpl^2}$);
consequently, all moduli are neutral with respect to the entire
four-dimensional gauge symmetry.
Note that although the moduli originate at the unification scale $O(\mpl)$,
they are massless and thus should be retained in the low-energy EFT,
unlike the genuinely superheavy fields that simply decouple from the low-energy
physics.
The `matter' superfields contain the ordinary scalars (\eg, Higgses)
and fermions (\eg, quarks and leptons) of the theory
and their superpartners.
Scalar superfields that are charged with respect to a `hidden' gauge
symmetry  are also treated as `matter'.
The kinetic-energy terms for the scalar fields and their superpartners
are encoded in the K\"ahler function $K$ --- a gauge-invariant real analytic
function of all the chiral superfields of the theory.
Expanding $K$ in powers of the matter superfields, we have
$$
K(\modul,\modulb,\matter,\matterb)\ =\ \Kmod(\modul,\modulb)\
+\ Z_{\Ib J}(\modul,\modulb)\, \matterb^\Ib e^{2V} \matter^J\ +\ \cdots,
\eqn\Kexpansion
$$
where the $\cdots$ stand for higher-order terms, which are irrelevant
for the present investigation.
$Z_{\Ib J}$ is the moduli-dependent normalization matrix
of the kinetic-energy  terms for the matter fields;
the normalization matrix for the moduli fields themselves is given by
$$
G_{\ib j}\ =\ {\partial^2 \Kmod \over \partial \modulb^\ib\,
\partial \modul^j} \, .
\eqn\modulimetric
$$
$W$ is the superpotential --- a gauge-invariant holomorphic function
of the chiral
superfields $\modul^i$ and $\matter^I$ controlling their masses and
Yukawa interactions.
Similar to the K\"ahler function, it expands into
$$
W(\modul,\matter)\ =\ \half M_{IJ}(\modul)\, \matter^I \matter^J\
+\ \coeff13 Y_{IJK}(\modul)\, \matter^I \matter^J \matter^K\ +\ \cdots,
\eqn\Wexpansion
$$
where $M_{IJ}$ is the un-normalized mass matrix for the matter fields
while $Y_{IJK}$ are their un-normalized Yukawa couplings.
A non-vanishing $M_{IJ}$ matrix only arises in
EFTs with an explicit mass scale well below $\mpl$ while
fields with $O(\mpl)$ masses are not part of a low-energy EFT.
Perturbatively, there is no such scale in the string theory and hence
generically no $M_{IJ}$ arise in EFTs based upon the superstring.
\foot{Certain string states do have $\modul$-dependent masses that
    can vary continuously between zero and $\mpl$.
    In section~4.5 we discuss their role in the low energy EFT.}
However, since we expect some kind of a non-perturbative mechanism to
generate the weak scale, we also allow for the possibility
that the same mechanism generates an intermediate scale.
(Section 3.3 is devoted to
the effects of those mass terms on the low-energy gauge couplings.)

The superfield $V\equiv V^{(a)} T_{(a)}$ in eq.~\Kexpansion\
contains the gauge fields $A_m^{(a)}$ and their superpartners,
\foot{For notational simplicity, we are suppressing the gauge indices of
    the matter superfields, so $\matter^I$ and $\matterb^\Ib$ actually stand
    for the entire irreducible multiplets of the gauge group.
    In more explicit notations, the second term on the right hand side
    of eq.~\Kexpansion\ becomes
    $$
    Z_{\Ib J}(\modul,\modulb)\, \matterb^{\Ib}_s
    \left[\exp\left(2 V^{(a)} T_{(a)}\right)\right]^s_{\,s'} \matter^{Js'} ,
    $$
    where $s,s'$ are the gauge indices and
    $[T_{(a)}]^s_{\>s'}$ are the explicit hermitian matrices
    realizing the Lie-algebra generators $T_{(a)}$ in the appropriate
    representation of the gauge group.
    In either notation, the matrix $Z_{\Ib J}$ is block diagonal ---
    the `flavor' indices $\Ib$ and $J$ must agree with respect to the
    kind of a gauge-group multiplet formed by respective matter fields,
    but different `generational' copies of the same kind of multiplet
    may freely mix with each other.}
while
${\cal W}^{(a)}_\alpha$ in eq.~\SUGRACT\ are the
superfield analogues of the  field strengths $F_{mn}^{(a)}$;
specifically
${\cal W}_\alpha\equiv {\cal W}^{(a)}_\alpha T_{(a)}=
\left(\coeff18 \overline{\cal D}^2-{\cal R}\right)
\mkern 2mu e^{-2V} {\cal D}_\alpha e^{2V}$.
The coefficients $f_{(a)(b)}$ are the supersymmetric gauge couplings
of the theory;
in the component-field expansion of \SUGRACT,
the kinetic-energy terms for the gauge fields are
$$
- \coeff14\ \int\!\!d^4x\, \sqrt{-g} \left(
\Re f_{(a)(b)}\,  F^{(a)}_{mn} F^{(b)mn}
  -\,\coeff{1}{2}  \Im f_{(a)(b)}\,
     \epsilon^{mnpq} F^{(a)}_{mn} F^{(b)}_{pq} \right) \, .
\eqn\gaugeact
$$
Gauge invariance
and local supersymmetry require the functions
$f_{(a)(b)}(\modul,\matter)$ to be both gauge-covariant and holomorphic.
Furthermore, in the low-energy limit,
$$
\eqalign{
f_{(a)(b)}(\modul,\matter)\ &
=\ f_{(a)(b)}(\modul)\ +\ \cdots\cr
&=\ \delta_{(a)(b)}\cdot f_\ind(\modul) +\ \cdots,\cr
}\eqn\deindexingf
$$
where the $\cdots$ stand for the irrelevant terms and the index $\ind$
(without parentheses) refers
to a particular simple factor of the gauge group $G = \prod_\ind G_\ind$
that contains the generator $T^{(a)}$; were the entire gauge group simple,
we would have only one $f(\modul)$.
The second equality here follows from the gauge-covariance of $f_{(a)(b)}$
and the neutrality of the moduli with respect to all gauge symmetries that
remain unbroken below the unification scale.
The relation \deindexingf\ makes for a particularly simple interpretation
of the gauge-field Lagrangian~\gaugeact\ in terms of the ordinary
moduli-dependent couplings $g_\ind(\modul,\modulb)$ and
moduli-dependent vacuum angles $\theta_\ind(\modul,\modulb)$,
which are related to the supersymmetric couplings $f_\ind(\modul)$ via
$$
g_\ind^{-2}(\modul,\modulb)\ =\ \Re f_\ind\ (\modul) \, , \qquad
\theta_\ind(\modul,\modulb)\ =\ -8\pi^2\Im f_\ind\ (\modul)\, .
\eqn\couplingrel
$$
Note that holomorphicity of the functions $f_\ind(\modul)$ implies that both
$1/g_\ind^2$ and $\theta_\ind$ are real harmonic functions of the complex
moduli fields and that each of the two functions determines the other
up to an overall constant.

Finally, the gravitational fields and their superpartners comprise the
vielbein $E_M^A$.
These superfields appear in the Lagrangian~\SUGRACT\ both explicitly,
through the densities ${\bf E}=\det( E^A_M)$ and
${\cal E}={\bf E}/2{\cal R}$
($\cal R$ is the scalar superspace-curvature superfield),
and also implicitly, through the commutation relations of
the supercovariant derivatives ${\cal D}_A=E_A^M{\cal D_M}$.
\foot{$M$ here is the curved superspace index:
    $z^M=(x^m,\Theta^\mu,\bar\Theta^{\dot\mu})$.
    $A$ is the flat local super-Lorentz index,
    running over both Lorentz vectors and Lorentz spinors.
    Elsewhere in this article, $\alpha,\beta,\ldots$ and
    $\dot\alpha,\dot\beta,\ldots$ are local Lorentz spinor indices
    while $a,b,\ldots$ are local Lorentz vector indices {\bf or}
    indices labelling simple factors of the gauge group
    (as in eq.~\couplingrel);
    we apologize for the coincidence.
    $\overline{\cal D}_{\dot\alpha}$ is the same as ${\cal D}_{\dot\alpha}$
    and $\overline{\cal D}^2$ is short-hand for
    $\overline{\cal D}^{\dot\alpha}\overline{\cal D}_{\dot\alpha}$ while
    ${\cal D}^2$ is short-hand for ${\cal D}^\alpha {\cal D}_\alpha$.}
The latter commutation relations are subject to the so-called
torsion constraints\refmark{\WB, \GGRS};
as the result of these constraints, the first term in \SUGRACT\
gives not only the kinetic-energy terms for the scalar fields and
their superpartners, but also for the gravitational fields themselves.

Unfortunately, extracting those kinetic-energy terms from eq.~\SUGRACT\
is not so easy.
A naive interpretation of the lowest  components
$\hat e_m^a\equiv \left. E_m^a\right|_{\Theta=\bar{\Theta}=0}$
of the vielbein as the ordinary vierbein
and the corresponding metric tensor $\hat g_{mn}\equiv \hat e_m^a
\eta_{ab}\hat e_n^b$ as the physical (Einsteinian) metric of spacetime
results in a field-dependent gravitational coupling
$\hat G_N=(\kappa^2/8\pi)\cdot\exp\left(-{1\over3}\kappa^2
    K\right)$,
as well as non-K\"ahlerian kinetic-energy terms for the scalar fields
and a host of other complications.
The usual procedure
for disentangling supergravity from other fields is known as the
Weyl rescaling because it involves a rescaled spacetime
metric\refmark{\CFGP,\WB}
$$
g_{mn}\ =\ \hat g_{mn}\cdot e^{-{1\over3}\kappa^2 K} .
\eqn\weylmetric
$$
All the fermionic fields of the theory are also rescaled
by  appropriate exponentials of $\kappa^2 K$; the purpose of this
rescaling is to assure that the bosons and the fermions belonging
to the same supermultiplet are normalized in the same way.

Let us close this section with a few technical remarks.
The action \SUGRACT\ is not completely generic.
In principle, the superpotential $W$ could contain an
O'Raifeartaigh term in addition to the terms listed in eq.~\Wexpansion;
similarly,
the K\"ahler function $K$ could also contain a Fayet-Iliopoulos
term for one of the abelian gauge superfields.
In superstring-based unified theories, whenever such terms occur,
they are of the order $O(\mpl^2)$ and thus cause major rearrangements of
the vacuum family right at the unification scale.
Consequently, once the theory is expanded around the right vacuum family
and the particles with
$O(\mpl)$ masses are excluded from the low-energy EQFT,
the O'Raifeartaigh and the Fayet-Iliopoulos terms disappear.
For other unified theories, such terms might conceivably be present
and have magnitudes much less than $\mpl^2$;
however, they would not have any effect
on the subject matter of this article.
Thus, to avoid unnecessary complication of our notations, we chose
to omit the O'Raifeartaigh and the Fayet-Iliopoulos terms
{}from eqs.~\Kexpansion\ and \Wexpansion\
and from all subsequent discussion.
For the same reason,
we have assumed that in theories with several {\sl abelian}
gauge symmetries $G_a$, the abelian gauge superfields $V^{(a)}$
do not mix with each other.
If they do, the supersymmetric gauge coupling matrix $f_{(a)(b)}$
has non-diagonal terms in addition to $f_a$;
generalizing our arguments to such non-diagonal couplings
is completely straightforward.

\ldf\ABGG{P.~Binetruy, G.~Girardi and R.~Grimm,
	\plt B265 (1991) 111;\brk
    P.~Adamietz, P.~Binetruy, G.~Girardi and R.~Grimm,
	\nup 401 (1993) 257.}
\ldf\KLc{V.~Kaplunovsky and J.~Louis, in preparation.}
On the other hand, we lose no generality by assuming that all scalars
belong to chiral rather than linear superfields.
In four spacetime dimensions, $N=1$ linear multiplets
are always dual to chiral multiplets
(with very specific couplings).
For example, the dilaton of
string theory can be
treated  as just another chiral modulus superfield $S$.
(The special properties of the dilaton are discussed
elsewhere.\refmark{\DFKZa, \CLO, \ABGG, \KLc})

\section{The Weyl Compensator Formalism}
\ldf\compensator{T.~Kugo and S.~Uehara, \nup 222 (1983) 125.}
In the usual formalism of  minimal supergravity, the Weyl rescaling
is done in terms of  component fields.
However, in order to understand the anomalous quantum corrections
to the classical action \SUGRACT, we need a manifestly supersymmetric
formalism, in which the Weyl rescaling is also
supersymmetric.\refmark{\compensator}

\ldf\HT{P.~Howe and R.~Tucker, \plt 80B, 138 (1978).}
\ldf\BGG{P. Binetruy, G. Girardi and R. Grimm,
    LAPP preprint LAPP--TH--275--90 and references therein.}
The simplest formalism of this kind makes use of the invariance of the torsion
constraints of  minimal supergravity and of the chirality constraints
${\cal D}_{\dot\alpha}\modul^i={\cal D}_{\dot\alpha}\matter^I=0$ under a local
super-Weyl symmetry\refmark{\HT, \BGG, \WB}
$$
\openup 1\jot
\displaylines{
\!\! E^\alpha_M\mapsto e^{(2\bar\tau-\tau)}\cdot \left( E^\alpha_M
	-\coeff{i}{2}\bar\sigma^{\dot\beta\alpha}_a
	({\cal D}_{\dot\beta}\bar\tau)\,E^a_M \right), \quad\!
E^{\dot\alpha}_M\mapsto e^{(2\tau-\bar\tau)}\cdot \left( E^{\dot\alpha}_M
	-\coeff{i}{2}\bar\sigma^{\dot\alpha\beta}_a
	({\cal D}_{\beta}\tau)\,E^a_M \right),\!\! \cr
E^a_M\mapsto e^{(\tau+\bar\tau)}\cdot E^a_M,\qquad
{\bf E}\mapsto e^{2(\tau+\bar\tau)}\cdot {\bf E},\qquad
{\cal E}\mapsto e^{6\tau}\cdot {\cal E} +\cdots,\qquad
\overline{\cal E}\mapsto e^{6\bar\tau}\cdot \overline{\cal E} +\cdots,\cr
{\cal D}_\alpha\mapsto e^{(\tau-2\bar\tau)}\left( {\cal D}_\alpha
	-2({\cal D}^\beta\tau)\,L_{\alpha\beta} \right),\qquad
{\cal D}_{\dot\alpha}\mapsto e^{(\bar\tau-2\tau)}\left( {\cal D}_{\dot\alpha}
	-2({\cal D}^{\dot\beta}\bar\tau)\,L_{\dot\alpha\dot\beta} \right),\cr
\modul^i\mapsto\modul^i,\qquad \matter^I\mapsto\matter^I,\qquad
\modulb^\ib\mapsto\modulb^\ib,\qquad \matterb^\Ib\mapsto\matterb^\Ib,\cr
V^{(a)}\mapsto V^{(a)} ,\qquad
{\cal W}_\alpha^{(a)}\mapsto e^{-3\tau}\cdot {\cal W}_\alpha^{(a)},\qquad
\overline{\cal W}_{\dot\alpha}^{(a)}\mapsto
	e^{-3\bar\tau}\cdot \overline{\cal W}_{\dot\alpha}^{(a)},\cr
\noalign{\hrule height 0pt\vskip 1\jot}
\hfill \eqname\superWeyl \cr }
$$
parametrized by an arbitrary chiral superfield $\tau(z)$
and its anti-chiral conjugate $\bar\tau(z)$.
Here $L_{\alpha\beta}$ and $L_{\dot\alpha\dot\beta}$
are the generators of the local Lorentz symmetry,
the transformation rule for the ${\cal D}_a$ derivatives is not displayed as
too complicated and not germane to the present discussion and
the `$\cdots$' stand for inhomogeneous terms in the chiral measures $\cal E$
and $\overline{\cal E}$ that do not contribute to
integrals such as in \SUGRACT.
Although the scalar and the vector superfields remain
invariant under \superWeyl,
the definitions of their fermionic components change due to changing
covariant super-derivatives ${\cal D}_\alpha$ and ${\cal D}_{\dot\alpha}$.
In particular, the gauginos $\lambda^{(a)}_\alpha$ transform like
$$
\lambda^{(a)}_\alpha\
\equiv\ \left.{\cal W}^{(a)}_\alpha\right|_{\Theta=\bar\Theta=0}\
\mapsto\ \lambda^{(a)}_\alpha\cdot e^{-3\tau} ,
\eqn\gauginoWeyl
$$
while the matter fermions $\Psi^I_\alpha$ transform like
$$
\Psi^I_\alpha\
\equiv\ \left.{\cal D}_\alpha\matter^I\right|_{\Theta=\bar\Theta=0}\
\mapsto\ \Psi^I_\alpha\cdot e^{\tau-2\bar\tau} .
\eqn\matterWeyl
$$

As written, the action functional \SUGRACT\ is not super-Weyl invariant.
However, this lack of symmetry can be easily remedied
with the help of a chiral superfield $\ccf $ (${\cal D}_{\dot\alpha}\ccf=0$)
that transforms under \superWeyl\ like\refmark{\compensator}
$$
\ccf\ \mapsto\ e^{-2\tau}\cdot \ccf \,.
\eqn\ccfWeyl
$$
This field is known as the Weyl compensator
because formula~\ccfWeyl\ allows us to compensate for `wrong' super-Weyl
scaling properties of any term in the action functional by multiplying
it by an appropriate power of $\ccf$ and/or $\ccfb$.
Specifically, for the {\it classical} action~\SUGRACT,  the
K\"ahler function $K$,
the superpotential $W$ and the supersymmetric
gauge couplings $f_\ind$ are modified according to the
following rules:
$$
\eqalignno{
K(\modul,\modulb,\matter,\matterb)\ &
\mapsto\ \widetilde K(\ccf,\ccfb,\modul,\modulb,\matter,\matterb)\
    =\ K\ -\ 6\kappa^{-2}\Re\log\ccf &
\eqname\Ktilde\cr
\noalign{\vskip 5pt}
W(\modul,\matter)\ &
\mapsto\ \widetilde W(\ccf,\modul,\matter)\
    =\ \ccf^3\cdot W(\modul,\matter), &
\eqname\Wtilde\cr
\noalign{\vskip 5pt}
f_\ind(\modul)\ &
\mapsto\ \tilde f_\ind(\ccf,\modul)\
    =\ \ccf^0\cdot f_\ind(\modul). &
\eqname\Ftilde\cr
}
$$
\count255=\equanumber \advance\count255 by -2
\xdef\tildext{(\chapterlabel.\number\count255--\number\equanumber)}%
Note that the compensating superfields $\ccf$ and $\ccfb$ are non-dynamical
--- according to eq.~\Ktilde, their components appear in the extended
action without derivatives --- and play purely auxiliary roles, which are two:
To make the super-Weyl symmetry manifest,
and also to break it spontaneously down to nothing.
The breakdown is provided by $\vev\ccf\neq0$; the actual value of this VEV
is irrelevant since any $\ccf\neq0$ can
be super-Weyl-transformed into any desired function of $(x^m,\Theta^\alpha)$,
as long as it is chiral and does not vanish.
In particular, it is possible to set $\ccf\equiv1$ and completely
trivialize the extension~\tildext;
for other values of the $\ccf$ superfield, the action of the theory
has a somewhat different form, but the physical content remains exactly
the same --- identical to the unextended theory defined by eq.~\SUGRACT.

The super-Weyl symmetry may be combined with the general supercoordinate
invariance into the superconformal symmetry;
accordingly, the $\ccf$ and the $\ccfb$ superfields are often called
the superconformal compensators.
However, viewed by itself, the transformation
\superWeyl\&\ccfWeyl\ is an internal
spontaneously-broken supersymmetric abelian gauge symmetry with an unusual
feature that the vielbein components are charged under~\superWeyl.
Another unusual feature is that the gauge superfield associated
with this symmetry is not an independent vector superfield but a composite
$\widetilde K\bigl(\ccf,\ccfb,\modul,\modulb,
\matter,\matterb\bigr)$ of the scalar superfields of the theory.
Indeed, according to eqs.~\superWeyl, \ccfWeyl\ and \Ktilde, $\widetilde K $
transforms under  super-Weyl transformations as
$$
\widetilde K \ \mapsto\ \widetilde K \ +\ {6\over\kappa^2}\,
\bigl(\tau +\bar\tau \bigr) ,
\eqn\KWeyl
$$
\ie, exactly like an abelian vector superfield $V$.

The particular super-Weyl parameters $\tau$ and $\bar\tau$ that
reproduce the Weyl rescaling \weylmetric\ of the metric and appropriate
rescalings of the fermionic fields are
$$
\tau \,+\,\bar\tau \ \equiv\ -{\kappa^2\over6}\,
\left.K\bigl(\modul ,\modulb ,\matter ,\matterb \bigr)
	\right|_{\rm harmonic}\ ,
\eqn\tauWeyl
$$
where the subscript `harmonic' indicates that we are extracting the
harmonic part of the real superfield on the right hand side,
\ie, the components that can be put together into a sum of a chiral and an
anti-chiral superfields.
{}From the gauge-transformation point of view~\KWeyl, eq.~\tauWeyl\
corresponds to starting from the `unitary' gauge $\ccf\equiv1$
and changing it into the Wess-Zumino gauge for the composite $\widetilde K $
superfield:
$$
\displaylines{
\hfill \left.\widetilde K \right|_{\rm harmonic}\ =\ 0,
	\hfill \eqname\WZK \cropen{1\jot}
\ie,\qquad \log\ccf \ +\ \log\ccfb \
    =\ \coeff13\kappa^2\,\left.
	K\bigl(\modul ,\modulb ,\matter ,\matterb \bigr)
	\right|_{\rm harmonic}\,.\cr }
$$
Similarly to ordinary gauge symmetries, the
super-Weyl gauge $\ccf\equiv1$
is manifestly supersymmetric but not quite physical
while the Wess-Zumino gauge
\WZK\ is more physical but not quite supersymmetric.
{}From the EQFT point of view, this means that
the Wilsonian couplings of the theory
defined in the $\ccf\equiv1$ gauge have clear analytic properties
due to manifest SUSY --- for example,
the Yukawa couplings $Y_{IJK}(\modul)$ must be holomorphic
functions of the moduli $\modul^i$, ---
but the physical meaning of those couplings is obscured by the non-physical
normalization of the metric and of the fermionic fields.
On the other hand, in the WZ gauge~\WZK, the normalization of the matter
fields is clear, but the analytic properties of the Yukawa couplings
are obscured by the fact that the holomorphic $Y_{IJK}(\modul)$
have to be multiplied by a non-holomorphic factor $\exp(\half\kappa^2 K)$.
Similarly, the Wilsonian gauge couplings $g_\ind^{-2}$ are harmonic functions
of the moduli in the $\ccf\equiv1$ gauge,
but have more complicated moduli dependence
 in the Wess-Zumino gauge.
This effect is anomalous --- {\sl classically}, eq.~\couplingrel\ holds in
any super-Weyl gauge, --- and is best understood in terms of a manifestly
supersymmetric EQFT.
It is for the sake of such manifest SUSY that we are using the Weyl compensator
formalism in this article instead of the component-field Weyl rescaling
of the usual formalism.

\indent\hskip 0pt minus 3pt
Actually, there is an alternative manifestly supersymmetric
formalism which eliminates the Weyl rescaling altogether.
In the so-called \K\ supergravity formalism, the local Lorentz symmetry is
supplemented by an additional local $U(1)$ symmetry.\refmark{\BGG}
This enlargement of the superspace structure group relaxes the torsion
constraints
of the ordinary minimal supergravity, which in turn leads to an extension of
the super-Weyl symmetry \superWeyl.
Consequently, as far as the $e^{-\kappa^2\widetilde K/3}$ factor
in the supergravity action is concerned, it becomes possible to impose at the
superfield-level a gauge condition $\widetilde K=0$ instead of the
component-level Wess-Zumino gauge \WZK.
For the purpose of this article  we found
the compensator formalism makes for a more transparent
correspondence between local and rigid supersymmetries,
and so we use
this formalism throughout this article.
Obviously, any other correct
formalism should yield exactly the same physical
results.\refmark{\CLO}

\section{Supersymmetric Wilsonian Couplings of a Quantum EFT\brk
    and the Super-Weyl Anomaly.}
The Wilsonian couplings of an EQFT are the moduli-dependent
coefficients of quantum operator products appearing in the action functional
of the quantum theory.
\foot{At energies well below $\mpl$,
    neither moduli quanta nor gravitons nor their superpartners have any
    renormalizable interactions with each other or any other light particles.
    In the Wilsonian approach to EQFTs with a cutoff $\Lambda \lsim \mpl$,
    this means that all quantum operators
    involving those quanta are irrelevant and may be simply discarded from the
    action functional.
    Therefore, we treat the moduli and the gravitational superfields
    in the quantized version of the classical action~\SUGRACT\ as purely
    classical background fields; in the compensator formalism,
    the same applies to $\ccf$ and $\ccfb$.
    Functions of those superfields appear in the quantized actions as mere
    coefficients --- moduli-dependent coupling parameters --- of the
    quantum operator products comprised of the matter and gauge superfields
    $\matter^I$, $\matterb^\Ib$ and $V^{(a)}$ and their derivatives.
    Perturbatively, this means that $\modul^i$, $\ccf$ and the gravitational
    superfields may correspond to the external legs of 1PI Feynman graphs
    of the EQFT but not to their internal propagators.}
Similar to its classical counterpart, the Wilsonian action is spacetime-local,
\ie, it is a $\int\!d^4x$ integral
of a convergent power series in quantum fields and
their derivatives.
Consequently, manifest symmetries of an EQFT restrict the Wilsonian
couplings of the theory in exactly the same way as the classical couplings
are restricted by the classical symmetries.
In particular, for an EQFT that maintains manifest local SUSY,
manifest super-Weyl invariance and manifest background gauge invariance
with respect to the ordinary gauge symmetries,
the Wilsonian Yukawa couplings $\widetilde Y^W_{IJK}(\modul,\ccf)$
and the Wilsonian gauge couplings $\tilde f_\ind^W(\modul,\ccf)$
are required to be holomorphic functions of the chiral moduli $\modul^i$
and of the chiral compensator $\ccf$.
Moreover, the $\ccf$-dependence of $\widetilde Y^W_{IJK}$ and
of $\tilde f^W_\ind$ is completely determined
by the super-Weyl scaling dimensions of the corresponding terms in the
Wilsonian action functional.
The canonical scaling dimensions of those terms give us the classical eqs.\
\Wtilde\ and \Ftilde; the goal of this section is to see how these
equations are affected by the anomalous scaling dimensions characteristic
of the quantum theory.

When studying the moduli-dependence or the $\ccf$-dependence of the Wilsonian
couplings of an EQFT, one has to keep in mind that a complete description of
an EQFT includes not only its action functional but also
the ultraviolet cutoff scale $\Lambda$
and the exact manner in which the UV cutoff is implemented.
Actually, only the relevant terms in the action functional are important,
and the specifics of the cutoff can be absorbed into a finite renormalization
of those terms; nevertheless, the `bare' couplings of the theory have no
precise meaning unless the cutoff is fully specified.
In particular, {\sl it makes no sense to discuss moduli dependence
of the bare gauge coupling of the theory without also discussing moduli
dependence of the ultraviolet cutoff.}
Therefore, to give precise meaning to $\tilde f^W_\ind(\modul,\ccf)$
and other Wilsonian couplings of the theory, we {\it define} them as the
bare couplings of an explicitly UV-regulated theory whose cutoff does not
depend on either the moduli or on the Weyl compensators and can be completely
specified in terms of a single overall scale $\Lambda$.
The difference between these Wilsonian couplings and the
bare couplings of the same theory regulated in some other way is nothing but
a finite renormalization of the theory.

\ldf\JJW{I.~Jack and D.~R.~T.~Jones, \plt B258 (1991) 382;
    P.~West, \plt B258 (1991) 375.}
\ldf\SSCS{S.~Cecotti, S.~Ferrara, L.~Girardello and M.~Porrati
	\plt 164B (1985) 64.}
In a rigidly supersymmetric EQFT with a manifestly supersymmetric
ultraviolet cutoff, the Wilsonian superpotential does not renormalize.
\foot{In EQFTs with massless matter superfields,
   some two-loop contributions to the three-point Green's functions
   appear to renormalize the effective (but not the Wilsonian!) Yukawa
   couplings.\refmark{\JJW}
   The exact physical meaning of these two-loop effects is controversial
   since they have singular dependence on the off-shell momenta of the
   three particles involved.}
Moreover, if the cutoff also preserves manifest {\it four-dimensional}
background gauge invariance,
\foot{Note that a dimensionally-reduced background gauge invariance
    is insufficient.
    In the DR regularization scheme, not only the gauge couplings
    renormalize to all orders of perturbation theory, but the loop
    corrections do not even respect the classical $(N=1,d=4)$-supersymmetric
    relation $g^{-2}_\ind(\modul,\modulb)=\Re f_\ind(\modul)$.
    The reason for this failure is that the dimensionally-reduced gauge
    invariance allows for counter-terms that contain
    the Chern-Simons tensor $\omega_{\ell mn}=\tr\bigl(A_{[\ell}F_{mn]}-
    {2i\over3}A_{[\ell}A_mA_{n]}\bigr)$ in such a manner that they cannot
    be re-expressed in terms of $\tr(F\tilde F)$.
    Manifest supersymmetry does not forbid such counter-terms either;
    instead, it relates them to corrections to the bare gauge couplings
    $g_\ind^{-2}(\modul,\modulb)$ whose dependence on the moduli fields
    $\modul^i$ and $\modulb^\ib$ is non-harmonic.
    \refmark{\SSCS}
    }
then the Wilsonian supersymmetric gauge couplings $f^W_\ind$
renormalize only at the one-loop level.\refmark{\SVa,\nilles}
These no-renormalization theorems apply equally well to the
locally supersymmetric EQFTs (the proof of this assertion is presented
in  Appendix~A), which suggests the following strategy for calculating
the quantum corrections to eqs.\ \Wtilde\ and \Ftilde:

\pointbegin
Start with an explicitly UV-regulated EQFT whose bare action and the cutoff
are both manifestly invariant under local SUSY, as well as under other local
symmetries, namely Lorentz, background gauge and super-Weyl.
An example of such a cutoff is described in Appendix~B.

\point
Now consider the same theory with a different, $\ccf$-independent cutoff.
The new cutoff should remain locally supersymmetric, Lorentz invariant
and background gauge invariant,
but in order to allow for the $\ccf$-independence, we give up
on the super-Weyl invariance.

\point
Because of the manifest super-Weyl symmetry, the $\ccf$-dependence of the
bare couplings corresponding
to the {\sl first} cutoff is given by the unmodified classical eqs.~\tildext.
However, the Wilsonian couplings of the theory are the bare couplings
corresponding to the {\sl second} cutoff, and they differ from the first set
of bare couplings by a finite renormalization.

Now is the time to make use of the no-renormalization theorems.
Both cutoffs 1) and 2) being supersymmetric, the Yukawa couplings
$\widetilde Y_{IJK}(\ccf,\modul)$ do not renormalize at all and thus
take exactly the same values for both cutoffs.
Hence, {\it the Wilsonian Yukawa couplings obey unmodified eq.~\Wtilde:}
$$
{\widetilde Y}^W_{IJK}(\ccf,\modul)\ =\ \ccf^3\cdot Y^W_{IJK}(\modul).
\eqno\eq
$$
On the other hand,
the supersymmetric gauge couplings $\tilde f_\ind(\ccf,\modul)$ do renormalize,
but only at the one-loop level.
Thus, it becomes a straightforward exercise in superfield Feynman rules
to calculate the entire finite renormalization of those couplings due to
changing from a super-Weyl invariant cutoff to a $\ccf$-independent one;
we present this calculation in the Appendix~C.
In terms of the $\ccf$-dependence of the Wilsonian gauge couplings,
the result is
\foot{For matter-less supersymmetric Yang-Mills theories,
    a similar result was obtained
    in ref.~[\DFKZa] from a somewhat different point of view.}
$$
\tilde f_\ind^W(\ccf,\modul)\
=\ f_\ind^W(\modul)\ +\ {3\cano_\ind\over 8\pi^2}\,\log\ccf \,,
\eqn\Wilsanom
$$
where
$$
\cano_\ind\ =\ \sum_r n_r T_\ind(r)\ -\ T(G_\ind).
\eqn\canodef
$$
Here the sum is over representations $r$ of the gauge group $G$,
$n_r$ is the number of multiplets of $\matter^I$ transforming like $r$,
$T_\ind(r)=\Tr_r\bigl(T_{(a)}^2\bigr)$ for $T_{(a)}\in G_\ind$ and
$T(G_\ind)$ stands for $T_\ind({\rm adjoint\ of}\ G_\ind)$.
After fixing the WZ gauge \WZK, the $\log\ccf$ terms
in eqs.~\Wilsanom\ translate into K\"ahler terms for
the Wilsonian couplings of the ordinary gauge fields:
$$
(g^W_\ind)^{-2}(\modul,\modulb;{\rm WZ})\
\equiv \Re \tilde f_\ind^W\
=\ \Re f_\ind^W(\modul)\ +\ \cano_\ind\,{\kappa^2\over
16\pi^2}\,K(\modul,\modulb),
\eqn\gWilson
$$
which have a rather different dependence on the moduli fields than the
classical couplings $g_\ind^{-2}=\Re f_\ind(\modul)$.
(Note that only the $O(\mpl^2)$ part of $K$ is relevant in this formula, so
it does not pay to discriminate between the
classical and the quantum K\"ahler functions or between $\Kmod(\modul,\modulb)$
and $K(\modul,\modulb,\matter,\matterb)$.)

Technically, the origin of the $\log\ccf$ term in eq.~\Wilsanom\ is in the
unavoidable $\ccf$-dependence of the super-Weyl invariant cutoff.
{}From the Wilsonian EQFT point of view, this means that the super-Weyl
invariance
of the quantum theory is anomalous (there is no field independent
cutoff that respects the super-Weyl symmetry),
but it can be restored with the help
of explicit local counter-terms
such as the $\log\ccf$ term in $\tilde f^W_\ind$.
In that sense
the $\log\ccf$ can be viewed as  a local Wess-Zumino counter-term
cancelling the potential super-Weyl anomaly.
Like other anomalies, the  anomaly of the super-Weyl symmetry
can be calculated in many ways, and one does not have to rely on the particular
cutoff described in Appendix~B to reproduce eq.~\Wilsanom.
We  conclude this section of the article with an alternative
derivation of eq.~\Wilsanom\ based upon the  Adler-Bell-Jackiw
anomaly of the axial symmetry of charged fermions.

The super-Weyl transformations~\superWeyl\ contain an R-symmetry,
under which the scalar, the vector and the gravitational fields
of the theory remain invariant, but the spinor fields and some of the
auxiliary fields change their phases.
Indeed, for $\tau$ that is purely bosonic, imaginary and constant,
$\tau(z) \equiv i\nu$,
the matter fermions and the gauginos transform like
$$
\Psi^I_\alpha\ \mapsto\ e^{+3i\nu}\cdot\Psi^I_\alpha\,,\qquad
\lambda^{(a)}_\alpha\ \mapsto\ e^{-3i\nu}\cdot\lambda^{(a)}_\alpha
\eqn\Rsymmetry
$$
(\cf\ eqs.~\matterWeyl\ and \gauginoWeyl).
As far as the gauge fields are concerned, \Rsymmetry\
is a classically-exact but anomalous axial symmetry.
Moreover, this R-symmetry
is spontaneously broken by $\vev\ccf\neq0$, whose phase plays a role
of a Goldstone boson, so its anomaly translates into  anomalous effective
couplings
$$
\sum_\ind{-3\cano_\ind\over 32\pi^2} \arg(\ccf)\cdot
(F_{mn}\tilde F^{mn})_\ind
\eqn\ABJanomaly
$$
of the phase of $\ccf$ to the gauge fields;
note that the coefficients $\cano_\ind$ of these couplings are
exactly as in eq.~\canodef.
The couplings \ABJanomaly\ are generated by the one-loop graphs
$$
\checkex Triangle.eps
\iffigureexists
\setbox0=\vbox{\epsfxsize 2in \epsfbox{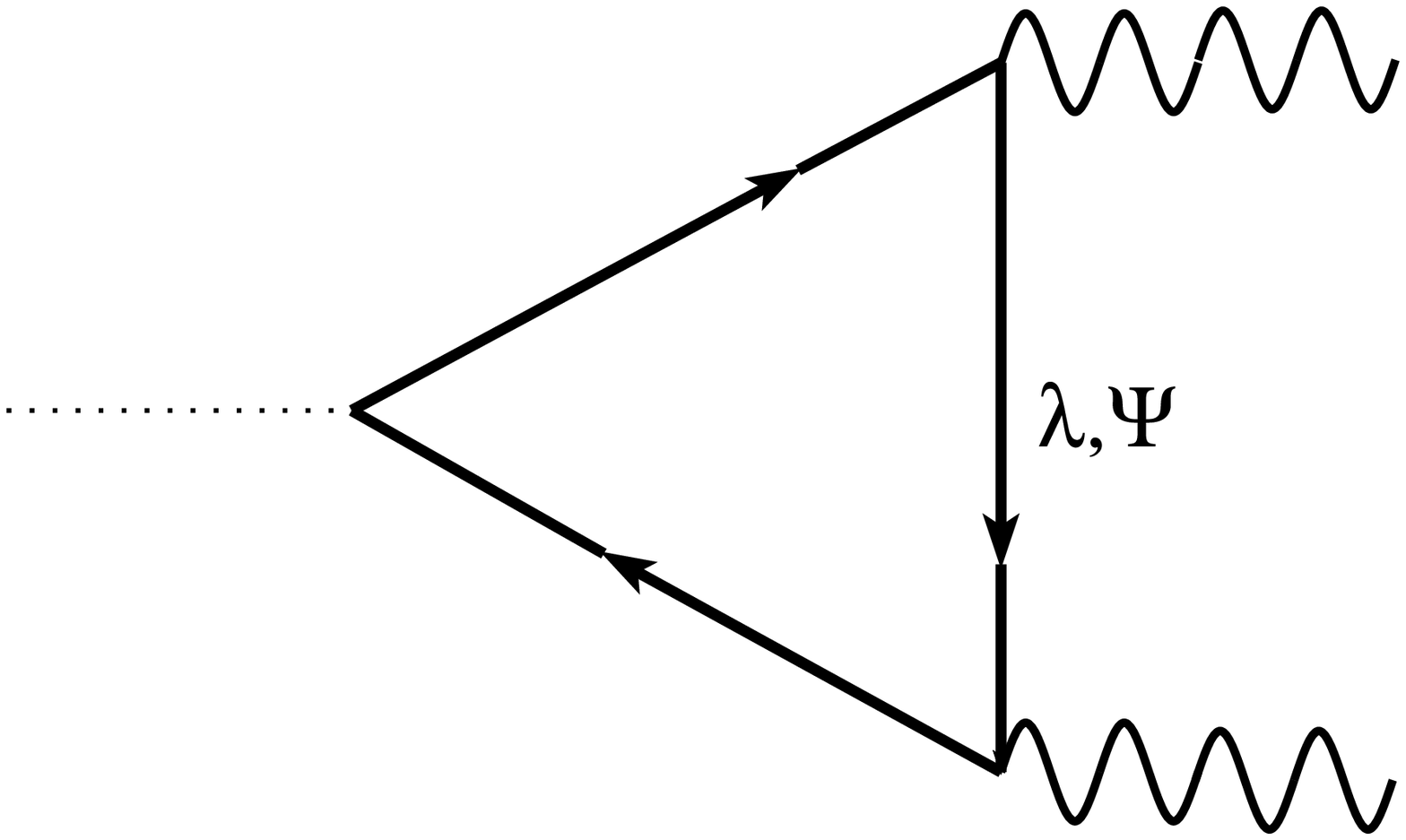}}
\dimen0=\ht0 \advance\dimen0 \dp0
\ccf \vcenter{\kern 4pt \box0}
\vcenter to \dimen0 {\hbox{$F_{mn}^{(a)}$}\vss\hbox{$\tilde F^{mn\,(a)}$}}
\else \missbox{Missing Diagram}\fi
\eqn\ABJtriangle
$$
and are completely analogous to the anomalous effective coupling responsible
for the $\pi^0\to\gamma\gamma$ decay in the standard model.
However, unlike $\pi^0$,
$\arg(\ccf)$  does not corresponds to any physical particle
that can decay, propagate or have {\sl any} interactions whatsoever.
In fact, any physical coupling of the Weyl compensator would
destroy the super-Weyl symmetry and prohibit the Weyl rescaling, leaving
the theory with an irreparably non-Einsteinian gravity.
To avoid such a calamity, the anomalous contribution of
the triangle graphs~\ABJtriangle\ must be cancelled by an explicit counter-term
in the Wilsonian action of the EQFT:
instead of the original $\theta_\ind^W(\modul,\modulb)$ --- the
moduli-dependent
coefficients of the $(1/16\pi^2)\,\tr_\ind F\tilde F$ operators
--- we should substitute
$$
\tilde\theta_\ind^W(\ccf,\ccfb,\modul,\modulb)\
=\ \theta_\ind^W(\modul,\modulb)\ -\ 3\cano_\ind\,\arg(\ccf) .
\eqn\thetaW
$$
Note that according to the Adler-Bardeen theorem, these counter-terms
cancel not just the one-loop Feynman diagrams~\ABJtriangle\ but
also all the higher-loop corrections to the anomalous couplings \ABJanomaly.

Formula~\thetaW\ should hold in any background gauge invariant
formulation of the theory, manifestly supersymmetric or not.
However, given manifest SUSY, we can identify
${-1\over 8\pi^2}\theta_\ind^W$ with the imaginary
part of the supersymmetric gauge couplings $f_\ind^W$,
which must be holomorphic functions of the chiral superfields;
for $\tilde f_\ind^W$ this includes holomorphic dependence on
the compensating superfield $\ccf$.
Hence, there is only one way to obtain  $\tilde\theta_\ind^W$ from a
supersymmetric gauge coupling $\tilde f_\ind^W(\ccf,\modul)$;
the explicit form of such $\tilde f_\ind^W$ is given in eq.~\Wilsanom.

Clearly, there is more to the super-Weyl symmetry~\superWeyl\ than
just the axial symmetry~\Rsymmetry, and in general, one should
consider the anomalies of all the currents of \superWeyl.
However, in a manifestly supersymmetric EQFT, all such currents
are superpartners of each other and form a single supercurrent
suffering from a single `super-axial'  anomaly.
The easiest way to compute this  anomaly and its effects
is to notice that there is
a unique gauge-invariant rigid-superfield expression whose bosonic part
includes the coupling \ABJanomaly\  of $\Im(\ccf)$ to $F\tilde F$,
namely
$$
\sum_\ind{-3\cano_\ind\over 32\pi^2} \int\!\!d^2\Theta\,
\log\ccf\cdot (W^\alpha W_\alpha)_\ind\ +\quad {\rm h.~c.};
\eqn\superABJ
$$
extending this expression to local SUSY is completely straightforward.

Before we discuss the relationship between the Wilsonian gauge couplings
and the physical gauge couplings of an EQFT,
let us summarize the important points of this section.
We established that the moduli dependent Wilsonian couplings
of a supersymmetric EQFT have to be defined with respect
to a moduli independent cutoff $\Lambda$.
Such a cutoff renders the super-Weyl symmetry anomalous.
However, this super-Weyl anomaly can be cancelled by an
explicit local Wess-Zumino counter-term, which introduces a
compensator dependence into the Wilsonian gauge couplings as specified by
eq.~\Wilsanom.
The coefficient of the counter-term can be computed rather simply
by utilizing the axial $U(1)$ part of the super-Weyl symmetry which
re-phases the fermions of the theory.

\chapter{Physical Gauge Couplings\break
    and the Perturbative Effects of the Super-Weyl Anomaly}
\section{Effective Gauge Couplings in Massless EQFTs.}
The Wilsonian couplings we have discussed so far are the parameters
of a local Wilsonian action of an EQFT, from which the high-energy degrees
of freedom have been integrated out, but the relevant low-energy degrees of
freedom are still present in the matter and gauge superfields $\matter^I$,
$\matterb^\Ib$ and $V^{(a)}$.
In this section, we turn our attention to the physical  effective
couplings --- the parameters of the S-matrix or of the closely-related
generating functional, which results from integrating out
{\sl all} the quantum degrees of freedom, both high-energy and low-energy.
(Perturbatively, the generating functional is the sum of  all
1PI Feynman graphs).
For example, given a background-covariant generating functional, the easiest
definition of the effective gauge coupling $\{g_\ind\}$
\foot{Following the notations of ref.~[\DKLb], we use the curly braces
    `$\{\ \}$' to distinguish the effective couplings of a quantum
    theory from its Wilsonian couplings or their classical counterparts.
    In ref.~[\SVa], the square brackets `[\ ]' were used for the same
    purpose.}
is via the two-point function
$$
{\cal A}\bigl(A_m^{(a)}(-p),A_n^{(a)}(p)\bigr)\
=\ (p^2 g_{mn}-p_mp_n)\,{1\over\{g_\ind(p)\}^2}
\eqn\geffdef
$$
for two un-renormalized gauge fields belonging to $G_\ind$.
Generally, the effective couplings are momentum dependent (and thus are often
called the ``running effective couplings''), and this dependence is
often non-local, \ie, becomes singular when some of the relevant momenta
approach zero.
Apart from possible thresholds,
the dependence of the effective couplings
on the overall momentum scale $p$ is logarithmic and  governed
by the Callan-Symanzik equations, such as
$$
\partder{\{g_\ind(p)\}}{\log p}\ =\ \beta_\ind(\{g\},\ldots)\
=\ {\cren_\ind\{g_\ind\}^3\over 16\pi^2}\ +\ {\secondb_\ind\{g_\ind\}^5\over
256\pi^4}\
+\ \cdots .
\eqn\CS
$$
Here
$$
\cren_\ind\ = \sum_r n_r T_\ind(r)\ -\ 3 T(G_\ind) ,
\eqn\crendef
$$
whereas the expressions for the $\secondb_\ind$ are more complicated
and also depend on the Yukawa couplings $\{Y_{IJK}\}$.
(The higher-order terms  also depend on the
choice of the renormalization scheme.)

\ldf\NSVZ{V.~Novikov, M.A.~Shifman, A.I.~Vainshtein and V.I.~Zakharov,
\nup 229 (1983) 381.}
Renormalization of the effective gauge couplings of a rigidly
supersymmetric EQFT was studied by Novikov, Shifman, Vainshtein
and Zakharov,\refmark{\NSVZ, \SVa}
who found exact relations between $\beta_\ind$ and the wave-function
renormalization of the charged matter fields $\matter^I$.
These relations are integrable and lead to
$$
\displaylines{
F_\ind\ \equiv\ \{g_\ind(p)\}^{-2}\
    -\ {T(G_\ind)\over 8\pi^2}\,\log\{g_\ind(p)\}^{-2} \hfill
\eqname\SV\cr
\hfill{} +\ \sum_r {T_\ind(r)\over 8\pi^2}\,\log\det\{Z^{(r)}(p)\}\
    +\ {\cren_\ind\over 8\pi^2}\,\log p\qquad\cr }
$$
being exact integrals of the renormalization group, ${d F_\ind \over dp}=0$;
the $\{Z^{(r)}(p)\}$ matrix here is the diagonal block of
the renormalized kinetic-energy matrix $\{Z_{\Ib J}(p)\}$
for the matter fields $\matterb^\Ib,\matter^J$
that transform like $r$ under $G$.
The integrals $F_\ind$ of the Callan-Symanzik equations for the effective gauge
couplings are related to the invariants of the Wilsonian renormalization
group for  $f^W_\ind$, and because the Wilsonian supersymmetric gauge
couplings do not renormalize beyond one loop, the relation reads simply
$$
F_\ind\ =\ \Re f_\ind^W\ +\ {\cren_\ind\over 8\pi^2}\,\log\Lambda ,
\eqn\rigidSV
$$
{}$\Lambda$ being the  ultraviolet cut-off scale.
Note that the effective gauge couplings $\{g(p)\}$ generally
do renormalize to all orders, including the two-loop order responsible
for the $\secondb_\ind\{g\}^5$ terms in eqs.~\CS;
correspondingly, eqs.~\SV\ have logarithmic singularities
as $\{g_\ind\}\to 0$.

\ldf\VKa{V. Kaplunovsky, in preparation.}
\ldf\ANT{I.~Antoniadis, K.~Narain and T.~Taylor, \plt B267 (1991) 37;\brk
    I.~Antoniadis, E.~Gava and K.~Narain,  \nup383 (1992) 93.}
\ldf\FKLZ{S.~Ferrara, C.~Kounnas, D.~L\"ust and F.~Zwirner,
	\nup365 (1991) 431.}
\ldf\MSr{P.~Mayr and S.~Stieberger, \nup 407 (1993) 725
	and Max Planck preprint  MPI--Ph/93-14.}
\ldf\GT{M.~K.~Gaillard and T.~Taylor, \nup 381 (1992) 577.}
Eqs.~\SV\ and \rigidSV\
can be used to relate the dependence of $\{g_\ind(p;\vev\modul,\vev\modulb)\}$
on the moduli VEVs to the moduli-dependence
of the Wilsonian couplings $f_\ind^W(\modul)$.
The UV cutoff scale $\Lambda$ is field-independent, and
the holomorphicity of $f^W_\ind(\modul)$ translates
via eqs.~\SV\ into a very specific non-harmonicity
of the moduli-dependence of the effective gauge couplings, which is
further related by SUSY to very specific non-integrability of the
effective axionic couplings of the moduli scalars.\refmark{\DKLb,\VKa}
The analysis of ref.~[\SVa]
was limited to rigidly supersymmetric EQFTs;
its extension to locally supersymmetric theories is the main subject
of this section.
\foot{Non-harmonic gauge couplings were first encountered in
    string theory in ref.~[\DKLb] and further expanded
    upon in refs.~[\ANT--\MSr].
    They also appear in the context  of the linear multiplet; this aspect
    is discussed in refs.~[\DFKZa,\CLO,\ABGG,\GT].
    A preliminary version of this article was presented in ref.~[\JLpascos].}

The first step towards such an extension is to notice that as far as the
renormalizable interactions of a low-energy EQFT are concerned,
there is no difference between local and rigid supersymmetries.
Hence, as long as the spacetime background of a locally supersymmetric
EQFT is flat,
the Callan-Symanzik equations for the relevant couplings of the theory
are identical to those of its rigidly supersymmetric counterpart.
In particular,  eqs.~\SV\ hold verbatim ---
$F_\ind$ are exact integrals of the renormalization group, \ie,
do not depend on the momentum scale~$p$.

Extending eqs.~\rigidSV\ to locally supersymmetric theories is more tricky:
relations involving both physical and Wilsonian couplings would carry over
only if the background of a locally supersymmetric theory is both physically
and formally compatible with rigid SUSY.
In other words, we need both the physical spacetime metric $g_{mn}$
and the superspace's vielbein $E_M^A$ to be flat at the same time:
$$
g_{mn}(x)\ =\ \eta_{mn}\qquad{\bf and}\qquad
E^A_M(x,\Theta,\bar\Theta)\ =\ E^A_M[{\rm flat}](\Theta,\bar\Theta)\, .
\eqn\simulflat
$$
Otherwise the relation between the Wilsonian couplings of the globally
supersymmetric and the locally supersymmetric theory are modified by the
super-Weyl anomaly.
In the compensator formalism, the two conditions~\simulflat\ are compatible
only in the Wess-Zumino gauge~\WZK\ for the super-Weyl symmetry.
Hence,
$$
\eqalign{
F_\ind(\modul,\modulb)\ &
=\ (g^W_\ind)^{-2}(\modul,\modulb;{\rm WZ})\
    +\ {\cren_\ind\over 8\pi^2}\,\log\Lambda\cr
&=\ \Re f_\ind^W(\modul)\ +\ {\cren_\ind\over 8\pi^2}\,\log\Lambda\
    +\ \cano_\ind\,{\kappa^2\over 16\pi^2}\,K(\modul,\modulb)\cr
}\eqn\Katerm
$$
(\cf\ eq.~\gWilson)
or, in terms of the super-Weyl covariant Wilsonian couplings \Wilsanom\
and \Ktilde,
$$
F_\ind\ =\ \Re\tilde f^W_\ind\ +\ {\cren_\ind\over 8\pi^2}\,\log\Lambda\
+\ \cano_\ind\,{\kappa^2\over 16\pi^2}\,\widetilde K
\eqn\KFterms
$$
(the $\ccf$-dependent terms of the super-Weyl invariant couplings
$\tilde f^W_\ind$ and  $\widetilde K$ cancel in this equation).

The argument we have just used to derive eqs.~\Katerm\ and \KFterms\
is rather heuristic and may appear to the reader as not too trustworthy.
To eliminate all doubts, consider the origin of possible supergravitational
corrections to the rigid-SUSY formula~\rigidSV.
The only source of these corrections is in the anomalies of the super-Weyl
rescaling we discussed in the last section, and like the ordinary axial
anomalies, these are
exhausted at the one-loop level.
Therefore, what we need to do is to calculate the effective gauge couplings
$\{g_\ind(p)\}$ to the one-loop order and to interpret the result as the
relation between $F_\ind$ and the Wilsonian gauge couplings $f_a^W$.
If this one-loop calculation agrees with eqs.~\Katerm\ and \KFterms,
then they automatically hold true to all orders of the perturbation theory
and no further verification of their validity would be necessary.

The most direct way of calculating $\{g_\ind (p)\}$ in terms of the
manifestly super\-symmetric Wilsonian couplings of the theory
is to provide the theory with a super\-symmetric, \etc, ultraviolet cutoff
and then use superfield Feynman rules; this calculation is presented
in the Appendix~C.
Alternatively, similar to the derivation of eq.~\superABJ\ from the axial
anomaly
of the R-symmetry, we can relate the moduli dependence of $\{g_\ind\}$ to
the effective axionic couplings of the moduli scalars
to the CP-odd combinations of the gauge bosons.
At the one-loop level, those effective axionic couplings can be easily computed
using component-field Feynman rules and a background-covariant UV cutoff
(however, there is no need for the cutoff to be supersymmetric, \etc).
The rigid-SUSY analogue of this calculation is presented in ref.~[\VKa];
the locally supersymmetric case is discussed in
refs.~[\JLpascos--\CLO] and also
presented in Appendix~D of this article.
This calculation of $\{g_\ind\}$ is less direct, but it is also less
formalism dependent;
needless to say, the results of both calculations confirm eqs.~\Katerm\
and \KFterms.

\section{\K\ Transformations and Global Symmetries.}
Describing non-gauge interactions of scalar superfields in terms of
a separate real-analytic K\"ahler function $K(\modul,\modulb,\matter,\matterb)$
and a separate complex-analytic superpotential $W(\modul,\matter)$
is somewhat redundant.
Classically, the so-called K\"ahler transformations\refmark{\CFGP-\GGRS}
$$
\eqalignno{
K(\modul,\modulb,\matter,\matterb)\ &
\to\ K(\modul,\modulb,\matter,\matterb)\ +\ {\cal J}(\modul,\matter)\
	+\ {\cal J}^*(\modulb,\matterb) ,&
\eqname\KahlerK \global\count255=\equanumber\cr
W(\modul,\matter)\
&\to\ W(\modul,\matter)\cdot
	\exp\left(-\kappa^2{\cal J}(\modul,\matter)\right),
&\eqname\KahlerW
 \xdef\Kahlerdef{(\chapterlabel.\number\count255--\number\equanumber)}\cr
f_\ind(\modul,\matter)\
&\to\ f_\ind(\modul,\matter),
&\eqname\KahlerF
 \xdef\Kahlerfull{(\chapterlabel.\number\count255--\number\equanumber)}\cr }
$$
leave all physical quantities invariant.
In the compensator formalism, these transformations are accompanied by the
$\ccf$ rescaling\refmark{\compensator}
$$
\ccf\ \to\ \ccf\cdot \exp\left(\coeff13\kappa^2
	{\cal J}(\modul,\matter)\right) ,
\eqn\ccfKahler
$$
which renders the functions $\widetilde K$, $\widetilde W$ and $\tilde f_\ind$
defined in eqs.~\Ktilde--\Ftilde\ --- and thus the action ---
manifestly invariant at the superfield level.
(Note that \ccfKahler\ is not a part of any super-Weyl transformation
--- the gravitational superfields are inert under \K\ transformations).
The $\cal J$ in eqs.~\KahlerK\ through \ccfKahler\
is an arbitrary holomorphic function of the
chiral superfields $\modul^i$ and $\matter^I$, and `holomorphic' here
really means complex-analytic {\sl and non-singular}.
Hence, according to eq.~\KahlerW, the only K\"ahler-invariant features
of the superpotential $W$ are the location of its zeros; a superpotential
without zeros can be completely K\"ahler-transformed away into a harmonic
term in ${\cal G}\equiv\kappa^2 K+\log|W|^2$ (\cf\ ref.~[\CFGP]).
However, the superpotential of an effective field theory usually has
continuous varieties of zeros (including $W(\modul,\matter=0)=0$ for all values
of the moduli $\modul^i$), so in spite of the K\"ahler invariance of the
theory, we have to describe it in terms of separate $W$ and $K$,
even classically.

For locally supersymmetric EQFTs there is an even better reason
to treat $W$ and $K$ as separate functions, namely the K\"ahler anomaly
of the Wilsonian gauge couplings $f^W_\ind $.
This anomaly is fairly obvious in the compensator formalism:
Combining the demand for K\"ahler-invariance of the action with
eqs.~\Ftilde\ for the Wilsonian gauge couplings, we immediately see that
the classical transformation  law
\KahlerF\ cannot be maintained at the quantum level. Instead,
eq.~\KahlerF\
has to be replaced with
$$
{\tilde f^W_\ind\ \to\ \tilde f^W_\ind}\quad \Longleftrightarrow\quad
{f^W_\ind\ \to\ f^W_\ind\ -\ {\cano_\ind\over 8\pi^2}\,\kappa^2{\cal J}} .
\eqn\FWKahler
$$
It is worth noting that this new transformation law
can also be inferred from eqs.~\SV\ and~\Katerm.
The effective gauge couplings $\{g_\ind(p)\}$ are physical quantities and
thus remain invariant under all symmetries of an EQFT.
The renormalized kinetic-energy
matrices $\{Z_{\Ib J}(p)\}$
do not change unless the matter fields
themselves are rescaled or mixed with each other.
No such mixing or rescaling is involved in the K\"ahler transformations,
\foot{Not even implicitly, through a Weyl rescaling, because the physical
    super-Weyl gauge \WZK\ is preserved by the combined effect of
    eqs.~\KahlerK\ and \ccfKahler.}
so a quick glance upon the right hand side of eq.~\SV\ tells us that $F_\ind$
are K\"ahler invariant.
An equally quick glance at the right hand side of eq.~\Katerm\ then tells
us that $f_\ind^W$ cannot possible be K\"ahler invariant but must instead
transform in accordance with eq.~\FWKahler.
\foot{In refs.~[\DFKZa,\CLO], a different mechanism was proposed for
    cancelling the \K\ anomaly of the gauge couplings:
    Instead of the Wess-Zumino $\log\ccf$ terms in $\tilde f^W_\ind$,
    one can couple the gauge superfields to a linear multiplet
    with appropriate transformation properties.
    Generically, this mechanism needs a separate linear multiplet
    for each independent gauge coupling.}

By themselves, the K\"ahler transformations~\Kahlerdef\ do not constitute
a physical symmetry of an EQFT since they leave all the physical fields
unchanged.
However, they often appear in the context of
a global non-linear internal symmetry of an EQFT, which transforms
the chiral superfields $\modul^i$, $\matter^I$ into some holomorphic
functions of each other while leaving
the superpotential and the K\"ahler function of the theory invariant
modulo a K\"ahler transform, that is,
$$
\eqalignno{
K(\modul',\modulb',\matter',\matterb')\ &
=\ K(\modul,\modulb,\matter,\matterb)\
    +\ {\cal J}(\modul,\matter)\ +\ {\cal J}^*(\modulb,\matterb),&
\eqname\symmK\cr
W(\modul',\matter')\ &
=W(\modul,\matter)\cdot\exp\bigl(-\kappa^2{\cal J}(\modul,\matter)\bigr)&
\eqname\symmW\cr }
$$
for some holomorphic ${\cal J}(\modul,\matter)$.
In addition, a true symmetry should also preserve the gauge charges
of all chiral superfields (\ie, commute with the gauge group) and preserve
the classical gauge couplings:
$$
f_\ind(\modul',\matter')\ =\ f_\ind(\modul,\matter).
\eqn\symmClF
$$

Again, eqs.~\symmClF\ are subject to anomalous corrections.
In fact, there are two anomalies at play here: the K\"ahler anomaly~\FWKahler,
and the ordinary Konishi anomaly (present in global supersymmetry)
due to redefinition of the charged matter
superfields.\refmark{\konishi}
Both anomalies can be obtained from direct superfield calculations,
but by far the easiest way of extracting their combined effect is
through eqs.~\SV\ and \Katerm.
For simplicity, consider a symmetry that is
linear with respect to the charged matter fields,
or at least becomes linear in the $|\matter|\ll\mpl$ limit:
$$
\matter^I\ \mapsto\ \matter^{\prime I}\
=\ \symmat^I_{\mkern 9mu J}(\modul)\,\matter^J\ +\ O(\matter^2/\mpl).
\eqn\mattertrans
$$
The effect of such a symmetry on the two-point Green's functions for
the matter fields is obvious: In terms of
$\{Z_{\Ib J}(p;\vev\modul,\vev\modulb)\}$,
we have, in matrix notations,
$$
\{Z(p;\modul',\modulb')\}\ =\ \bigl(\symmat^\dagger(\modulb)\bigr)^{-1}
\cdot \{Z(p;\modul,\modulb)\}\cdot \bigl(\symmat(\modul)\bigr)^{-1} .
$$
An exact symmetry of an EQFT should leave
$\{g_\ind(p;\vev\modul,\vev\modulb)\}$ invariant;
hence, according to eq.~\SV,  $F_\ind$ should transform according to
$$
F_\ind(\modul',\modulb')\ =\ F_\ind(\modul,\modulb)\
-\sum_r^{\rm matter}{T_\ind(r)\over 8\pi^2}\,
	\log\left|\det \symmat^{(r)}(\modul)\right|^2.
\eqn\symmBigF
$$
It remains to substitute eqs.~\symmBigF\ and \symmK\ into eq.~\Katerm\
and make use of the holomorphicity of $f^W_\ind(\modul)$; this implies
$$
f^W_\ind(\modul')\ =\ f^W_\ind(\modul)\
-\ {\cano_\ind\over 8\pi^2}\,\kappa^2{\cal J}(\modul)\
- \sum_r^{\rm matter}
{T_\ind (r)\over 4\pi^2}\,\log\det \symmat^{(r)}(\modul) .
\eqn\fLocal
$$

The power of the anomalous symmetry relations \fLocal\ is that
they impose analytic constraints on a holomorphic function
$f^W(\modul)$ relating its values at distant points in the moduli space.
\def\MS{M_0}
To illustrate the power of such constraints, consider what happens
when the global symmetry group $\bf S$ is discrete but big enough
to assure the compactness of $\MS/{\bf S}$ (we denote the moduli space by
$\MS$).
Classically, this immediately implies moduli-independence of all the
supersymmetric gauge couplings $f_\ind$:
An {\bf S}-invariant holomorphic function $f_\ind(\modul)$ is the
same as a single-valued holomorphic function on the compact complex manifold
$\MS/\bf S$, and therefore has to be  constant.
In a quantum theory, the anomalous right hand side of eq.~\fLocal\
guarantees that $f^W_\ind$ do depend non-trivially on the moduli fields;
however, the exact form of such dependence is completely determined
by the symmetry constraints.
Indeed, were there two holomorphic functions $f_1(\modul)$ and $f_2(\modul)$
satisfying the same anomalous symmetry relation of the form~\fLocal, then the
difference $f_1-f_2$ would be a holomorphic function that is invariant under
all the symmetries.
Note that although the above argument assumed that the moduli space becomes
compact after symmetry-related points are identified with each other,
the same result is obtained
 when such space is not quite compact, but some
physical reasons limit the growth of $f(\modul)$ along the non-compact
directions.
Thus, {\sl given a sufficiently big discrete symmetry group
\rm and some physical limits on $g_\ind^{-2}$,
\it the exact analytic form of the moduli dependence of the Wilsonian gauge
couplings can be completely determined from the anomalous symmetry relations
\fLocal,} up to moduli-independent overall constants.

\ldf\BCOV{M.~Bershadsky, S.~Cecotti, H.~Ooguri and C.~Vafa, \nup 405 (1993)
    279 and Harvard preprint HUTP--93--A025.}
One can show\refmark{\JLpascos, \DFKZa, \BCOV, \KLc}
that for many vacuum families
of the heterotic string, symmetry relations~\fLocal\ indeed completely
determine the moduli dependence of the $f^W$, or at least its dependence
on some of the moduli.
For example, for  orbifolds which have exact `duality' symmetries,
the Wilsonian gauge couplings $f^W_\ind$ depend on
the toroidal moduli $T^i$ of the orbifold according to
$$
f^W_\ind(T)\ = \sum_i {\alpha^\ind_i\over 8\pi^2}\,\log\eta(iT^i)\
+\ \rm const
\eqn\orbiform
$$
where $\alpha^\ind_i$ are integer coefficients computable at the tree level of
the string and the `constant' may be a function of other moduli,
but not of the $T^i$.
All one needs to know to derive eq.~\orbiform\
is the spectrum of the massless charged particles
of the theory and their classical couplings to the moduli $T^i$
and no string loop calculations are necessary,
although the result (including the coefficients $\alpha^\ind_i$) always agrees
with that of the one-string-loop calculation of ref.~[\DKLb].

\section{EQFT with a Threshold.}
In the previous discussion of the effective couplings we have assumed
that all light charged particles are exactly massless.
Now consider what happens when some gauge or matter particles do have
small masses of the order $M_I\ll\mpl$.
Clearly, in the energy-momentum range $M_I\ll p\ll\mpl$ we can
describe all interactions in terms of a massless EQFT while for
$p\ll M_I$ we can use another EQFT in which only the truly
massless particles appear as quantum fields.
Formally, the lower-energy EQFT can be obtained from the higher-energy
EQFT by integrating out the massive fields.
The goal of this subsection is to show the effect of integrating out
the massive modes
for the gauge couplings of  locally supersymmetric theories.

We presume that some part of the gauge symmetry  is
spontaneously broken at the threshold scale $M_I$ but
the supersymmetry itself remains unbroken.
Consider the renormalization of the effective gauge couplings
$\{g'_\ind(p)\}$ corresponding to simple factors $G'_\ind$ of the group $G'$
of the gauge symmetries that remain unbroken below the threshold.
Modifying the Shifman-Vainshtein arguments in order to account for the
$O(M_I)$ masses of some of the matter and gauge superfields, we find that
the exact renormalization-group integrals of the massive theory are
$$
\eqalign{
F^M_\ind\ \equiv\, {}&
\{g'_\ind(p)\}^{-2}\,
    +\sum_{r'} {T_\ind(r')\over 8\pi^2}\left(
	\log\det\{Z^{(r')}(p)\}
	-n_{r'}({\rm gauge})\log\{g_{r'}(p)\}^{-2}\right)\cr
&+\sum_{r'}{T_\ind(r')\over 8\pi^2}\left(
	\Tr\log h(p,{\cal M}^{(r')}_{\rm matter})
	-3\Tr\log h(p,{\cal M}^{(r')}_{\rm gauge})
	\right) ,\cr
}\eqn\MassiveF
$$
where the sums are over the irreducible representations $r'$ of the
unbroken gauge symmetry $G'$, $n_{r'}({\rm gauge})$ are the numbers
of the $r'$ multiplets formed by the gauge fields, massive or massless,
$\{g_{r'}\}$ are the gauge coupling associated with those gauge fields,
${\cal M}^{(r')}_{\rm matter}$ and ${\cal M}^{(r')}_{\rm gauge}$ are,
respectively,
the canonically-normalized mass matrices for the matter and gauge
superfields transforming like $r'$ under $G'$
and $h(p,{\cal M})$ is a complicated homogeneous
function of the momentum $p$ and the mass $\cal M$.
Fortunately, $h$ is only complicated near the threshold $p\sim\cal M$;
well-below the threshold $h(p\ll{\cal M})\approx {\cal M}$
while well above the threshold $h(p\gg{\cal M})\approx cp$,
$c$ being a numerical constant of order 1 whose value depends on
details of the renormalization scheme used to define the running gauge
couplings $\{g'_\ind(p)\}$.
For the renormalization scheme \geffdef, $c=e^{-1}$;
nevertheless, for the sake of notational simplicity,
we will henceforth put $c=1$; modifying all the formul\ae\ in this section
to accommodate $c\neq1$ is completely straightforward.

For simplicity, let us assume that each $G'_\ind$ is wholly contained in
a single simple factor $G_\ind$ of the full gauge group $G$, although
one can have the same $G_\ind$ for several $G'_\ind$ (as in GUTs).
For momenta $p$ so far above the threshold scale $M_I$ that the masses become
negligible, the last term in eq.~\MassiveF\ reduces to
${1\over 8\pi^2}\cren_\ind\log p$.
At the same time, we have $\{g_{r'}\}=\{g_\ind\}$ for all gauge bosons
belonging to $G_\ind$ (and for all others $T_\ind(r')\equiv0$)
while for the matter fields
$\{Z^{(r')}\}=\{Z^{(r)}\}$ for all $r'$ representing pieces of the
same multiplet $r$ of $G$.
Let us assemble all thus related terms in eq.~\MassiveF\ together;
comparing the result with the massless eq.~\SV, we immediately see that
$$
F^M_\ind\ =\ F_\ind\
=\ \Re f^W_\ind \ +\ {\cano_\ind\,\kappa^2\over16\pi^2}\, K\
+\ {\cren_\ind\over 8\pi^2}\,\log \Lambda ;
\eqn\FMK
$$
the second equality here follows from eq.~\Katerm, which applies
because masses do not affect the anomaly of super-Weyl transformations.

Now consider the low-energy limit of eq.~\MassiveF.
Segregating the contributions of the massless particles from those
of particles with masses $O(M_I)$, we have
$$
F^M_\ind\
=\ F'_\ind\ -\ \Delta F_\ind\,,
\eqn\Frel
$$
where
$$
\eqalign{
F'_\ind\ =\ {}&
\{g'_\ind(p)\}^{-2}\ -\ {T(G'_\ind)\over 8\pi^2}
	\left(\log\{g'_\ind(p)\}^{-2}\,+\,3\log p\right)\cr
&+\sum_{r'} {T_\ind(r')\over 8\pi^2}
	\left(\log\det\{Z^{(r')}(p)\}\,+\,n_{r'}\log p\right)
	    _{\rm massless\ matter}\cr
}\eqn\Flow
$$
and
$$
\eqalign{
\Delta F_\ind\ = \sum_{r'} {T_\ind(r')\over 8\pi^2} \biggl[ &
\left( n_{r'}\log\{g_{r'}\}^{-2}\,+\,3\log\det\{{\cal M}^{(r')}\}\right)
	_{\rm massive\ vectors}\cr
&-\left(\log\det\{Z^{(r')}\}\,+\,\log\det\{{\cal M}^{(r')}\}\right)
	_{\rm massive\ matter} \biggr].\cr
}\eqn\Fdiff
$$
Obviously, $F'_\ind$ are the exact analogues of $F_\ind$ for the
low-energy theory from which all massive particles have decoupled.
Since the same low-energy phenomenology can be obtained from
a locally supersymmetric EQFT that retains only the massless fields, and
since all the results of sections 2 and 3.1 apply verbatim to such an EQFT,
we conclude that its Wilsonian gauge couplings $f^{\prime W}_\ind$ are
related to the $F'_\ind$ via an exact analogue of eq.~\Katerm, namely
$$
F'_\ind\ =\ \Re f^{\prime W}_\ind\
+\ {\cano'_\ind\,\kappa^2\over16\pi^2}\, K\
+\ {\cren'_\ind\over 8\pi^2}\,\log\Lambda' .
\eqn\lowF
$$
The coefficient $\cano'_\ind$ in this formula is computed similar to
eq.~\canodef, but counting only the massless fields
and ditto for the coefficient $\cren'_\ind$;
this is different from the $\cano_\ind$ and $\cren_\ind$ appearing in eq.~\FMK,
which account for all the charged fields of the higher-energy EQFT,
regardless of whether they have $O(M_I)$ masses or not.
We show momentarily  that the difference between $\cano'_\ind$ and
$\cano_\ind$ is necessary for maintaining supersymmetry on both sides of the
threshold.

It is clear from eqs.~\FMK, \Frel\ and \lowF\ that the key to threshold
corrections to the Wilsonian gauge couplings is provided by $\Delta F_\ind$,
which depends only on the masses and other properties of the
massive particles.
Let us reorganize those particles into proper massive supermultiplets.
The supersymmetric Higgs mechanism makes the
massive vector supermultiplets from gauge and massless matter
supermultiplets, which acquire a common mass
${\cal M}=\{g\}\{Z_{\rm Higgs}\}^{1/2}\left|\vev{\rm Higgs}\right|$;
in our notations, all indices are suppressed and
$\vev{\rm Higgs}$ is the un-normalized VEV of the
Higgs field times the appropriate group-theoretical factor.
All other charged massive particles belong to non-Higgs massive scalar
supermultiplets and have their masses because of the $\half M_{IJ}(\modul)
\matter^I\matter^J$ terms in the superpotential $W(\modul,\matter)$ of the
EQFT;
the canonically-normalized mass-square matrix for these particles is
\penalty-1000
${\cal M}^2=e^{\kappa^2 K} \{Z\}^{-1/2}M^\dagger (\{Z\}^{-1})^\top
M\{Z\}^{-1/2}$.
Substituting these cumbersome-looking masses into an already cumbersome
eq.~\Fdiff\ actually results in a great simplification of the latter:
After some algebra, we arrive at
$$
\displaylines{
\Delta F_\ind\ =\sum_{r'} {T_\ind(r')\over 8\pi^2}\biggl(
2\sum_{\rm massive\atop vectors}^{(r')} \log\left|\vev{\rm Higgs}\right|
\hfill\eqname\MassVEV\cr
\hfill{}-\ \left[\log\left|\det M^{(r')}\right|\,
    +\half n_{r'}\kappa^2 K \vphantom{M^{(r')}_{mass}}
    \right]^{\rm non-Higgs}_{\rm massive\ matter} \biggr).\qquad\cr }
$$
The un-normalized mass matrices $M^{(r')}$ are holomorphic functions
of the chiral moduli fields $\modul^i$ and,
as argued in ref.~[\VKa], properly defined Higgs VEVs
$\vev{\rm Higgs}$ are also holomorphic;
the only non-harmonic terms here are those containing
the \K\ function~$K$.
These terms are absent from the rigid-SUSY analogue of
eq.~\MassVEV\refmark\VKa; just as the K\"ahler terms in eq.~\Katerm,
they are peculiar to local SUSY.
In fact the two kinds of K\"ahler terms are closely related to each other:
Combining eqs.\ \FMK, \Frel, \lowF\ and \MassVEV\ together and
separating holomorphic terms from terms proportional to $K$,
we see that {\it the Wilsonian gauge couplings of locally supersymmetric EQFTs
describing the two sides of the threshold are related to each other via}
$$
\eqalign{
\Delta f^W_\ind(\modul)\, &
\equiv\ \left(f^{\prime W}_\ind(\modul)
	+{\cren'_\ind\over8\pi^2}\log{\Lambda'\over M_I}\right)\
-\ \left(f^W_\ind(\modul)
	+{\cren_\ind\over8\pi^2}\log{\Lambda\over M_I}\right)\cr
&=\sum_{(r')}{T_\ind(r')\over 8\pi^2}\biggl(
	2\sum_{\rm massive\atop vectors}^{(r')}
	\log{\vev{\rm Higgs}\over M_I}\
	-\ \left.\log\det{M^{(r')}\over M_I}
		\right|_{\strut^{\rm non-Higgs}_{\rm massive\ matter}}
	\biggr) \cr
}\eqn\Wilthreshold
$$
($M_I$ here is the `official' threshold scale).
This formula is identical to its rigid-SUSY counterpart (\cf~[\VKa]),
but in the locally super\-symmetric case, we also need an exact agreement
between the non-holo\-morphic K\"ahler terms:
In terms of their coefficients, we must have
$$
\cano_\ind -\ \cano'_\ind
= \sum_{r'}T_\ind(r')\cdot n_{r'}(\hbox{massive, non-Higgs }\matter^I).
\eqn\Afixing
$$
Were eq.~\Afixing\ to fail, the higher-energy EQFT and the lower-energy EQFT
would become inconsistent with each other, and no supersymmetric correction
to eq.~\Wilthreshold\ could possibly repair this inconsistency.
Fortunately, simple algebra shows that
$$
\displaylines{
\sum_{r'}T_\ind(r')\cdot n_{r'}(\hbox{massive, non-Higgs
}\matter^I)\hfill\cr
{}=\sum_{r'}T_\ind(r')\left( n_{r'}({\rm massive}\ \matter^I)\,
	-\,n_{r'}({\rm massive\ vectors})\right)
    \hfill\eqname\Acheck\cr
{}=\Bigl( \sum_r T_\ind(r)\,n_r({\bf all}\ \matter^I)\,-\,T(G_\ind)\Bigr)\
    -\Bigl( \sum_{r'}T_\ind(r')\,n_{r'}({\rm massless}\ \matter^I)\,
	-\,T(G'_\ind)\Bigr),\hfill\cr }
$$
which indeed equals to $\cano_\ind-\cano'_\ind$ according to eq.~\canodef\
and to the spectra of the two EQFTs.
Conversely, eqs.~\Afixing\ and \Acheck\ can be used to justify the presence
of the K\"ahler term in eq.~\Katerm,
and also to derive its coefficient~\canodef,
without recourse to either the super-Weyl symmetry or to the axial anomaly
and relying solely on the locally supersymmetric formul\ae\ for the
particle's masses.

\ldf\SW{S.~Weinberg, \plt B91 (1980) 51.}
\ldf\VKb{V. Kaplunovsky, \nup307 (1988) 145.}
Formula \Wilthreshold\ for the threshold corrections to the Wilsonian
gauge couplings is holomorphic and completely determined at the one-loop
level of integrating out the heavy fields.
Threshold corrections to the effective gauge couplings behave quite
differently.
To be precise, let us define $\Delta_\ind$ as the difference between
the low-energy effective couplings $16\pi^2/\{g'_\ind(p)\}^2$,
extrapolated up to $p=M_I$
using the Callan-Symanzik equations of the lower-energy effective theory
in which particles with $O(M_I)$ masses do not appear,
and between the appropriate higher-energy effective couplings
$16\pi^2/\{g_\ind(p)\}^2$,
extrapolated down to $p=M_I$ using the Callan-Symanzik equations of the
higher-energy effective theory in which all the $O(M_I)$ masses are ignored.
\foot{At the one-loop level, this definition is equivalent to
    $$
    {16\pi^2\over\{g'_\ind(p'<M_I)\}^2}\
    =\ {16\pi^2\over\{g_{\ind}(p>M_I)\}^2}\
    +\ 2b_\ind\,\log{p\over M_I}\ +\ 2b'_\ind\,\log{M_I\over p'}\
    +\ \Delta_\ind\,;
    $$
    at higher loop levels, one needs a much more complicated formula.}
Using appropriate versions of eq.~\SV\ for both theories, we find
$$
\eqalign{
\Delta_\ind\ =\ 16\pi^2\,\Delta F_\ind\ &
+\ 2\biggl[ T(G'_\ind)\log\{g'_\ind(p\nearrow M_I)\}^{-2}\,
    -\ \cren'_\ind\,\log M_I\cr
&\qquad-\sum_{r'}T_\ind(r')\,
    \log\det\{Z^{(r')}_{{\rm massless}\ \matter}(p\nearrow M_I)\}\biggr]\cr
&-\ 2\biggl[ T(G_\ind)\log\{g_{\ind}(p\searrow M_I)\}^{-2}\,
	-\ \cren_\ind\,\log M_I\cr
&\qquad-\sum_{r'}T_\ind(r')\,
    \log\det\{Z^{(r')}_{{\rm all}\ \matter}(p\nearrow M_I)\}\biggr] ,\qquad \cr
}\eqn\genDelta
$$
where $\Delta F_\ind$ is as in eqs.~\Fdiff\ and \MassVEV.
Formula~\genDelta\ is true to all orders of perturbation theory, but its
application to higher orders requires care in using the right renormalization
scheme for each appearance of the wave-function normalization matrix $\{Z\}$.
Fortunately, at the one-loop level analysis of $\Delta_\ind$,
this subtlety can be safely ignored, which reduces eq.~\genDelta\ to
a much simpler formula
$$
\,\eqalign{
\Delta_\ind^\ol =\ 16\pi^2\Re\Delta f^W_\ind &
-\ (\cano_\ind-\cano'_\ind)\kappa^2 K\
-2(T(G_\ind)-T(G'_\ind))\log\{g_\ind (M_I)\}^{-2}\cr
&+\sum_{r'} 2T_\ind(r')\,\log\det \{Z^{(r')}_{\rm massive}(M_I)\}\cr
}\,\eqn\olDelta
$$
in full agreement with the old result\refmark{\SW,\VKb}
$$
\Delta_\ind^\ol\
=\sum_{r'}T_\ind(r')\left( 4\Tr\left[\log{{\cal M}\over M_I}\right]^{(r)}_\MV\
-\ 2\Tr\left[\log{{\cal M}\over M_I}\right]^{(r)}_\NHMM \right).
\eqn\oldolDelta
$$
Note that the second, the third and the fourth terms on the right hand side
of eq.~\olDelta\ are non-harmonic functions of the moduli VEVs.
This is similar to the non-harmonicity of the string-threshold corrections
to the effective gauge couplings obtained in ref.~[\DKLb].

\ldf\flip{see J.~Lopez and D.V.~Nanopoulos, \prv D47 (1993) 2468
	and references therein.}
\ldf\oxford{B.~Greene, K.~Kirklin, P.~Miron and G.G.~Ross,
	\nup 278 (1986) 667 and \nup 292 (1987) 292.}
We conclude this section by relaxing the assumption that each $G'_\ind$
is wholly contained in a single $G_\ind$.
Many unified models have intermediate-energy thresholds for which this
assumption does not hold
(\eg, the `flipped' $SU(5)\otimes U(1)$\refmark{\flip} or the
Oxford $SU(3)^3$ model\refmark{\oxford})
and the mapping between the low-energy gauge couplings
$\{g'_\ind\}$ and the high-energy gauge couplings $\{g_\ind\}$ is rather
complicated.
However, this complication is purely notational while the physics remains
the same as in the case of single $G_\ind$'s.
In terms of the Wilsonian gauge couplings, we have
$$
f_a^{\prime W}(\modul)\ +\ {\cren'_a\over 8\pi^2}\log{\Lambda'\over M_I}\
=\sum_b C_{ab}\left( f_b^W(\modul)\,+\,{\cren_b\over 8\pi^2}
\log{\Lambda\over M_I}\right)\quad +\ \Delta f^W_a(\modul) ,
\eqn\GeneralThreshold
$$
where $C_{ab}$ are group-theoretical factors characterizing the threshold
and $\Delta f^W_a$ are exactly as in eq.~\Wilthreshold.
Note that since the Wilsonian gauge couplings renormalized only at one loop,
eqs.~\Wilthreshold\ and \GeneralThreshold\ are {\sl exact}, \ie, accurate
to all orders of the perturbation theory.

\chapter{Gaugino Condensates and Effective Superpotentials}
In most theories of  string unification, the four-dimensional EQFT
describing very high energies has
an unbroken gauge group that is a lot bigger than the
$SU(3)\otimes SU(2)\otimes U(1)$ of the standard model.
At some intermediate energy scale --- well below $\mpl$
but well above the weak scale --- the excess gauge symmetry decouples
{}from the ordinary particles via spontaneous gauge symmetry breakdown,
confinement or some combination of the two mechanisms.
Implications of  local SUSY for the spontaneous breakdown are discussed
in section 3.3; the present section is devoted to a manifestly supersymmetric
treatment of the confinement and associated dynamical effects:
Formation of the gaugino condensate
$\vev{\lambda\lambda}$ and breakdown of the perturbative
degeneracy of the vacua with different moduli VEVs.
The techniques we  present here are purely field-theoretical and do not
in any way depend on the string nature of the ultimate unification;
all we presume is that the unified theory, whatever its nature, gives rise
to a non-abelian asymptotically-free hidden gauge group $G_\ind$, which we
shall henceforth denote as simply $G$.

Our discussion starts with locally supersymmetric Yang-Mills theories
with\-out charged matter fields.
Pure SSYM theories have been extensively studied
{}from different points of view;
section 4.1 summarizes some generally known results and puts them in
context of a locally supersymmetric pure Yang-Mills theory with
a moduli-dependent gauge coupling.
In section 4.2 we use EQFT techniques to calculate the effective
superpotential for the moduli $\Wmod(\modul)$
that is induced by the gaugino condensation
in a pure-SSYM hidden sector,
and in section 4.3 we extend these techniques to hidden sectors with
charged matter scalars.
We show how to calculate $\Wmod(\modul)$ for any hidden sector,
with or without charged scalars, as long as it has a stable,
supersymmetric vacuum state.
In section 4.4 we illustrate this technology on a few examples of hidden
sectors one often encounters in string-unified theories.
Finally, in section 4.5 we address the questions of the overall
vacuum stability and of the supersymmetry breaking.

\section{Hidden Sectors Without Matter.}
\ldf\EW{E.~Witten, \nup 188 (1981) 513.}
By itself, gaugino condensation does not break supersymmetry.
Indeed, as shown by E.~Witten,\refmark{\EW}
in a rigidly supersymmetric theory of only gauge bosons and gauginos
there is no spontaneous
SUSY breakdown regardless of confinement, gaugino condensation and
other non-perturbative phenomena.
Implications of this result for locally supersymmetric Yang-Mills theories
or for SSYM theories with moduli-dependent gauge couplings have been widely
discussed in the
literature.\refmark{\dindrsw-\BG, \JLpascos}
This section is the summary of the relevant results put together.

\pointbegin
In rigid SUSY, the gaugino bilinear  $\lambda^\alpha\lambda_\alpha$ is
the lowest component of the composite gauge-invariant {\sl chiral}
superfield $U\equiv W^\alpha W_\alpha$.
In local SUSY, there are two ways to generalize this superfield,
 namely\refmark{\FMTV}
$$
\uu\
\equiv  {\cal W}^{\alpha}{\cal W}_\alpha
\qquad {\rm and}\quad \uuh\ \equiv\ \uu/\ccf^3.
\eqn\uudef
$$
Both $\uu$ and $\uuh$ are chiral and gauge invariant,
but $\uu$ is $\ccf$-independent and hence invariant under the K\"ahler
transformations~\Kahlerfull\ while $\uuh$ is not;
on the other hand, it is $\uuh$ and not $\uu$ that is invariant under the
super-Weyl transformations.
The exact relation between these chiral superfields and
the conventionally normalized gaugino bilinear is
$$
\lambda^{\alpha}\lambda_\alpha\
=\ \left. e^{\kappa^2 \widetilde K/2}\, \uu\right|_{\Theta=\bar\Theta=0}\
=\ \left. \left(\ccfb/\ccf\right)^{3/2}\,e^{\kappa^2 K/2}
\, \uuh\right|_{\Theta=\bar\Theta=0}\ .
\eqn\uuGG
$$
By conventional normalization we mean that the gauginos $\lambda_\alpha^{(a)}$
have the same normalization as the gauge bosons $A_m^{(a)}$;
this normalization is obtained after the Weyl rescaling, and indeed,
in the Wess-Zumino gauge \WZK, eq.~\uuGG\ reduces to
$\lambda^\alpha\lambda_\alpha=\left.\uu\right|_{\Theta=\bar\Theta=0}$.

\point
The confinement scale $\mu$ of an asymptotically-free theory is the
momentum scale below which perturbative renormalization of the running
effective gauge coupling $\{g(p)\}$ no longer makes sense.
(We presume $\mu\ll\mpl$.)
For a supersymmetric Yang-Mills theory without 
charged scalars,
eq.~\SV\ breaks down at
$$
\!\eqalign{ \mu &
= \left(8\smash{\pi^2}e/ T(G)\right)^{1/3}\, p\,\{g(p)\}^{-2/3}
    \exp\left( -{8\pi^2\over 3T(G)\,\{g(p)\}^2}\right),
    \quad ({{\rm any}\ p\ge\mu })\cropen{1\jot}
&
= \left(8\smash{\pi^2}e/ T(G)\right)^{1/3}\,\Lambda\,
    \exp\left( -{8\pi^2\over 3T(G)}\,\Re f^W\
	+\ \coeff16\kappa^2 K\right),\cr }\!
\eqn\confscale
$$
where the second equation follows from \Katerm.
Note that the first equation here has a pre-exponential factor $\{g\}^{-2/3}$
while in the second equation the Wilsonian gauge coupling $f^W$ appears
only in the exponential --- this reflects the fact that the effective gauge
coupling is renormalized in all orders of perturbation theory, but the
renormalization of the supersymmetric Wilsonian coupling $f^W$ stops at one
loop.\refmark{\SVb}
The K\"ahler term in the last exponential is peculiar to local SUSY;
its coefficient does not depend on the size of the gauge group $G$.

\point
Because of its relation \uuGG\ to the lowest component of a chiral superfield,
the gaugino-bilinear operator $\lambda\lambda$ receives
no anomalous corrections to its canonical scaling dimension~3.
Consequently, $|\vev{\lambda\lambda}|\propto\mu^3$, with a
coupling-independent proportionality coefficient of order 1.
At the same time, the phase of the gaugino condensate equals to
the Wilsonian $\theta$-angle for $G$, divided by $T(G)$.
With the help of eqs.~\confscale\ and \uuGG\ we can combine these
two results into a single exact formula for
the VEV of the composite superfield $\uu$ in terms of
the Wilsonian supersymmetric gauge coupling:
$$
\vev{\uu}\ =\ \Lambda^3
\exp\left( -{8\pi^2\over T(G)}\,\tilde f^W\right)\times
[\hbox{an $O(1)$ coupling-independent constant}].
\eqn\uuVEV
$$
This relation is holomorphic --- the K\"ahler factors in
eqs.~\uuGG\ and \confscale\ cancel each other, ---
which reflects the fact the SUSY is not spontaneously broken
($\uu$ and $\tilde f^W(\modul,\ccf)$ are both composite chiral superfields).
In terms of the super-Weyl invariant quantities, eq.~\uuVEV\ becomes
$$
\vev{\uuh}\ =\ \Lambda^3
\exp\left( -{8\pi^2\over T(G)}\, f^W(\modul)\right) \times{\rm const}\ ;
\eqn\uuhVEV
$$
note that the Wess-Zumino $\log\ccf$ term in eq.~\Wilsanom\ is absolutely
necessary for the consistency of this formula, as is the fact that
$\cano=-T(G)$ for a matter-less Yang-Mills theory.

\point
{}From the low-energy ($O(\mu)$) point of view, the $\uuh$
supermultiplet of the pure SSYM theory
is analogous to the $\pi K\eta\eta'$ meson nonet of the ordinary QCD ---
it describes the lightest composite particles of the confined theory,
and its VEV serves as the order parameter of the chiral symmetry breakdown.
In QCD, integrating out all hadronic degrees of freedom except
for the order parameters leads to a sigma model;
a similar treatment of the SSYM theory leads to an effective
locally supersymmetric theory for  $\uuh$.
The vacuum structure of this effective theory is described by
an effective superpotential $\Weff(\uuh,\modul)$ where
the VEV $\vev{\uuh}$ is the SUSY-preserving solution of
$\partial \Weff/\partial\uuh=0$.
The specific formula \uuVEV\ 
is reproduced by the effective superpotential\refmark{\TVY}
$$
\Weff(\uuh ,\modul)\ =\ {1\over 4}\,\uuh\, f^W (\modul)\
+\ {\uuh \over 32\pi^2}\left(T(G)\,\log{\uuh \over\Lambda^3}\,
	+\,{\rm const}\right)
\eqn\uuhWeff
$$
or, in terms of the K\"ahler-invariant superfield $\uu$,\refmark{\FMTV}
$$
\eqalignno{
\widetilde\Weff(\uu,\modul,\ccf)\ &
\equiv\ \ccf^3 \Weff(\uu,\modul) &
\eqname\uuWeff\cr
& =\ {1\over 4}\, \uu\, \tilde f^W(\modul,\ccf)\
    +\ {\uu\over 32\pi^2} \left(T(G)\,\log{\uu\over\Lambda^3}\,
        +\,{\rm const}\right).\qquad\cr }
$$
The first term in this superpotential corresponds
to the Wilsonian Lagrangian for the gauge superfields;
the second term is a purely low-energy non-pertur\-bative effect
and thus involves only  $\uu $
itself but does not depend on the gauge coupling, the moduli or even
the Weyl compensator $\ccf$.

\point
There is an alternative argument for eqs.~\uuhWeff\ and \uuWeff\
for the effective superpotential that does not involve eq.~\uuVEV.
Instead, one {\it assumes} that a \K-invariant effective superpotential
should look like
$\widetilde\Weff=\coeff14\tilde f^W\ \uu -\Xi(\uu)$
where  $-\Xi(\uu)$ is a `dynamical' superpotential that does
not depend on any superfields other than $\uu$.
The super-Weyl transformations should leave the
$\Weff\equiv\widetilde\Weff/\ccf^3$ invariant;
in light of eqs.~\gauginoWeyl\ and \Wilsanom,
this means that $\Xi(\uu)$ should satisfy
$$
\Xi(e^{-6\tau}\uu)\
=\ e^{-6\tau}\left[\Xi(\uu)\,
-\,{6\cano\tau\over 32\pi^2}\uu\right] ;
\eqn\Xitrans
$$
the solution to this constraint gives eq.~\uuWeff\
(for a SSYM theory without charged matter $\cano=-T(G)$).
For rigidly supersymmetric pure Yang-Mills hidden sectors, the same
argument can be made in terms of the scale and
R-symmetries instead of super-Weyl;
this is how the superpotential \uuhWeff\ was first derived in ref.~[\TVY].
\subpar
Naturally, there are many other ways to obtain the effective superpotential.
For example, for EQFTs based upon string orbifolds,
the analytic form of the $\Weff$ can be deduced from the
modular invariance.  \refmark{\FMTV,\BG,\JLpascos}

\section{Effective Potential for the Moduli.}
Classically, moduli VEVs $\vev{\modul^i}$ parametrize a continuous family
of exactly degenerate vacua of the unified theory.
Because of the no-renormalization theorem for the
superpotential~\Wexpansion, this exact degeneracy persists
to all orders of perturbation theory.
However, the non-perturbative effects associated with the confinement
and with the
formation of the gaugino condensate break this degeneracy,
and once {\sl all} of the strongly interacting fields are integrated out,
the resulting effective theory acquires a non-trivial effective potential
$\Vmod(\vev\modul,\vev\modulb)$ for the moduli fields.

\ldf\CFILQ{M.~Cveti\v c, A.~Font, L.E.~Ib\'a\~nez, D.~L\"ust and
	F.~Quevedo, \nup361 (1991) 194.}
Given the effective superpotential $\Weff(\uuh,\modul)$, the
calculation of  the effective potential $\Vmod$ is rather simple:
Starting with the formula \uuhWeff\ and the solution \uuhVEV\
for $\partial\Weff(\uuh,\modul)/\partial\uuh=0$,
one can integrate out the $\uuh$ superfield and derive
an effective theory just for the moduli superfields.
The superpotential of this effective theory is simply
$$
\Wmod(\modul)\
\equiv\ \Weff(\vev\uuh(\modul),\modul)\
=\ -\, {T(G)\over 32\pi^2}\ \vev\uuh(\modul) .
\eqn\Wmoduli
$$
A similar formula describes the contribution of the hidden matter
fields to the effective K\"ahler function for the moduli.
However, this contribution is of the order $O(\mu^2)$ and thus negligible
compared to the perturbative $\Kmod=O(\mpl^2)$;
for all practical purposes, the K\"ahler function of the
effective theory is simply $\Kmod(\modul,\modulb)$.
Together, $\Wmod(\modul)$ and $\Kmod(\modul,\modulb)$
yield an effective scalar potential
according to the usual (classical) rules of the local SUSY
\refmark{\CFGP-\GGRS}:
$$
\!\eqalign{
\Vmod(\modul,\modulb) = &
\left[ G^{\ib j}\left(\partder{\Wmod^*}{\modulb^\ib}
	+ \kappa^2 \partder{\Kmod}{\modulb^\ib} \Wmod^*\right)
\left(\partder\Wmod{\modul^j}
	+ \kappa^2 \partder{\Kmod}{\modul^j} \Wmod\right)\right.\cr
&\mskip 45mu{}-\ 3\kappa^2\left|\Wmod\right|^2 \biggr]\,
	\times e^{\kappa^2 \Kmod} .\cr }\!
\eqn\SUSYVeff
$$
This potential must be invariant under all the exact symmetries of the EQFT
and often this fact can be used to constrain the analytic form of the
$\Wmod(\modul)$, for example in EQFTs based upon string orbifolds with
duality symmetries.
\refmark{\FILQ,\NO,\CFILQ,\JLpascos}

In this section we follow a different approach
(pioneered by Ferrara, Girardello and Nilles)\refmark{\FGN}
and derive the effective scalar potential starting from a
component-field Wilsonian action for the gauge fields, the moduli and
their superpartners.
Using the most general rules of the quantum field theory,
one can integrate out the strongly interacting fields and thus obtain
$\Vmod(\modul,\modulb)$ without ever using an effective
superpotential.
\foot{The authors thank Steven Weinberg for suggesting this approach
    and Lance Dixon and Michael Peskin
    for collaboration at an early stage of this calculation.}
However, as long as the strong interactions do not break  local SUSY,
either explicitly or spontaneously, the result must have the form \SUSYVeff\
for {\it some} $\Wmod(\modul)$;
in other words, the {\it existence} of an effective superpotential
$\Wmod(\modul)$ is inevitable, and the only question is its exact form,
which may agree or disagree with eqs.~\uuhWeff\ and \Wmoduli.

In this subsection we  prove that the $\Wmod$ induced by
a pure-SSYM hidden sector indeed agrees with \uuhWeff\ and \Wmoduli.
Moreover, our arguments can be extended to any hidden sector, with
or without matter fields, that has a stable supersymmetric vacuum.
This extension is presented in the following section 4.3;
for the present section we limit our attention to
pure SSYM theories coupled to supergravity and to
moduli fields.

In order to keep our notations simple, let $\schi$ denote all strongly
interacting component fields, which for the theory at hand are simply
the components of the Yang-Mills supermultiplet
$(A_m^{(a)}, \lambda_\alpha^{(a)}, \bar\lambda_{\dot{\alpha}}^{(a)},
 D^{(a)})$ of the hidden sector $G$,
plus ghosts due to the quantum gauge fixing.
Similarly, let $\phi$ denote all the weakly interacting component fields,
namely the moduli, the gravitational fields and all their superpartners,
{\sl including the auxiliary fields} as well as the components of the
compensator $\ccf$.
In the same notations, the Lagrangian can be written as
$$
{\cal L}\ =\ {\cal L}_{\rm SSYM}(\schi;\phi)\
+\ {\cal L}^0_\phi(\phi) ,
\eqn\twoLag
$$
where the first term is the SSYM Wilsonian Lagrangian for
the $\schi$ fields in a background of  $\phi$ fields
and the second term is the Lagrangian for the $\phi$ fields themselves.
Ideally, the ${\cal L}^0_\phi$ should also be a Wilsonian Lagrangian,
but because the interactions between quantum $\phi$ fields are badly
non-renormalizable, we would not know how to regulate the resulting
theory.
Instead, we take ${\cal L}^0_\phi$ to be the ``effective classical Lagrangian''
\ie, the generating
functional of an effective theory that consists only of the $\phi$ fields;
formally, this generating functional is the sum of all 1PI Feynman graphs
involving only $\phi$, but practically it should be calculated directly
{}from the perturbative string theory.

Let us formally integrate out the strongly-interacting $\schi$ fields;
this gives us an effective Lagrangian for the $\phi$ fields,
$$
{\cal L}^{\rm eff}(\phi)\ =\
{\cal L}^0_\phi(\phi)\ +\ {\cal L}^1_\phi(\phi),
\eqn\phiLag
$$
where
$$
{\cal L}^1_\phi\ =\ \vev{{\cal L}_{\rm SSYM}}\
+\ \vev{\delta{\cal L}_{\rm SSYM}\over \delta\phi}
    \left(\phi-\vev\phi\right)\
+\ \cdots
\eqno\eq
$$
is the effective Lagrangian for $\phi$ induced by the strong interactions;
below the confinement scale, this effective Lagrangian is local.
The Feynman rules of the EQFT defined by eq.~\phiLag\ are as follows:
The propagators and the vertices are given by
expanding both $ {\cal L}^0_\phi$ and ${\cal L}^1_\phi$ into powers of
$(\phi-\vev\phi)$; diagrammatically, we have:
$$
\checkex 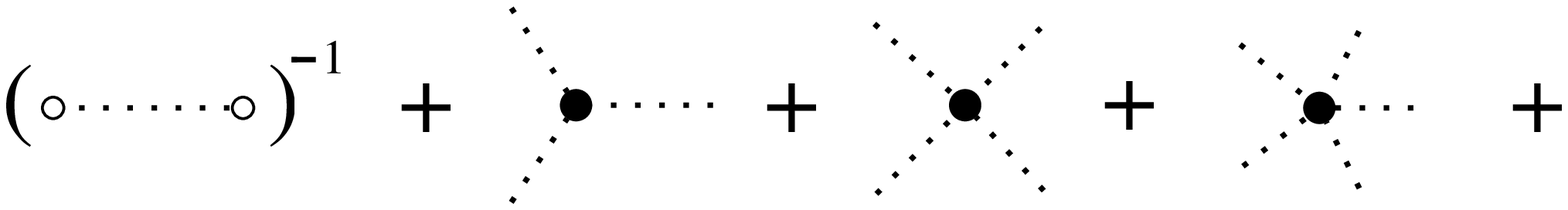
\iffigureexists \checkex 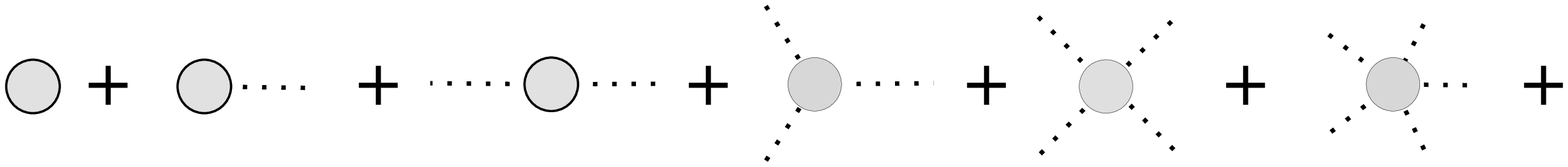 \fi
\iffigureexists
\eqalign{
 {\cal L}^0_\phi\ & =\ \diagram TreeVertices.eps \>  \cdots ,\cr
{\cal L}^1_\phi\ & =\ \diagram InducedVertices.eps \>  \cdots ,\cr
}\else \missbox{Missing Diagrams},\fi
\eqn\FeynmanRules
$$
\looseness=1
where the dotted lines are the perturbative propagators of $(\phi-\vev\phi)$,
the solid circles are their perturbative vertices,
and the bubbles
$\checkex Bubble.eps \iffigureexists \diagram Bubble.eps \fi $
are the non-perturbative vertices induced by the strong interactions.
Since $ {\cal L}^0_\phi$ is a generating functional and not a Wilsonian
Lagrangian,
the only Feynman diagrams we should consider are those in which all loops
contain at least one induced vertex%
\checkex Bubble.eps \iffigureexists $\diagram Bubble.eps $\fi ;
the loops involving only the classical vertices are already included
in $ {\cal L}^0_\phi$.
Using the $ {\cal L}^0$ rather than the full ${\cal L}^{\rm eff}$
to define the propagators means that the induced vertices
can have any number of external legs, $n=0,1,2,3,\ldots$.
(as opposed to the perturbative vertices that always have $n\ge3$
external legs).
On the other hand, this makes for easy counting
of the powers of the gravitational coupling $\kappa^2=8\pi/\mpl^2$:
The propagators are proportional to $\kappa^2$,
the perturbative vertices to $\kappa^{-2}$,
and the induced vertices to $\kappa^0$;
this reflects the facts that the entire $ {\cal L}^0_\phi$
is proportional to $\mpl^2$ while the induced Lagrangian
${\cal L}^1_\phi$ is $\mpl$-independent.
Moreover, because all loops have to include an induced vertex, they are
effectively cut off at the confinement scale $\mu$ of the SSYM theory;
consequently, the loops and the associated momentum integrals carry
no extra powers of $\mpl$.

In any perturbative EQFT without a classical potential,
the effective potential is the sum of
all connected Feynman graphs with no external legs.
For the problem at hand, the $ {\cal L}^0_\phi$ contains
no potential, and so, expanding in powers of $\kappa\mu$, we have
$$
\def\diatest#1 {\checkex #1.eps \iffigureexists \diagram #1.eps \else
	\pmatrix{\rm Missing\cr \rm Diagram\cr }\fi }
{\cal V}_{\rm eff}(\modul,\modulb)\ =\ \diatest Bubble \
+\ \left\{ \diatest BigLoop \ +\ \diatest Dumbell \right\}\ +\ \cdots .
\eqn\MainDiagrams
$$
Naively, the first diagram here contributes an $O(\mu^4)$ term to the
effective potential while the contributions of the other two diagrams are
$O(\kappa^2\mu^6)$;
the `$\cdots$' stand for terms of still higher order in $(\kappa\mu)^2$.
The first diagram corresponds to
the effective potential generated by the
SSYM theory, which vanishes since there is no spontaneous SUSY breaking
in pure SSYM\refmark{\EW};
hence, the whole ${\cal V}^{\rm eff}$ is at most of the order
$O(\kappa^2\mu^6)$.

The second diagram in \MainDiagrams\ gives us
$$
\int\!{d^4p\over (2\pi)^4}\,\mathop{\hbox{\fourteenrm Str}}
\left[ \Pi_{\phi_1\phi_2}(p)\times
\vev{{\delta{\cal L}_{\rm SSYM}\over\delta\phi_1(p)}
{\delta{\cal L}_{\rm SSYM}\over\delta\phi_2(-p)}
}_{\rm SSYM}\right]\,,
\eqn\BigLoopZero
$$
where the supertrace is taken over all the
weakly interacting component fields $\phi$,
$\Pi_{\phi_1\phi_2}$
is the classical propagator matrix for  those fields, and
$\vev{\cdots}_{\rm SSYM}$ is the expectation value of the
appropriate operator in the non-perturbative vacuum of
the strongly interacting SSYM theory.
In the absence of spontaneous SUSY breakdown in
the strongly interacting sector, this $\vev{\cdots}_{\rm SSYM}$
is supersymmetric,
and as long as the auxiliary fields are included among $\phi_1,\phi_2$,
this supersymmetry holds even for the off-shell momenta $p$.
Since the propagator matrix $\Pi$ is also supersymmetric,
the supertrace in eq.~\BigLoopZero\ vanishes ---
the contributions of the bosonic and the fermionic components of any
complete off-shell supermultiplet exactly cancel each other.
Thus,  unbroken supersymmetry of the SSYM theory also implies the
vanishing of the second diagram in \MainDiagrams.

The third diagram in \MainDiagrams\ yields
$$
\sum_{\phi_1\phi_2} \Pi_{\phi_1\phi_2}(p=0)
\vev{\delta{\cal L}_{\rm SSYM}\over\delta\phi_1(p=0)}
\vev{\delta{\cal L}_{\rm SSYM}\over\delta\phi_2(p=0)}\,.
\eqn\Dumbell
$$
Clearly, only the spinless fields $\phi_1,\phi_2$ contribute
to this sum; these include the moduli scalars $\modul^i$,
the Weyl compensator $\ccf$, their auxiliary superpartners,
and also the spinless auxiliary member $M$ of the
supergravity multiplet.
Moreover, the
moduli scalars do not contribute in
eq.~\Dumbell\ because
the corresponding tadpoles vanish:
$$
\vev{\delta{\cal L}_{\rm SSYM}\over\delta\modul^i(p=0)}\
\checkex 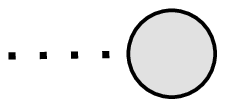
\iffigureexists =\> \modul^i \diagram Tadpole.eps \> \fi
=\ \partder{\vev{{\cal L}_{\rm SSYM}}}{\vev{\modul^i}}\
\equiv\ \partder{{\cal V}^0_{\rm eff}(\vev\modul,\vev\modulb)}{\vev{\modul^i}}\
=\ 0.
\eqn\Tadpole
$$
(In a SSYM theory ${\cal V}^0_{\rm eff} \checkex Bubble.eps
    \iffigureexists =\,\diagram 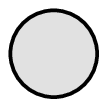 \fi$
vanishes regardless of the moduli VEVs.)
{\it A~fortiori}, there is no tadpole for the Weyl compensator $\ccf$.
Thus the only contributions to eq.~\Dumbell\ come from the auxiliary fields,
which is just as well since their propagators are momentum-independent and do
not diverge at $p=0$.

Finally, two of the complex auxiliary fields, namely $M$ and
$F^0\equiv\left.\log\ccf^3\right|_{\Theta\Theta}$,
transform inhomogeneously under super-Weyl transformations,
which allows us to fix or eliminate one of them by means of a suitable
super-Weyl gauge conditions.
For simplicity, we choose to set $M=0$ and leave $F^0$ unconstrained
(however, the other components of the $\ccf$ superfield remain fixed by the
truncated \WZK\ condition $\left.\widetilde K\right|_{{\rm no}\,\Theta}=
\left.\widetilde K\right|_\Theta=\left.\widetilde K\right|_{\bar\Theta}=0$).
The propagator matrix for the $F^0$ and the
$F^i=\left.\modul^i\right|_{\Theta\Theta}$ is easily read
{}from eqs.~\SUGRACT\ and \Ktilde;
hence, in light of ${\cal V}^0_{\rm eff}=0$ and eq.~\Tadpole, we have
$$
\eqalign{
{\cal V}^{\rm eff}\ ={} &
 G^{\ib j} \left( \vev{\partder{{\cal L}_{\rm SSYM}}{F^{*\ib}}}
    + \kappa^2\partder{\Kmod}{\modulb^\ib}
        \vev{\partder{{\cal L}_{\rm SSYM}}{F^{*0}}}\right)\times{}\cr
&\omit\hfill$\displaystyle{ \left( \vev{\partder{{\cal L}_{\rm SSYM}}{F^j}}
    + \kappa^2\partder{\Kmod}{\modul^j}
        \vev{\partder{{\cal L}_{\rm SSYM}}{F^0}}\right)}$\cr
&-\ 3\kappa^2 \left| \vev{\partder{{\cal L}_{\rm SSYM}}{F^0}}\right|^2\
    +\ O(\kappa^4\mu^8).\cr }
\eqn\VeffFormula
$$
Note that the right hand side of this formula is a bilinear product of
the operators' VEVs and not a VEV of the bilinear product of the same
operators;
for example, the gaugino condensate's contribution to the effective potential
\VeffFormula\ is proportional to $|\vev{\lambda\lambda}|^2$ and not
to $\vev{\lambda\lambda\bar\lambda\bar\lambda}$.
\foot{This has to do with the fact that the $F^i$, $F^0$ and $M$
    auxiliary fields contribute to the effective potential via both
    the second and the third diagrams \MainDiagrams, but the contributions
    via the second diagrams cancel against those of the other components
    of the $\modul^i$, $\ccf$ and gravitational supermultiplets.
    The distinction between the `connected' parts of the VEVs such as
    $\vev{\lambda\lambda\bar\lambda\bar\lambda}$ and their `disconnected'
    parts such as $\vev{\lambda\lambda}\vev{\bar\lambda\bar\lambda}$
    corresponds to the distinction between the second and the third
    diagrams \MainDiagrams, both with an auxiliary propagator.
    It is important to keep track of this distinction; this is precisely
    why we did not integrate out the auxiliary superfields at the beginning
    of our calculations.}
Therefore, eq.~\VeffFormula\
may be re-written in terms of an effective superpotential
if and only if the SSYM expectation values agree with
$$
\eqalign{
\vev{\partder{{\cal L}_{\rm SSYM}}{F^j}}(\modul,\modulb,\ccf,\ccfb)\ &
=\ \partder{\widetilde \Wmod(\modul,\ccf)}{\modul^j}\cr
{\rm and}\quad
\vev{\partder{{\cal L}_{\rm SSYM}}{F^0}}(\modul,\modulb,\ccf,\ccfb)\ &
=\ \widetilde \Wmod(\modul,\ccf)\ \equiv\ \ccf^3 \Wmod(\modul)\cr }
\eqn\Wcondition
$$
for some holomorphic $\Wmod(\modul)$.

Up until this point we essentially followed the arguments of ref.~[\FGN]
while using only the most general properties of a
confining SSYM theory.
Extracting the explicit component form of ${\cal L}_{\rm SSYM}$ from the
relevant superspace integral in eq.~\SUGRACT, we have
$$
\partder{{\cal L}_{\rm SSYM}}{F^j}\
=\ {\uu\over4}\,\, \partder{\tilde f^W(\modul,\ccf)}{\modul^i}
\quad{\rm and}\quad
\partder{{\cal L}_{\rm SSYM}}{F^0}\
=\ {\uu\over4}\,\, \partder{\tilde f^W(\modul,\ccf)}{\log\ccf^3}\,,
\eqn\SSYMoperators
$$
which reduces eqs.~\Wcondition\ to
$$
\partder{\Wmod(\modul)}{\modul^i}\
=\ {\vev\uuh\over4} \,\, \partder{\tilde f^W}{\modul^i}
\quad{\rm and}\quad \Wmod(\modul)\
=\ {\vev\uuh\over4} \,\, \partder{\tilde f^W}{\log\ccf^3}\,.
\eqn\WmodEQS
$$
Note that these equations would be inconsistent
without the $\ccf$-dependent Wess-Zumino term in
the Wilsonian gauge coupling $\tilde f^W$ and
therefore would fail to produce
an effective potential for the moduli that is consistent
with eq.~\SUSYVeff\ and  local SUSY;
this is precisely the problem encountered in ref.~[\FGN].
On the other hand, given the correct Wilsonian coupling \Wilsanom,
eqs.~\WmodEQS\ are consistent, but only if $\vev\uuh(\modul)$ satisfies
eq.~\uuhVEV\ (note that $\cano=-T(G)$ for a pure SSYM theory),
and they have a unique solution for $\Wmod(\modul)$, namely eq.~\Wmoduli.

Usually, eq.~\uuhVEV\ is derived in the way presented in section~4.1
--- from the renormalization group equation for the gauge coupling,
supplemented by the phase formula for the gaugino condensate.
The moduli fields do not enter into this derivation, they are simply
parameters that affect the Wilsonian gauge coupling $f^W(\modul)$.
Here we obtain exactly the same equation for $\vev\uuh$ from the argument
that has nothing to do with the renormalization group and everything
to do with the moduli:
Eq.~\uuhVEV\ must hold, or else the effective potential for the moduli
would not be locally supersymmetric.
We find it remarkable  that the two totally unrelated arguments
produce the same result.
This agreement confirms that
eq.~\uuhVEV\ {\it must be exact}; there can be no corrections
that are not suppressed by higher powers of $\mu/\mpl$.
Moreover, the fact that the only solution to eqs.~\WmodEQS\ is the same
$\Wmod(\modul)$ that is obtained from eq.~\uuhWeff\ by integrating
out the $\uuh$ field gives us a much better confidence in the superpotential
\uuhWeff\ than  eq.~\uuhVEV\ alone.
Indeed, eq.~\uuhWeff\ gives the only form for the $\Weff(\modul,\uuh)$
that lead to correct formul\ae\ \uuhVEV\ and \Wmoduli\ for the
$\vev\uuh(\modul)$ and for the $\Wmod(\modul)$ and does so without
any unnatural fine tuning of its parameters.
In our opinion, this proves that the superpotential~\uuhWeff\
is exact, modulo corrections suppressed by negative powers of $\mpl$.

\section{Supersymmetric Hidden Sectors with Matter.}
\ldf\CHSW{P.~Candelas, G.~Horowitz,
    A.~Strominger and E.~Witten, \nup258 (1985) 46.}
Calabi-Yau compactifications of the ten-dimensional heterotic
string give rise to hidden sectors with $G\subset E'_8$ and no
matter multiplets charged under~$G$.\refmark{\CHSW}
For more general four-dimensional string vacua, absence of the hidden
matter is the exception rather than the rule.
In order to understand what happens in string models of this kind,
we now turn our attention to hidden sectors with charged
matter.\refmark{\TVY-\Amati, \hiddenmatter}
In this section, we assume that the vacuum state of the hidden
sector is stable and supersymmetric;
these assumptions allow us to extend the analysis of section 4.2 to
the more generic case at hand.
Hidden sectors that break SUSY or have unstable vacua will be dealt with
in section~4.5.

Let us recall the arguments we used to derive
eqs.~\VeffFormula\ and \Wcondition.
In those arguments, we made no use of any features that are peculiar
to matter-less SSYM theories but used only
the general properties of their strong dynamics,
namely the stability of the vacuum,
the confinement and the unbroken supersymmetry.
Hence, for any hidden sector with matter that has the same general properties,
eqs.~\VeffFormula\ and \Wcondition\ should hold true.
The only difference is that the ${\cal L}_{\rm SSYM}$ should be
extended to ${\cal L}_{\rm hidden}$ that also
accommodates the hidden matter multiplets $\matter^I$.
Thus, instead of eqs.~\SSYMoperators\ we now have
$$
\openup 1\jot
\eqalign{
\partder{{\cal L}_{\rm hidden}}{F^i}\ &
=\ {\uu\over4}\,\partder{\tilde f_W}{\modul^i}\
    +\ \partder{\widetilde W}{\modul^i}\
    +\ \overline{F^I}\matter^J\,\partder{Z_{\Ib J}}{\modul^i}\cr
{\rm and}\quad\partder{{\cal L}_{\rm hidden}}{F^0}\ &
=\ {\uu\over4}\,\partder{\tilde f_W}{\log\ccf^3}\
    +\ \partder{\widetilde W}{\log\ccf^3}\
    +\ \overline{F^I}\matter^J\,\partder{Z_{\Ib J}}{\log\ccf^3}\,.\cr
}\eqn\GHSoperators
$$
Fortunately, the unbroken supersymmetry of the hidden sector prevents
the operators $\overline{F^I}\matter^J$ from acquiring non-zero
expectation values, so when we substitute eqs.~\GHSoperators\ into
eqs.~\Wcondition\ for the effective superpotential, we have only
$$
\partder{\Wmod(\modul)}{\modul^i}\
=\ {1\over 4} \partder{f_W}{\modul^i}\,\vev\uuh\
+\ \vev{\partder{W(\matter,\modul)}{\modul^i}}
\eqn\WmodDer
$$
and
$$
\Wmod(\modul)\
=\ {\cano\over 32\pi^2}\,\vev\uuh\
+\ \vev{W(\matter,\modul)}
\eqn\WmodVal
$$
to worry about.
The precise meaning of the $\vev{\partial W/\partial\modul^i}$
here is
$$
\eqalign{
\vev{\partder{W(\matter,\modul)}{\modul^i}}\ &
=\ {1\over2} \partder{M_{IJ}}{\modul^i}\,\vev{\matter^I\matter^J}_{\rm hid}\
+\ {1\over3}
\partder{Y_{IJK}}{\modul^i}\,\vev{\matter^I\matter^J\matter^K}_{\rm hid}\
+\cdots\cr
&\equiv\ \sum_t \partder{y_t(\modul)}{\modul^i}\,\vev{\QQ^t}_{\rm hid} \,,\cr }
\eqn\ycouplingdef
$$
where the `$\cdots$' correspond to the non-renormalizable terms in the
superpotential that may become relevant because of unusually large VEVs
of the hidden matter scalars $\matter^I$.
On the second line in \ycouplingdef, $\QQ^t$ runs over all the relevant
gauge-invariant polynomials of the hidden matter superfields~$\matter^I$
and the $y_t$ are the corresponding masses or couplings.

The operators $\uuh$ and $\QQ^t$ are chiral
and their expectation values in a supersymmetric vacuum depend on the
holomorphic couplings of the hidden sector, namely the $f_W(\modul)$
and the $y_t(\modul)$, but not on the non-holomorphic couplings such as
the $Z_{\Ib J}$ matrices.
Furthermore, the functional form of $\vev\uuh(f_W,y)$
and $\vev{\QQ^t}(f_W,y)$
\foot{$y$  without a subscript stands for the whole set of $y^t$.}
is strongly constrained by the eqs.~\WmodDer.
Indeed, once we write eqs.~\WmodDer\ in terms of
$\vev\uuh\bigl(f_W(\modul),y(\modul)\bigr)$ and
$\vev{\QQ^t}\bigl(f_W(\modul),y(\modul)\bigr)$
it becomes obvious that they are mathematically consistent
with each other if and only if
$$
\left.\eqalign{
\Wmod(\modul)\ &
=\ {\bf W}(f_W,y)\cropen{1\jot}
\coeff14 \vev\uuh(\modul)\ &
=\ \partder{{\bf W}(f_W,y)}{f_W}\cr
\vev{\QQ^t}(\modul)\ &
=\ \partder{{\bf W}(f_W,y)}{y_t}\cr
}\right|_{\textstyle{f_W(\modul),\,y(\modul)}}
\eqn\Wconsistency
$$
for some holomorphic function $\bf W$ of the gauge and the Yukawa
couplings.
The form of eqs.~\Wconsistency\ immediately suggests a Legendre transform
that replaces a holomorphic function ${\bf W}$ of the couplings $f_W$ and
$y_t$ with a holomorphic function $\Xi$ of the condensates $\uuh$ and
$\QQ^t$ (but not of the $f_W$ or $y_t$).
After the transform, all of the eqs.~\Wconsistency\ can be
expressed in terms of a single effective superpotential
$$
\Weff(\modul,\uuh,\QQ)\
=\ \coeff14 f_W(\modul)\,\uuh\ + \sum_t y_t(\modul)\QQ^t\
-\ \Xi(\uuh,\QQ) .
\eqn\WeffW
$$
Specifically, the expectation values $\vev\uuh$ and $\vev{\QQ^t}$
are the supersymmetric solutions of
$$
\openup 1\jot
\eqalign{
\partder{\Weff(\modul,\uuh,\QQ)}{\strut\uuh}\ &
=\ \partder{\Weff(\modul,\uuh,\QQ)}{\QQ^t}\ =\ 0,\cr
\ie,\quad \partder\Xi{\strut\uuh}\ =\ \coeff14 f_W(\modul) &
\qquad {\rm and}\quad \partder\Xi{\QQ^t}\ =\ y_t(\modul)\,,\cr
}\eqn\uuQQsolns
$$
and the moduli superpotential is simply
$$
\Wmod(\modul)\
=\ \Weff\left(\modul, \vev\uuh(\modul), \vev{\QQ}(\modul)\right),
\eqn\WmodW
$$
exactly as in eq.~\Wmoduli.
Physically, the first two terms in the superpotential $\Weff$
have obvious origins at the tree level of the hidden sector, while
the $-\Xi(\uuh,\QQ)$ term should be thought of as dynamically induced
at the non-perturbative level.
Notice that by construction, this dynamical term does not depend
on any couplings of the hidden sector;
its exact form, therefore, should be completely determined by the
gauge group $G$ and the spectrum of the charged matter fields
$\matter^I$.
It is this coupling-blindness that will allow us to derive exact formul\ae\
for the $\Weff$ for many EQFTs.

In the previous section, we saw that $\Weff$ is determined
by requiring the consistency of eqs.~\WmodEQS.
Similarly,  eqs.~\WmodVal,
\WeffW\ -- \WmodW\  imply
$$
\partder{\Xi(\uuh,\QQ)}\uuh\ -\ {\Xi(\uuh,\QQ)\over\uuh}\
=\ {\cano\over 32\pi^2}\,,
\eqn\diffeq
$$
and since the $\Xi(\uuh,\QQ)$ is blind to the moduli and to the couplings,
eq.~\diffeq\ must be satisfied identically, \ie, for all possible
values of $\uuh$ and $\QQ^t$.
Solving eq.~\diffeq\ and substituting the solution into eq.~\WeffW\
gives us
$$
\Weff(\modul,\uuh,\QQ)\
=\ \Wtree(\modul,\QQ)\ +\ {\uuh\over 4}\, f_W(\modul)\
-\ {\uuh\over 32\pi^2}\,\left( \cano\log{\uuh\over\Lambda^3}\
    -\ P(\QQ) \right) ,
\eqn\BigW
$$
where $\Wtree(\modul,\QQ)\equiv\sum_t y_t(\modul)\QQ^t$ is
the Wilsonian superpotential of the perturbative theory re-expressed
in terms of the gauge-invariant matter condensates $\QQ^t$ instead
of the hidden matter fields $\matter^I$ themselves.
The $P(\QQ)$ in eq.~\BigW\ is a holomorphic function
of the hidden matter condensates $\QQ^t$; it does not depend
on the gaugino condensate $\uuh$ or on the moduli.
In the limiting case of a hidden sector without  matter
fields, $P$ is simply a constant and eq.~\BigW\
 reduces to eq.~\uuhWeff.

Notice that both the general form of the superpotential \BigW\ and
the way it determines $\vev\uuh(\modul)$, $\vev{\QQ^t}(\modul)$
and $\Wmod(\modul)$ (\cf~eqs.\ \uuQQsolns\ and \WmodW)
follow from a single physical demand:  Whenever the vacuum state
of a hidden sector is stable and supersymmetric, the effective potential
for the moduli generated by that hidden sector must be consistent
with the local supersymmetry.
This general argument even fixes the value of the coefficient $\cano$
to be exactly as in eq.~\canodef.
Consequently, regardless of the exact spectrum of the
hidden matter fields, we can always
re-write the effective superpotential in a manifestly \K-invariant
form
$$
\eqalignno{
\widetilde\Weff(\ccf,\modul,\uu,\QQ)\ &
\equiv\ \ccf^3 \Weff(\modul,\uuh,\QQ) &
\eqname\BigWuu\cr
&=\ \widetilde\Wtree(\ccf,\modul,\QQ)\
    +\ {\uu\over 4}\,\tilde f_W(\ccf,\modul)\
    -\ {\uu\over 32\pi^2}\,\left( \cano\log{\uu\over\Lambda^3}\
	-\ P(\QQ) \right)\cr }
$$
($\widetilde\Wtree\equiv\ccf^3\Wtree$).

\ldf\nsei{N.~Seiberg, Rutgers preprint RU-93-45.}
The function $P(\QQ)$ is not determined by the above arguments;
nevertheless, knowing that the non-perturbative part of the effective
superpotential is completely blind to the couplings
of the theory helps us to turn symmetry considerations into severe
constraints upon the form of that function.\refmark{\TVY-\Amati,\nsei}
Indeed, eq.~\BigWuu\
must be invariant under all exact symmetries of the EQFT
(for spontaneously broken symmetries, the superpotential
is invariant but the solutions of eqs.~\uuQQsolns\ are not);
in particular, the function $P(\QQ)$ must be invariant.
However, because $P(\QQ)$ is blind to the non-gauge couplings of the theory,
it would remain invariant even if we were to change those couplings in
an asymmetric way.
Therefore,
{\it $P$ must respect any `flavor' symmetry of
the strong gauge interactions, no matter how badly this symmetry may
be broken by the other interactions}.

The R-symmetries and even the anomalous flavor symmetries
of the gauge interactions also impose constraints on the $P(\QQ)$.
For such symmetries we may change the continually-adjustable couplings
of the hidden sector until the flavor symmetry becomes an exact
\K-symmetry and the Adler-Bell-Jackiw-Konishi anomaly of the gauge
interactions is exactly cancelled by the Wilsonian gauge coupling
transforming as in eq.~\fLocal.
\foot{This is precisely what happens to the symmetries arising from
    the target-space modular invariance in string theory.
    \refmark{\hiddenmatter, \KLc}}
After such deformation, we must have an invariant $\widetilde\Weff$;
according to eq.~\BigWuu, this requires the invariance of
the combination $8\pi^2 \tilde f_W(\ccf,\modul)+P(\QQ)$.
(We are using eq.~\BigWuu\ rather than \BigW\ because the
 $\widetilde\Weff$ and the $\uu$ are \K-invariant while the $\Weff$
 and the $\uuh$ are not.)
The transformation rule for the $\tilde f_W$ is just  eq.~\fLocal\
without the \K\ term; hence, $P(\QQ)$ should transform according to
$$
P(\QQ')\ =\ P(\QQ)\ + \sum_r^{\rm hidden\atop matter} 2T(r)\,
\log\det \symmat^{(r)} .
\eqn\Ptrans
$$
Again, because of the coupling-blindness of the $P(\QQ)$,
{\it eq.~\Ptrans\ must hold regardless of how badly the flavor symmetry
in question may be broken by the non-gauge interactions.}

In many cases, eq.~\Ptrans\ completely determines the form of $P(\QQ)$
(except for an additive constant).
For example, consider a SQCD-like hidden sector with
$N_C$ colors and $N_F$ flavors.
For $N_F<N_C$, the only independent relevant chiral condensates of
the theory are $\QQ^{IJ}=\matter_L^I\cdot\matter_R^J$.
(This is not true for theories with $N_F\ge N_C$,
which are discussed  later in this section.)
Under the $SU(N_F)\times SU(N_F)$ flavor symmetry of the theory, the $\QQ^{IJ}$
form a single irreducible $(N_F,N_F)$ representation;
therefore, all $SU(N_F)\times SU(N_F)$ invariant holomorphic functions of
these condensates depend only on the determinant of the $\QQ^{IJ}$ matrix.
For the vector $U(1)$ symmetry and for the pure-R symmetry
eq.~\Ptrans\ is trivially satisfied,
\foot{Because the \K\ factor cancels out of eq.~\Ptrans, this equation
    is automatically satisfied by the pure-R symmetry of
    the hidden sector, regardless of whether this symmetry is anomalous or not.
    Hence, for any supersymmetric hidden sector, one should only check
    eq.~\Ptrans\ for the flavor symmetries that commute with SUSY.}
but the constraint imposed by eq.~\Ptrans\ for the anomalous axial symmetry
has a unique solution, namely $P(\det\|\QQ^{IJ}\|)=\log\det\|\QQ^{IJ}\|
+\rm const$.
Therefore,
$$
\eqalignno{
\Weff\ =\ {}&
\Wtree(\modul,\QQ)\ +\ {\uuh\over4}\, f^W(\modul) &
\eqname\WVYT\cr
&+\ {\uuh\over 32\pi^2}\left( (N_C-N_F)\log{\uuh\over\Lambda^3}\
    +\ \log{\det\|\QQ^{IJ}\| \over\Lambda^{2N_F}}\ +\ C\right) ,\cr }
$$
where $C$ is a constant, presumably of order $O(1)$.
{}From the low energy point of view,  the Wilsonian gauge coupling $f^W$,
the ultraviolet cutoff $\Lambda$ and the unknown constant $C$ can all be
combined into a single parameter, which can be identified with the
confinement scale of the theory.
In terms of that parameter, eq.~\WVYT\ becomes identical to
the well-known Taylor-Veneziano-Yankielowicz superpotential for SQCD.
\refmark{\TVY}

In the arguments leading to eq.~\WVYT, the assumption was
that the confinement scale
$\mu$ of the hidden sector is much smaller than $\Lambda$ or $\mpl$,
or, rather, that all the condensates are much smaller than $\Lambda$ or $\mpl$
to the appropriate power.
However, there is no need to assume $\vev\uuh=O(\mu^3)$ or
$\vev{\QQ^{IJ}}=O(\mu^2)$, and indeed,
the superpotential \WVYT\ suffers no corrections when
some of the condensates are much larger or much smaller than the others.
This observation
 can be used to establish a  relationship between
the  constants $C$ of hidden sectors with different numbers of colors and
flavors.\refmark{\SVc,\Amati}
Suppose a `squark' $\matter_L^1$ and an `antisquark' $\matter_R^1$ somehow
acquire Higgs-like VEVs that are much larger than the confinement scale $\mu$
(this can be easily arranged by a judicious choice of the $\Wtree$).
The superpotential \WVYT\ does not care that the theory has a threshold,
all it sees is that
$\vev{\QQ^{11}}\approx\vev{\matter_L^1}\vev{\matter_R^1}\gg\mu^2$.
On the other hand, the physics of sub-threshold energies can also be
described in terms of a low-energy effective theory,
which for the problem at hand has $N'_C=N_C-1$, $N'_F=N_F-1$
and a Wilsonian gauge coupling
$f'_W=f^W+(1/8\pi^2)\log(\vev{\matter_L^1}\vev{\matter_R^1}/2\Lambda^2)$
(\cf~eq.~\Wilthreshold).
It is easy to see that both points of view lead to exactly the same
effective superpotential for the $\uuh$ and the $\QQ^{IJ}$ ($I,J\ge2$),
provided $C(N_C-1,N_F-1)=C(N_C,N_F)+\log 2$.
Now suppose that one of the quark flavors, \eg,~$\matter^1$ is very heavy.
Again, there are two ways to analyse this case:
Either first integrate out the heavy
flavor from the perturbative low-energy EQFT and then use
eq.~\WVYT\ for that theory,
or first write down eq.~\WVYT\ for the theory with all $N_F$ flavors
and then integrate out all the condensates involving the heavy fields.
Fairly straightforward algebra shows that again the perturbative
eq.~\Wilthreshold\ and the non-perturbative eq.~\WVYT\ are entirely
consistent with each other, provided
$C(N_C,N_F-1)=C(N_C,N_F)-1-\log(-32\pi^2)$, and hence
$$
\eqalignno{
\Weff\ =\ {}&
\Wtree(\modul,\QQ)\ +\ {1\over4}\,
\uuh\, f^W(\modul) &
\eqname\CVYT\cr
&+\ {\uuh\over 32\pi^2}\left( (N_C-N_F) \log{\uuh\over-64\pi^2e\Lambda^3}\
    +\ \log{\det\|\QQ^{IJ}\| \over(2\Lambda^2)^{N_F}}\ +\ C_0\right) ,\qquad\cr
}
$$
where $C_0$ is a universal constant, the same for all $N_F<N_C$.
The same constant $C_0$ also appears in the effective superpotentials
of other
hidden sectors, for example when the hidden gauge
group $G=SO(N_C)$ rather than $SU(N_C)$.\refmark{\SVc}
For the hidden matter that forms $N_F$ vector representations of the $SO(N_C)$
we can repeat our previous argument almost verbatim.
The result is:
$$
\eqalignno{
\Weff\ =\ {}&
\Wtree(\modul,\QQ)\ +\ {1\over4}\,
\uuh\, f^W(\modul) &
\eqname\SONVM\cr
&+\ {\uuh\over 32\pi^2}\left( (N_C-2-N_F) \log{\uuh\over-64\pi^2e\Lambda^3}\
    +\ \log{\det\|\QQ^{IJ}\| \over(2\Lambda^2)^{N_F}}\ +\ C_0\right) ,\qquad\cr
}
$$
where $\QQ^{IJ}$ is now a symmetric matrix originating in
$\matter^I\cdot\matter^J$
and $C_0$ is exactly the same constant as in eq.~\CVYT\ because
for the case of $SO(6)$ with $N_F=0$ the hidden sector is identical to the
$SU(4)$ SSYM theory (also with $N_F=0$).
The actual computation of $C_0$  requires an honest calculation of a
non-perturbative VEV in a strongly interacting theory
which is outside the scope of this paper.
What we find remarkable is that {\it one} such calculation would be sufficient
to evaluate {\it all} the relevant non-perturbative VEVs in a whole class of
strongly interacting theories!\refmark{\SVc}

The methods just presented (relating different supersymmetric gauge theories)
can also be used
to {\it completely} specify the effective superpotential
for $SU(N_C)$ gauge theories with $N_F\ge N_C$.
The well known difficulty with such theories is that they have
more independent relevant matter condensates than just the
$\QQ^{IJ}\equiv\matter^I_L\cdot\matter^J_R$;\refmark{\MPRV, \Amati}
for example, in the $N_F=N_C$ case there are two additional condensates,
$\QQ^L\equiv\det\matter_L$ and $\QQ^R\equiv\det\matter_R$.
\foot{In fully indexed notations, $\QQ^L\equiv\epsilon^{s_1,\ldots, s_N}
    \matter^1_{s_1}\cdots\matter^N_{s_N}$ where $N=N_F=N_C$,
    $s_1,\ldots,s_N$ are the gauge indices, and ditto for the $\QQ^R$.}
Although the operators $\QQ^L$, $\QQ^R$ and $\QQ^{IJ}$ are related to
each other via $\QQ^L\QQ^R=\det\|\QQ^{IJ}\|$,
the non-perturbative effects in the quantum theory lead to
$\vev{\QQ^L}\vev{\QQ^R}\neq\det\|\vev{\QQ^{IJ}}\|$.
This is why in the effective superpotential~\BigW, the classical variables
$\QQ^L$, $\QQ^R$ and $\QQ^{IJ}$ should be regarded as independent.
Moreover, the condensates $\QQ^L$
and $\QQ^R$ are always relevant (even for $N>3$ when the couplings associated
with the operators $\QQ^L$ and $\QQ^R$ are non-renormalizable).
This follows from the observation that the theory may have SUSY-preserving
Higgs-like VEVs $\vev{\matter^I_{L,s}}\propto\delta^I_s$, $\vev{\matter_R}=0$
or vice verse, and the simplest holomorphic, gauge-invariant order parameters
for this behavior of the theory are the $\QQ^L$ and the~$\QQ^R$.

Our first step towards constructing the $P(\QQ^{IJ},\QQ^L,\QQ^R)$
function for a hidden sector with $N_F=N_C=N$ is to use eqs.~\Ptrans\
for the $SU(N)_L\times SU(N)_R\times U(1)_V\times U(1)_A$ flavor symmetry,
under which the condensates $\QQ^{IJ}$, $\QQ^L$ and $\QQ^R$ transform
as $({\bf N},\overline{\bf N},0,+2)$, $({\bf1},{\bf1},+1,+N)$
and $({\bf1},{\bf1},-1,+N)$.
Under these circumstances, eq.~\Ptrans\ yields
$$
P(\QQ^{IJ},\QQ^L,\QQ^R)\
=\ \log {h(\det\|\QQ^{IJ}\| , \QQ^L\QQ^R) \over (2\Lambda^2)^N}\,,
\eqn\PFEC
$$
where $h$ is a  homogenous function of degree one,
but its specific form cannot be determined from the symmetry arguments alone.
\refmark{\Amati}
(For obvious reasons, the product $\QQ^L\QQ^R$ has exactly the same quantum
 numbers as the determinant $\det\|\QQ^{IJ}\|$.)
Instead, let us compare the $h$ functions for theories with different values
of $N$.
As before, we  consider the limit of large Higgs-like VEVs
$\vev{\matter_L^1}$ and $\vev{\matter_R^1}$, in which
$\QQ^L=\vev{\matter_L^1}\QQ^{\prime L}$,
$\QQ^R=\vev{\matter_R^1}\QQ^{\prime R}$ and
$\det\|\QQ^{IJ}\|=\vev{\matter_L^1}\vev{\matter_R^1}\det'\|\QQ^{IJ}\|$,
where the primes refer to the condensates of the sub-threshold theory
with $N'_F=N'_C=N-1$.
Substituting these values into eq.~\BigW\ and using the homogeneity of $h$ in
eq.~\PFEC, we find that the sub-threshold theory has
$$
\! \Weff'\, -\, \Wtree'\, =\, \coeff14 f^W\,\uuh\,
+\, {\uuh\over 32\pi^2}\,\log
	{\vev{\matter_L^1}\vev{\matter_R^1})\over2\Lambda^2}\,
+\, {\uuh\over 32\pi^2}\, \log{ h( \det'\|\QQ^{IJ}\| ,
	\QQ^{\prime L}\QQ^{\prime R}) \over (2\Lambda^2)^{N-1}}
\eqn\WFECi
$$
(there is no $\uuh\log\uuh$ term because $\cano={N'_C-N'_F}={N_C-N_F}=0$).
The first two terms on the right hand side of this formula can be
combined into $\coeff14 f^{\prime W}\uuh$ and according to
eq.~\Wilthreshold,
the $f^{\prime W}$ is precisely the Wilsonian gauge coupling of the
sub-threshold effective theory.
Similarly, the remaining $\log(h/(2\Lambda^2)^{N-1})$ term
can be identified with the non-perturbative superpotential
$\log(h'/(2\Lambda^2)^{N'})$ of the subthreshold theory,
which means that both $h$ functions
(above and below the threshold) have exactly the same analytic form.
An immediate corollary  is that the $h$ functions has
 the same analytic form for all $N_F=N_C=2,3,4,\ldots$.

Now consider the special case $N_F=N_C=2$.
For this theory $\QQ^L$ is a squark bilinear,
$\QQ^R$ is an antisquark bilinear,
and because  $SU(2)$ does not distinguish between
squarks and  antisquarks,
it also does not distinguish between these two condensates and the
squark-antisquark bilinears $\QQ^{IJ}$.
In fact, the full flavor symmetry of this gauge theory is $SU(4)\times U(1)$
rather than $SU(2)\times SU(2)\times U(1)\times U(1)$
and the bilinear condensates $\QQ^L$, $\QQ^R$ and $\QQ^{IJ}$
form an irreducible six-dimensional representation of the $SU(4)$.
This representation has only one independent invariant,
which in our notations looks like
$\QQ^{11}\QQ^{22}-\QQ^{12}\QQ^{21}-\QQ^L\QQ^R$.
Comparing this expression to eq.~\PFEC, we immediately conclude that
for $N_F=N_C=2$,
$$
h\ \propto\ \det\|\QQ^{IJ}\|\ -\ \QQ^L\QQ^R .
$$
However, we have just argued that the $h$ functions
are universal for all theories with $N_F=N_C$;
hence, for all such theories
$$
\Weff\ =\ \Wtree(\modul,\QQ)\ +\ {1\over4}\, \uuh\, f^W(\modul)\
+\ {\uuh\over 32\pi^2}\,\left( \log{ \det\|\QQ^{IJ}\| - \QQ^L\QQ^R
    \over(2\Lambda^2)^N }\ +\ C_0 \right) .
\eqn\WFEC
$$
(The constant $C_0$ in this formula is exactly as in eqs.~\CVYT\ and
\SONVM;  this equality follows from the decoupling of massive flavors
and can be proven in  the same way as \CVYT.)

The theories with $N_F>N_C$ can be analysed in exactly the same manner.
For the sake of notational simplicity, we  only discuss
the case $N_F=N_C+1$, in which the independent relevant matter condensates
are comprised of $\QQ^{IJ}$, $\QQ^L_I$ and $\QQ^R_J$.
\foot{%
    The definition of the $\QQ^L_I$ is $(1/N_C!)
    \epsilon^{s_1, \ldots, s_{N_C}}\epsilon_{I_1,\ldots I_{N_C},I}\,
    \matter^{I_1}_{L,s_1}\cdots\matter^{I_{N_C}}_{L,s_{N_C}}$;
    the $\QQ^R_J$ is defined  likewise.
    For $N_F>N_C+1$ the $\QQ^L$ and the $\QQ^R$ are $(N_F-N_C)$-index
    tensors of the respective $SU(N_F)$ flavor groups and the
    notations become somewhat unwieldy.
    The reasons why all of the $\QQ^L_{\ldots}$, $\QQ^R_{\ldots}$ and
    $\QQ^{IJ}$ should be regarded as independent variables in
    the effective superpotential are exactly as in the $N_F=N_C$ case;
    the same goes for the relevancy of all these condensates.}
The $SU(N_F)_L\times SU(N_F)_R\times U(1)_V\times U(1)_A$ flavor symmetry
of the theory now implies $P(\QQ)=
 \log\bigl( h(\det\|\QQ^{IJ}\|,\QQ^L_I\QQ^{IJ}\QQ^R_J)/\Lambda^{2N_F})$,
where as before $h$ is a homogeneous function of degree one.
Again, the arguments using the decoupling of large Higgs-like VEVs
tell us that $h$ should have  the same analytic
form for all $N_C\ge2$ as long as $N_F=N_C+1$,
and the specific form of $h$ is fixed by the special case
of $N_C=2, N_F=3$ where the flavor symmetry is extended to
the $SU(6)\times U(1)$.
The result is
$$
\eqalignno{
\Weff\ =\ {}&
\Wtree(\modul,\QQ)\ +\ {1\over4}\, \uuh\, f^W(\modul) &
\eqname\WFBC\cr
&+\ {\uuh\over 32\pi^2}\left( -\log{\uuh\over-64\pi^2e\Lambda^3}\
    +\ \log{\det\|\QQ^{IJ}\|\,-\,\QQ^L_I\QQ^{IJ}\QQ^R_J
	\over(2\Lambda^2)^{N_F}}\ +\ C_0\right) .\cr }
$$
Note that the coefficient $(N_C-N_F)$ of the $\uuh\log\uuh$ term
is now negative,
and because of this sign, the superpotential \WFBC\ leads to a stable
vacuum even in the absence of a classical superpotential $\Wtree$.
Indeed, eliminating $\uuh$ from eq.~\WFBC\ by its equation of motion
leads to an effective superpotential for the
matter condensates that looks like
$$
\Weff(\QQ)\ \propto\ \QQ^L_I \QQ^{IJ} \QQ^R_J\  -\ \det\|\QQ^{IJ}\|
\eqn\Wstability
$$
and has no run-away directions:
All eqs.~\uuQQsolns\ are satisfied by  $\vev{\QQ}=0$.
Furthermore,  the superpotential \Wstability\ contains no mass terms.
Hence, as conjectured in ref.~[\MPRV],
the confinement in $N_F=N_C+1$ theories gives rise to exactly
massless composite supermultiplets whose quantum numbers match those
of the $\QQ^L_I$, $\QQ^R_J$ and $\QQ^{IJ}$ condensates,
which is precisely the massless spectrum required by the 't~Hooft's
flavor anomaly matching conditions for the {\sl unbroken} chiral symmetry
$SU(N_F)\times SU(N_F)\times U(1)\times U(1)$.

Let us summarize the results of this section.
The superpotential \WVYT\ was derived  before, using
a variety of techniques  such as the
leading-order instantonic calculations or renormalization-group
equations.\refmark{\TVY-\Amati}
Here we obtain exactly the same superpotential from a
different approach, namely the consistency
of the effective locally supersymmetric theory for the moduli.
Our main result is that  this  superpotential is exact
(modulo corrections suppressed by the negative powers of $\mpl$)
and so are other superpotentials derived in this manner.
It is precisely the exactness of these superpotentials that legitimizes
the techniques we used  to relate different gauge theories to each other
and thus to derive the superpotentials \WFEC\ and \WFBC\ for the theories
with $N_F\ge N_C$.
In addition, we confirmed that  integrating
out the entire hidden sector --- namely, using eq.~\WmodW\
and inserting the result into eq.~\SUSYVeff, --- indeed gives the correct
effective potential for the moduli.

\section{Hidden Sectors in String Context.}
In the context of string unification, the main result
of the previous section is that for a generic
supersymmetric hidden sector, the general form and all the parameters
of the effective superpotential \BigWuu\
are exactly calculable in analytic form.
The calculation is done in terms of the low-energy EQFT and
needs very little input from the string theory, namely
the spectrum of the light particles, the relevant superpotential couplings
$y_t(\modul)$, and the Wilsonian gauge coupling $f^W(\modul)$.
The spectrum and the superpotential are determined
at the tree level of the heterotic string
whereas $f^W$ requires a one-string-loop calculation
(plus some tree-level data, to separate
the $f^W$ from the non-holomorphic terms in $\{g^{-2}\}^\ol$).
\refmark{\JLpascos, \DFKZa, \KLc}
Consequently, without ever performing  higher loop calculations,
we can obtain an analytic form of the non-perturbative effective
superpotential $\Wmod$ for the moduli fields and be sure that it is
exact.

In this section, we present several examples of analytically computing
$\Wmod$ in terms of the holomorphic Wilsonian  couplings of the theory.
In our examples, the hidden matter fields  are exactly massless
(perturbatively), but the chiral symmetry
is broken by the Yukawa couplings or by the non-renormalizable terms
in $\Wtree$.
Such behavior is quite rare in ordinary GUTs and thus did not receive
much attention in the literature.
In string unification, however, masslessness without a chiral symmetry
is the norm, and we believe this phenomenon deserves a closer look from
the effective superpotential point of view.\refmark{\hiddenmatter}

Our first example is a $SU(3)$ gauge theory with
three flavors of triplets $\matter^I_L$ and anti-triplets $\matter^J_R$
and with `baryon' number violating Yukawa self-couplings of the matter fields,
\foot{This type of a hidden sector often shows up in string models
    whose construction involves modding out by discrete $Z_3$
    symmetries.}
$$
\Wtree\ =\ Y_L(\modul)\,\det\|\matter_L\|\
+\ Y_R(\modul)\,\det\|\matter_R\|\ .
\eqn\exoneWtree
$$
Substituting \exoneWtree\ into eq.~\WFEC\ for $N_F=N_C$
and solving the resulting eqs.~\uuQQsolns\ and \WmodW\
result in
$$
\eqalign{
\Wmod(\modul)\ =\ 2\Lambda^3\,&
\left[ -8e^{-C_0}\, e^{-8\pi^2 f^W(\modul)}\,
	Y_L(\modul)\, Y_R(\modul)  \right]^{1/2} ,\cr
\vev\uuh\ =\ -32\pi^2\Lambda^3\,&
\left[ -8e^{-C_0}\, e^{-8\pi^2 f^W(\modul)}\,
	Y_L(\modul)\, Y_R(\modul)  \right]^{1/2} ,\cr
\vev{\QQ^L}\,\equiv\,\vev{\det\|\matter_L\|}\ =\ \Lambda^3\,&
\left[ -8e^{-C_0}\, e^{-8\pi^2 f^W(\modul)}\,
	Y_L^{-1}(\modul)\, Y_R(\modul)  \right]^{1/2} ,\cr
\vev{\QQ^R}\,\equiv\,\vev{\det\|\matter_R\|}\ =\ \Lambda^3\,&
\left[ -8e^{-C_0}\, e^{-8\pi^2 f^W(\modul)}\,
	Y_L(\modul)\, Y_R^{-1}(\modul)  \right]^{1/2} ,\cr
\omit\span \det\left\|\vev{\QQ^{IJ}}\right\|\,\equiv\,
	\det\left\|\vev{\matter_L^I\cdot\matter_R^J}\right\|\, =\ 0 .\cr
}\eqn\exoneres
$$
The individual VEVs $\vev{\QQ^{IJ}}$ are undetermined, but
adding quartic or other non-renormalizable couplings to the
tree-level superpotential \exoneWtree\ would force them
to vanish without affecting any of the results~\exoneres.
Without such non-renormalizable couplings, the theory has an accidental
$SU(3)\times SU(3)$ chiral symmetry, which remains unbroken by the
strong interactions;
instead, the theory forms a massless composite matter multiplet with
the quantum numbers of $\QQ^{IJ}$, in full agreement with the 't~Hooft's
flavor anomaly matching conditions.
On the other hand, the abelian flavor symmetries of the gauge theory are
explicitly broken by the Yukawa couplings $Y_L$ and $Y_R$, which are
necessary for the stability of the condensate VEVs~\exoneres:
In the limit $Y_L\to0$, $Y_R$ fixed, we have
$\vev{\QQ^L}\to \infty$ and  for $Y_R\to0$, $Y_L$ fixed, $\vev{\QQ^R}$
diverges;
in the double limit $Y_L,Y_R\to0$, both $\vev{\QQ^L}$ and $\vev{\QQ^R}$
become undetermined.

Unless the Yukawa couplings $Y_L$ and $Y_R$ are unusually small,
all the non-vanishing condensates~\exoneres\ are comparable in magnitude
and can be used as a crude estimate of the confinement scale of the theory:
$$
\vev{\QQ^L}\ \simeq\ \vev{\QQ^R}\ \simeq\ \vev\uuh\
\simeq\ \left[ \Lambda e^{-8\pi^2 f^W/6} \right]^3 \sim\ \mu^3,
\eqn\exonescales
$$
where the factor $1/6$ in the exponential is related to $\cren = -6$ for a
theory with $N_C=N_F=3$.
A more accurate formula for the physically normalized confinement
scale would be
$$
\mu\ \simeq\ \left[ \Lambda e^{-8\pi^2 f^W/6} \right]\,e^{\kappa^2 K/6} ,
\eqn\Kscale
$$
where the \K\ factor appears exactly as in eq.~\confscale\ and for exactly
the same physical reason.
Unfortunately, the numerical coefficient missing from eq.~\Kscale\
is not as easy to calculate as that for the matter-less SSYM theories.

The fact that all condensates (in eq.~\exonescales)
are proportional to a single scale  is characteristic
of hidden sectors in which all the important superpotential couplings
are renormalizable.
As a more generic example of this behavior, consider an $SU(N_C)$
with $N_F<N_C$ flavors and some gauge-singlet matter
fields with Yukawa coupling to the `quarks' and `antiquarks' and
also to each other.
For notational simplicity, we present the case of $N_F=1$ and only
one matter singlet $\matter_1$; thus,
$$
\Wtree\ =\ Y(\modul)\,\matter_1\matter_L\matter_R\
-\ \coeff13 Y_1(\modul)\,\matter_1^3 .
\eqn\extwoWtree
$$
Supersymmetric EQFTs with gauge-singlet matter fields are governed
by a simple rule:
{\it The only independent condensates involving the gauge singlet fields
are those fields themselves}.
\foot{%
    The physical reason for this rule is the fact that all of the physical
    couplings of the quantum degrees of freedom contained in the singlet
    fields can be rendered arbitrarily weak by adjusting the non-holomorphic
    parameters of the theory, namely the $Z^{(1)}$ matrix for the singlets.
    In the $Z^{(1)}\to\infty$ limit, the quantum nature of the singlets
    becomes irrelevant and only the Higgs-like VEVs remain to complicate the
    theory, but these VEVs obey the classical relations such as
    $$
    \vev{\matter_1^3}\ =\ \vev{\matter_1}^3 ,\qquad
    \vev{\matter_1\matter_L\matter_R}\
	=\ \vev{\matter_1}\vev{\matter_L\matter_R} ,
    \eqn\classrel
    $$
    \etc.
    In a supersymmetric vacuum, expectation values of chiral
    operators cannot depend on the non-holomorphic couplings.
    Therefore, equations such as \classrel\ must always hold true,
    even when the $Z^{(1)}$ factors are small and the physical Yukawa
    couplings of the theory are strong.}
Furthermore, {\it the non-perturbative part of the effective superpotential
cannot involve the gauge singlet fields}.
\foot{%
    This follows from the same argument:  In the $Z^{(1)}\to\infty$
    limit the singlets act like moduli rather than interacting matter
    fields and hence should only enter the effective superpotential
    \BigW\ through its classical part $\Wtree$.
    Since the entire effective superpotential is blind to the $Z$
    factors, $P(\QQ)$ remains singlet-independent even when the
    physical couplings of the singlets are strong.}
Consequently, the non-perturbative superpotential for the theory at hand
is exactly the Taylor-Veneziano-Yankielowicz superpotential for $N_F=1$.
Combining together eqs.~\extwoWtree\ and \CVYT, we find
$$
\!\eqalign{
\!\Wmod(\modul)\ =\ (N_C-\coeff13)\Lambda^3\,&
\left[ 2^{3N_C}e^{-3C_0}\, e^{-24\pi^2 f^W(\modul)}\,
	Y^3(\modul) Y_1^{-1}(\modul)  \right]^{1/(3N_C-1)} ,\cr
\vev\uuh\ =\ -32\pi^2\Lambda^3\,&
\left[ 2^{3N_C}e^{-3C_0}\, e^{-24\pi^2 f^W(\modul)}\,
	Y^3(\modul) Y_1^{-1}(\modul)  \right]^{1/(3N_C-1)} ,\cr
\vev{\QQ}\,\equiv\,\vev{\matter_L\cdot\matter_R}\ =\ \Lambda^2\,&
\left[ 2^{2N_C}e^{-2C_0}\, e^{-16\pi^2 f^W(\modul)}\,
	Y^{3-3N_C}(\modul) Y_1^{N_C-1}(\modul)  \right]^{1/(3N_C-1)} \cr
{\rm and}\quad\vev{\matter_1}\ =\ \Lambda\,&
\left[ 2^{N_C}e^{-C_0}\, e^{-8\pi^2 f^W(\modul)}\,
	Y(\modul) Y_1^{-N_C}(\modul)  \right]^{1/(3N_C-1)} .\cr
}\!\eqn\extwores
$$
Again, for  $O(1)$ Yukawa couplings $Y$ and $Y_1$
we find crude similarity between the magnitudes of the condensates and
the appropriate powers of the confinement scale:
$$
\vev{\uuh}^{1/3}\ \simeq\ \vev{\QQ}^{1/2}\ \simeq\ \vev{\matter_1}\
\simeq\ \Lambda e^{-8\pi^2 f^W/(3N_C-1)}\ \sim\ \mu
\eqn\extwoscales
$$
(for the theory at hand, $\cren=3N_C-1$;
a more accurate formula for the confinement scale $\mu$ involves
the \K\ factor, as in eq.~\exonescales).
Notice that both of the Yukawa couplings $Y$ and $Y_1$ are necessary
for the stability of the condensates~\extwores:
in the $Y\to\infty$ limit, the squark and the antisquark develop large
Higgs-like expectation values, while for $Y_1\to\infty$ it is the singlet
VEV $\vev{\matter_1}$ that grows out of control.

In our final example we dispense with both the singlet fields and
the Yukawa couplings;
instead, the Higgs-like VEVs of the massless squarks and antisquarks
are controlled by the non-renormalizable couplings.
The simplest theory of this kind is $SU(N_C)$ with $N_F=1$
and a tree-level superpotential
$$
\Wtree\ =\ {\gamma(\modul)\over 2\Lambda}\,(\matter_L\cdot\matter_R)^2;
\eqn\exthreeWtree
$$
in the context of string unification, $\Lambda$ is usually the same
as $\mpl$.
The non-renormalizable couplings are relevant only when the corresponding
operators have unexpectedly large VEVs;
for the problem at hand, the quartic coupling \exthreeWtree\ is relevant
if $\vev{(\matter_L\cdot\matter_R)^2}=O(\Lambda\mu^3)\gg\mu^4$.
Such large VEVs are characteristic of the Higgs limit of the theory,
$\vev\matter\gg\mu$.
However, because of the classical nature of the Higgs mechanism,
it does not contribute to expressions such as
$\vev{(\matter_L\cdot\matter_R)^2}-\vev{(\matter_L\cdot\matter_R)}^2$.
Consequently, the difference between the square of the squark-antisquark
bilinear condensate and the quartic condensate is irrelevant, and
in the effective superpotential \WeffW\ we do not need an independent
variable for the quartic condensate.
Thus $\Weff$ is given by  eq.~\CVYT,
with $N_F=1$ and $\Wtree=(\gamma/2\Lambda)\QQ^2$,
where $\QQ\equiv(\matter_L\cdot\matter_R)$ is the usual bilinear
condensate.
After going through
the usual algebra we arrive at
$$
\eqalign{
\Wmod(\modul)\ =\ (N-\half)\Lambda^3\,&
\left[ 2^{2N_C}e^{-2C_0}\, e^{-16\pi^2 f^W(\modul)}\,
	\gamma(\modul)  \right]^{1/(2N-1)} ,\cr
\vev\uuh\ =\ -32\pi^2\Lambda^3\,&
\left[ 2^{2N_C}e^{-2C_0}\, e^{-16\pi^2 f^W(\modul)}\,
	\gamma(\modul)  \right]^{1/(2N-1)} ,\cr
\vev{\QQ}\,\equiv\,\vev{\matter_L\cdot\matter_R}\ =\ \Lambda^2\,&
\left[ 2^{N_C}e^{-C_0}\,e^{-8\pi^2 f^W(\modul)}\,
	\gamma^{1-N}(\modul)  \right]^{1/(2N-1)} .\cr
}\eqn\exthreeres
$$
Note that the Higgs-like
VEVs of the squark and the antisquark are indeed very large compared to
the confinement scale of the theory.
To a crude approximation,
$$
\vev{\uuh}^{1/3}\ \sim\ \mu,\qquad {\rm but}\quad
\vev{\matter_L},\vev{\matter_R}\ \sim\ \root 4\of{\mu^3\Lambda}
\eqn\exthreescales
$$
as usual, this formula assumes $\gamma\sim O(1)$ and ignores the \K\ factor
in the confinement scale~$\mu$.
Below the Higgs threshold we effectively have a matter-less $SU(N_C-1)$
SSYM theory, so we can use eq.~\confscale\ to write down an exact
formula for the confinement scale of the model at hand;
all we need is eq.~\Wilthreshold\ for the threshold effect and eq.~\exthreeres\
for the un-normalized Higgs VEVs.
Combining all the factors together, we have
$$
\mu\ =\ \left[{8\pi^2 e^{1-C_0/(N_C-1)(2N_C-1)}\over N_C-1}\right]^{1/3}
\gamma^{1/(6N_C-3)}\, \Lambda
\exp\left( -{8\pi^2\Re f^W\over 3N_C-{3\over2}}\, +\,\coeff16 \kappa^2 K\right)
{}.
\eqn\exthreeconfscale
$$
The denominator of the $\Re f^W$ term in the exponential reflects the fact
that the theory has a threshold between $\mu$ and $\Lambda$:
$3N_C-{3\over2}$ is the appropriately weighted (\cf~eq.~\exthreescales)
average of the $\beta$-function coefficient $\cren'=3(N_C-1)$ for the
effective sub-threshold theory and of
$\cren=3N_C-1$ for the unbroken theory above the threshold.

\section{Families of EQFTs, Vacuum Instabilities and SUSY Breaking.}
In the previous sections, we discussed how
the hidden sectors of the unified theory generate a non-perturbative
effective potential for the moduli fields.
In this section, we  discuss the physical implications of this potential
but to put the discussion into a wide enough context, we
begin by considering the domains of validity of the low-energy EQFTs
and what happens when we try to push an EQFT outside its proper domain.

Consider a generic unified theory that has several alternative vacua.
Generally, expanding around each individual vacuum yields a different
spectrum of light particles and hence a different low-energy EQFT.
The situation becomes more complicated when the vacua of the unified theory
form continuous families.
In this case, the domain of validity of the EQFT obtained by expanding around
any particular vacuum also includes the
 immediate
neighborhood of that vacuum and
the moduli VEVs $\vev{\modul^i},\vev{\modulb^\ib}$ serve as coordinates
in this neighborhood.
However, when we go far away from the original vacuum, the low energy
physics may suffer more changes that could be handled by a mere change
of coordinates in the moduli space;
instead, re-expansion around the new vacuum may yield a different
spectrum of light particles and a different EQFT to describe their
low-energy behavior.
The key to this discontinuity is that although the particles' masses
vary continuously throughout the vacuum space,
at some point we have to make a decision whether a particular field
is light enough to include in the low-energy EQFT or is heavy enough
to integrate out.
This decision is necessarily arbitrary to some extent, but not completely
so; therefore, a continuous vacuum family of the heterotic string
(or other unified theory) may need several distinct EQFTs to describe
its low-energy behavior, and the domains of validity of those EQFTs
overlap at each other's edges, but cannot be amalgamated into a single
domain of a single EQFT.

When a low-energy EQFT approaches the limit of its domain of validity,
it usually gives some signals of its impending breakdown.
When the moduli VEVs enter a neighborhood of the vacuum space
where some normally heavy gauge or matter fields become light or
even massless, some Wilsonian couplings of the EQFT become singular.
For example, when the additional light fields are charged under
some gauge group $G_\ind$, the Wilsonian gauge coupling $f^W_\ind$
diverges like the logarithm of the low mass.
\foot{In string theory, and in particular in Kaluza-Klein compactifications
    of the ten-dimen\-sional heterotic string,
    an infinite number of heavy charged fields simultaneously become
    massless in the `decompactification' limit when the radius of the
    `internal' manifold becomes large.
    In this case, the  couplings of the four-dimensional
    EQFT diverge like powers (rather than the logarithm) of the
    appropriate modulus.
    We shall return to this scenario later in this section.}
Going in the opposite direction, we see light fields becoming superheavy,
and this phenomenon normally has a low-energy EQFT explanation in terms
of unexpectedly large VEVs of some matter scalars $\matter^I$:
The gauge fields of the symmetries broken by these large $\vev{\matter^I}$
become heavy via the Higgs mechanism while the non-Higgs matter fields
can also get superheavy masses via Yukawa couplings to the large VEVs.
When the large VEVs reach the Planck scale, the original EQFT breaks down,
not because of the heavy masses, but because we can no longer truncate
the expansions \Kexpansion, \Wexpansion\ and \deindexingf\ of the
Wilsonian couplings of the theory into powers of $\matter^I/\mpl$.

\ldf\ILLT{L.~Ib\'a\~nez, W.~Lerche, D.~L\"ust and S.~Theisen,
    \nup352 (1991) 435.}
Clearly, the VEVs of matter scalars can reach the Planck scale only if
they grow along an exactly flat direction of the classical scalar potential;
such directions are quite common in supersymmetric EQFTs, especially those
derived from the heterotic string.\refmark{\ILLT}
When a flat combination of matter fields does acquire a Planck-sized VEV,
it behaves exactly like a modulus, and once we integrate out all the fields
that become superheavy in the $\vev{\matter^{\rm flat}}=O(\mpl)$ regime,
the $\matter^{\rm flat}$ becomes a modulus since now it is both flat and
neutral (all the gauge symmetries that act upon the $\matter^{\rm flat}$
are too badly broken to retain in the low-energy EQFT).
On the other hand, whenever $\vev{\matter^{\rm flat}}$ happens to be small
compared to $\mpl$ and the gauge symmetries that act upon the
$\matter^{\rm flat}$ take part in the low-energy EQFT, the $\matter^{\rm flat}$
is not a modulus but a combination of charged matter fields $\matter^I$.
Thus, we see that the distinction between the moduli and the flat
combinations of the matter fields is as arbitrary as the boundaries
between the validity domains of EQFTs describing the same vacuum family
of the unified theory:
In each EQFT we know which scalar is a modulus and which is matter, but
as we cross from the domain of one EQFT into the domain of another,
a modulus may become a combination of matter fields and vice versa.

\ldf\DHVW{L.~Dixon, J.~Harvey, C.~Vafa and E.~Witten, \nup 261 (1985) 678;
    \nup 274 (1986) 185.}
For example, consider a string orbifold such as $Z_3$ and the
related Calabi-Yau compactification.\refmark{\DHVW, \CHSW}
In the orbifold limit, the gauge group is $SU(3)\otimes E_6\otimes E_8$
and the only neutral scalars are the nine toroidal moduli
and the dilaton.
However, among the 243 scalar superfields that transform like
81 triplets of the $SU(3)$ and are neutral with respect to
the $E_6\otimes E_8$, there are 27 exact flat directions.
To the orbifold-based EQFT, these flat directions are not moduli
but matter; in fact they are linear combinations of the charged
fields that cannot possibly be moduli.
However, because those particular combinations of matter scalar
happen to be exact flat directions of the potential, their
VEVs are indeterminate, and can take any values from zero to $\mpl$
and beyond.
But when the $SU(3)$-triplets acquire Planck-sized VEVs, the
$SU(3)$ gauge bosons become so heavy that there is no longer
any sense in including them among the light fields
(some charged scalars become super-heavy too).
Thus, within the same continuous family of string vacua,
we have to switch to a different EQFT, presumably corresponding
to an orbifold that is `blown-up' to a smooth Calabi-Yau manifold;
this new EQFT has a smaller gauge group --- just the $E_6\otimes E_8$ ---
and fewer scalars.
However, of the scalars that remain, the 27 flat combinations of the
scalars that used to be charged under the 
$SU(3)$ now become
moduli, precisely because any particles to whom these scalars give masses
are too heavy to appear in the new EQFT.

Note that of the two EQFTs --- one for the orbifold,
and one for the smooth compactification ---
neither is a special case of the other.
Although the orbifold-based EQFT has more fields,
the manifold-based EQFT has more moduli,
and that means that all the couplings of the theory --- the K\"ahler
function \Kexpansion, the superpotential \Wexpansion\ and the
gauge couplings \deindexingf\ ---
must be written as analytic functions of all the 36 moduli.
While formally we can always expand those functions into powers
of the twisted moduli, such an expansion is a power series in
${{\rm twisted\ moduli}\over\mpl}$ which are not small, and thus
truncating the expansion is quite illegitimate.
In contrast, the couplings of the orbifold-based EQFT need only be analytic
in the toroidal moduli, and it is perfectly legitimate to truncate the
expansion into powers of all other fields, including the twisted
matter scalars that combine into the twisted moduli.

The previous discussion makes it clear what to do when a hidden
sector of a low-energy EQFT does not have a stable vacuum,
supersymmetric or otherwise.
This problem is common in non-perturbative supersymmetric gauge
theories with matter:  When the classical scalar potential for
the hidden matter fields has a flat direction and the non-perturbative
effective potential happens to decrease along that direction,
the theory has no stable vacuum and a run-away expectation value
$\vev{\matter^{\rm flat}}\to\infty$.
In the previous section, we saw that when the would-be run-away
direction is only flat in the renormalizable approximation but
lifted by the non-renormalizable couplings, the run-away eventually
stops and the theory does have a stable vacuum.
A characteristic feature of this scenario
is an unusually large scalar VEV, which is nevertheless
very small compared to the Planck scale:
$\vev\matter\sim\mu^{1-\rho}\mpl^\rho$ (${0<\rho<1}$).

On the other hand, when the run-away happens along a truly flat
direction of the perturbative EQFT, it does not stop until
$\vev{\matter^{\rm flat}}\ge O(\mpl)$,
at which point the original EQFT is no longer applicable.
Instead, we are now in the domain of some other low-energy EQFT
and the former run-away matter field is just a modulus.
Thus, our analysis must start over again:
Compute the spectrum and couplings
of the gauge, matter and moduli fields
of the new theory, and then use this
information to study the stability properties  the hidden sectors,
if they are still present.
If the new theory again gives rise to a run-away vacuum instability,
we must enter the domain of a yet another EQFT. 
To summarize, run-away matter VEVs in a strongly interacting hidden
sector tell us that we are expanding
around a wrong perturbative vacuum of the unified theory.

A different kind of vacuum instability is associated with the moduli
rather than with the hidden matter fields.
Indeed, let us assume that all the strongly interacting hidden sectors
 have perfectly stable supersymmetric vacua
as long as the moduli are frozen;
however, once  the hidden sectors are integrated out,
the effective potential \SUSYVeff\ might not have a stable minimum.
Instead, it may continuously decrease in the same direction
throughout the moduli space
of the EQFT, and in this case, it is the moduli VEVs that run away.
A distinct possibility is that the run-away moduli simply evolve away from
the validity domain of the original EQFT and towards the domain
of another;
the true effective potential does have a stable minimum, but it lies
within the validity domain of the second EQFT and not the first
(or perhaps within the domain of a third EQFT, \etc).
In this situation, once one uses the low-energy EQFT for the right
area of the vacuum family,
the existence of the stable non-perturbative vacuum becomes
apparent, but when one starts in a wrong domain, the true minimum of
the $\Vmod(\modul,\modulb)$ may become invisible and all one
can tell is that the moduli VEVs would rather lie elsewhere.

Ideally, one should have a global picture of the non-perturbative
effective potential for the entire continuous family of perturbative
vacua of the unified theory.
If this potential has a global minimum, then this minimum corresponds
to the true vacuum of the theory.
(Unless the unified theory allows tunneling into an entirely different
vacuum family.)
Alas, the global picture is seldom available, and in its absence the only
alternative is to start with some candidate vacuum, construct the corresponding
low-energy EQFT and use it to calculate the $\Vmod(\modul,\modulb)$.
After that, one follows the moduli
VEVs as they evolve towards lower values of this effective potential,
and whenever they run away from the validity domain of the original EQFT,
new low-energy EQFT comes into play and the cycle repeats
until one either reaches a minimum of the effective potential
or else runs into a genuine instability.
Unfortunately, this procedure is only good for finding local rather than
global minima of effective potentials and one should always be aware of
the possibility that the true non-perturbative vacuum of the theory lies
elsewhere.

\ldf\DS{M.~Dine and N.~Seiberg, \plt  162B (1985) 299.}
\ldf\twogaugino{N.V.~Krasnikov, \plt  193B (1987) 37;\brk
    L.~Dixon, in the Proccedings of the 15th A.P.S. D.P.F.
	Meeting, Houston, 1990;\brk
    J.~A.~Casas, Z.~Lalak, C.~Mu\~noz and G.G.~Ross, \nup347 (1990) 243;\brk
    T.~Taylor, \plt B252 (1990) 59.}
\ldf\MR{A.~de la Macorra and G.~Ross, \nup404 (1993) 321.}
Sometimes, a run-away direction has no end;
instead, the non-perturbative $\Vmod(\modul,\modulb)$ continually decreases
along some infinitely long trajectory in the moduli space and there
is no mechanism in sight that would stop the runaway.
In string unification, the dilaton field $\Re S$ often suffers
{}from this instability because large values of the $\Re S$ correspond
to the weak coupling regime of the EQFT, which guarantees that
for $\Re S\to{+\infty}$ the effective potential
$\Vmod(S,\overline S)$ always approaches
zero asymptotically.\refmark{\DS}
For many models, this effective potential is decreasing for large enough
$\Re S$, and as the result, the VEV of the dilaton
never stops growing.
\foot{For suggestions how to cure this problem see
   refs.~[\twogaugino,\MR,\KLc].}
In this scenario, the theory has a genuine low-energy instability.

\ldf\CGH{P.\ Candelas,  P.S.\ Green and  T.\ H\"ubsch, \nup330 (1990) 49.}
Another kind of a vacuum instability occurs when the
non-perturbative effective potentials push the moduli VEVs into a
domain where the low-energy physics cannot be described by
{\it any} four-dimensional EQFT.
For example, when the radius $R$ of a Calabi-Yau manifold
is pushed by the $\Vmod(R,\ldots)$ to larger and larger values,
the vacuum  decompactifies from four spacetime dimensions to ten;
in a general string-based theory, similar instabilities may correspond
to partial decompactifications (from four dimensions to five or six).
A generic sign of impending decompactification is an infinite
number of massive fields that become light and it is not always clear
which degrees of freedom correspond to the new spacetime dimensions;
an example of this behavior in a Calabi-Yau context
is the conifold limit of the quintic variety in $\rm CP^4$.
\refmark{\CGH}

Finally, the run-away moduli may drive the
theory into a strong-coupling regime that we do not know how to analyze;
in string-based theories this happens whenever the VEV  of the dilaton is
small.
To be precise, one has a two-fold problem whenever the running couplings
become strong  just below $\mpl$:
First, a perturbatively cut-off EQFT is ill-defined unless we can do
perturbation theory near the cutoff scale.
Second, the string theory itself is defined perturbatively and we
have no idea how a strongly-coupled four-dimensional string theory
may look and what are its low-energy implications.
The same is {\it a fortiori} true for all other presently known
candidate theories of the ultimate unification.

Let us now assume that the low-energy theory does have a stable vacuum
and turn our attention to the issue of non-perturbative
supersymmetry breaking.
In a theory with both moduli and hidden sectors, there are two distinct
mechanisms for such breakdown.
First, the vacuum state of a strongly-interacting hidden sector may
have positive energy density and thus spontaneously break SUSY;
this is known
to happen in some supersymmetric gauge theories with
{\sl chiral} matter, \eg, an $SU(5)$ with a single $\bf 10$ and a single
$\overline{\bf5}$.\refmark{\ADS}
In this scenario --- usually referred to as the `dynamical' SUSY breakdown
--- the effective potential for the moduli is not given  by eq.~\SUSYVeff\
but rather by the coupling-dependent vacuum energy of the hidden sector
\setbox0=\hbox{$\eqname\VnoSUSY$}\setbox0=\null
\foot{Strictly speaking, eq.~\VnoSUSY\ gives only the leading term
    in the expansion of the $\Vmod$ into powers of (confinement scale)$/\mpl$,
    but since it does not vanish, the sub-leading terms are unimportant.%
    }:
$$
\Vmod(\modul,\modulb)\ =\ \checkex Bubble.eps
\iffigureexists\diagram Bubble.eps \ \equiv\ \fi
\vev{{\cal L}_{\rm hidden}}(f^W,Y_{IJK},Z_{\Ib J})\ >\ 0.
\eqno\VnoSUSY
$$
When a symmetry is spontaneously broken in some sector of an EQFT,
integrating that sector out results in a new EQFT in which the
original symmetry appears to be broken explicitly;
this rule of quantum field theory
applies to all symmetries, and SUSY --- rigid or local ---
is no exception.
Consequently, the non-perturbative potential \VnoSUSY\ generally
does not have a supersymmetric form.

The detailed analysis of SUSY-breaking hidden sectors is outside
the scope of this article; nevertheless, one can generically estimate
 the magnitudes of various SUSY-breaking effects.
If we assume a stable minimum for the potential \VnoSUSY\
the moduli scalars generally have masses of the order
$O(\mu^2/\mpl)$ ($\mu$ is the confinement scale of the hidden sector).
At the same time, the gravitino mass is also $O(\mu^2/\mpl)$
and the cosmological constant is $O(\mu^4)$.
It is not clear,
how seriously one should take this cosmological constant since
at present, all unified theories suffer from this  problem.
However,
any mechanism that adjusts the cosmological constant by an $O(\mu^4)$
amount is liable to change the whole effective potential \VnoSUSY\
by a comparable amount and thus completely re-arrange its shape.
Consequently, any predictions based upon the scenario in which
SUSY is spontaneously broken in a hidden sector should be taken
with  grain of salt.

The second mechanism for a dynamically induced spontaneous breakdown
of local supersymmetry proceeds in two stages.
At stage one, one or more hidden sectors of the low-energy theory
produce gaugino condensates (and, possibly, some matter condensates
as well), but they do not break SUSY (\ie, their vacua are supersymmetric
and stable).
Instead, integrating out those hidden sectors yields an effective
superpotential $\Wmod(\modul)$ for the moduli fields
according to the rules we discussed in sections 4.1--4.
At stage two we thus have a locally supersymmetric effective theory
for the moduli, with a perfectly supersymmetric effective potential
\SUSYVeff.
Nevertheless, this effective potential may lead to spontaneous
supersymmetry breakdown among the moduli;
a signal for this kind of SUSY breakdown would be a
VEV $\vev{F^i}$ of an auxiliary component of
a modulus superfield $\modul^i$.\refmark{\FILQ-\hiddenmatter}

Just as there is no {\it a priori} guarantee for the effective
potential \SUSYVeff\ to have   a stable minimum, one cannot {\it a priori}
determine whether such a minimum is supersymmetric or not.
The only way to find this out is to calculate the $\Wmod(\modul)$
{}from the holomorphic couplings of the hidden sectors, evaluate
eq.~\SUSYVeff\ and see where it leads.
Similarly, it is  a matter of model-by-model calculations to find out
whether the value of the effective potential at its minimum
vanishes or not;
this question has a direct bearing on the cosmological constant problem.

Indeed, consider the magnitudes of the SUSY-breaking effects in this
scenario.
The natural scale of the non-perturbatively induced $\Wmod(\modul)$
is $O(\mu^3)$
($\mu$ again being the confinement scale of the hidden sector).
Hence, the natural scale of the scalar potential \SUSYVeff\
is $O(\mu^6/\mpl^2)$, the moduli masses are $O(\mu^3/\mpl^2)$
and if SUSY is spontaneously broken,
then $\vev{F^i}=O(\mu^3/\mpl^2)$ and the gravitino mass $m_{3/2}$
is also $O(\mu^3/\mpl^2)$.
Thus, in terms of the gravitino mass, the moduli mass is
$O(m_{3/2})$ and the natural scale of the effective potential
is $O(m_{3/2}^2\mpl^2)$, exactly like for SUSY broken in a hidden
sector.
Hence, if the minimal value of the effective potential \SUSYVeff\ does not
at least approximately vanish relative to its natural scale, we
have a cosmological constant of the order $O(m_{3/2}^2\mpl^2)$
and thus the usual danger that an unknown mechanism which eliminates
this cosmological constant from the observable Universe would also
completely obliterate all other predictions based upon eq.~\SUSYVeff.

On the other hand, the physics of the low-energy moduli is locally
rather than rigidly supersymmetric and thus the effective potential
\SUSYVeff\ is {\sl not} positive definite.
Hence, one can reasonably hope that for some theories there is at least
an approximate cancellation between the positive
and the negative terms in eq.~\SUSYVeff\
and the naive cosmological constant is therefore much smaller than
$m_{3/2}^2\mpl^2$.
For all the obvious reasons, this cannot be the final answer to the
cosmological constant problem,
but it at least makes it plausible that an unknown mechanism that
does solve this problem does not do too much damage to the potential
\SUSYVeff.

\ldf\poloni{G.D.~Coughlan, W.~Fischler, E.~Kolb, S.~Raby and G.G.~Ross,
	\plt 131B (1983) 59.}
\ldf\BKN{T.\ Banks, D.\ Kaplan and A.~Nelson, Rutgers preprint RU--37 (1993).}
\ldf\CCQR{B.~de Carlos,  J.A.~Casas, F.~Quevedo and E.~Roulet,
	CERN preprint CERN--TH.6958/93.}
In both scenarios, the moduli masses are of the order $O(m_{3/2})$,
which may cause severe difficulties for the standard history of the early
Universe.\refmark{\poloni-\CCQR}
This problem is generic to all unified theories with moduli originating at
the Planck scale that use the same field-theoretical mechanism for breaking
SUSY and for stabilizing the moduli VEVs.
One solution, suggested in ref.~[\CCQR], involves not quite standard cosmology
in which the inflation, the relaxation of the moduli scalars to their ultimate
VEV and the baryogenesis all happen in the same era.
Another solution is the hybrid scenario, in which
one or more hidden sectors become strongly interacting
at a rather high intermediate scale $\mu_1$ and stabilize all the moduli VEVs
but do not break SUSY either directly or through the $\Wmod(\modul)$ and do
not produce a cosmological constant.
Then, at a much lower scale $\mu_2$, another hidden sector or sectors become
strong and break SUSY.
The hybrid scenario is rather baroque and we do not seriously advocate it as
the ultimate answer to all the problems of the string unification,
but it definitely warrants further investigation.

\ldf\KN{E.J.~Chun, J.E.~Kim and H.P.~Nilles, \nup 370 (1992) 105.}
We conclude this article with a comment on how the two scenarios for
SUSY breaking affect the observable sector of the low-energy theory.
When the hidden sectors supply the $\Wmod$ that triggers a
spontaneous SUSY breakdown by the moduli superfields, it is
the moduli dependence of the couplings $f^W_a$, $Y_{IJK}$ and $Z_{\Ib J}$
of the observable sector that breaks SUSY among the ordinary particles.
Consequently, the superpartner masses are of the order
$O(\vev{F^i})=O(m_{3/2})=O(M_{\rm mod})$;
the detailed formul\ae\ for the masses and other soft SUSY-breaking parameters
are presented in ref.~[\KLa].
In this scenario, non-renormalizable (or even renormalizable)
cross-couplings between the hidden
and the observable sectors do not affect SUSY breaking.
It is possible however that such cross-couplings
may create an intermediate-energy supersymmetric threshold
in the observable sector.
Indeed, if some of the hidden matter fields are not quite hidden but
have non-trivial charges under the observable gauge group, then the
observable sector has a threshold near the confinement scale.
Alternatively, a threshold may result from the Wilsonian superpotential
of the EQFT containing a cross-coupling of the form $W\supset
 O(\mpl^{1-n})(\matter_{\rm obs})^2 (\matter_{\rm hid})^n$
(all indices suppressed).
\foot{The energy scale of such a threshold depends on $n$ and on the
    kind of a hidden sector involved.
    When all of the hidden scalar VEVs are controlled by the renormalizable
    Yukawa couplings, one generally has $\vev{(\matter_{\rm hid})^n}=O(\mu^n)$,
    which leads to the threshold scale $M_I=O(m_{3/2}^{n/3}\mpl^{1-n/3})$.
    In this scenario, the $n=1$ and $n=2$ cross-couplings produce
    thresholds well above the weak scale,
    the $n=3$ cross-couplings produce the so-called `$\mu$-terms' right
    at the weak scale \refmark{\KN,\KLa}
    while the cross-couplings with $n\ge4$ are too weak to have any physical
    effects.
    The situation is more complicated when it takes the non-renormalizable
    superpotential couplings to stabilize all the hidden VEVs.
    In this scenario, some of the hidden scalar VEVs are unusually large,
    of the order $\vev{\matter_{\rm hid}}=O(\mu^{1-\rho}\mpl^\rho)$
    for some $0<\rho<1$.
    The energy scale of the thresholds produced in this scenario depends
    not just on the $n$ but also on the particular hidden scalars participating
    in the cross-coupling.
    Generically, the observable sector may have a threshold
    above the weak scale whenever $n<3/(1-\rho)$.}
Since the hidden sector does not break SUSY by itself,
both kinds of thresholds are manifestly supersymmetric
and their effects upon the Wilsonian gauge
couplings of the observable sector are analytically computable with
the help of eq.~\Wilthreshold.

\ldf\DN{M.~Dine and
  A.~Nelson, \prv  D48 (1993) 1277.}
On the other hand, when the strong interactions in hidden sectors are
directly responsible for the breakdown of SUSY, it is the moduli that
play a  peripheral role of merely determining the couplings of the EQFT
whereas the SUSY breaking among the ordinary particles is controlled
by the direct cross-couplings between the hidden and the observable
sectors.
One scenario of this kind involves
matter fields that transform under both the observable gauge group
and the hidden gauge group that breaks SUSY.
In this case, SUSY breaking in the observable sector is caused by loops
of gauge fields and charged particles and has nothing to do with the
Planck-scale physics.
A similar effect can be achieved by having gauge singlets
with Yukawa couplings to both the observable and the hidden
sectors.\refmark{\DN}
Phenomenologically, this scenario makes for natural degeneracy among the
squarks or sleptons of the same charge, which is good for avoiding
the flavor-changing neutral currents.
It also implies that the confinement scale of the SUSY-breaking
sector is in the multi-TeV range and hence the gravitino and the
moduli are very light, perhaps a fraction of an eV.
Alternatively, SUSY breaking in the observable sector may be induced by
the non-renormalizable cross-couplings to the hidden sector.
We have not analyzed this scenario in any detail and can only say
that the formul\ae\ it yields are somewhat  unwieldy.

\par\smallskip
\noindent {\bf Acknowledgements: }
Major parts of this work were performed at SLAC and CERN.
We would like to take this opportunity to thank the members of both  theory
groups  for their warm hospitality, their encouragement and for innumerable
enlightening discussions we enjoyed over the past years.
Our special thanks go to Brian Warr, for teaching us the fully covariant
regularization techniques (\cf\ Appendix~B).
\subpar
The research of V.~K. is supported in part by the NSF,
under grant PHY--90--09850,
and by the Robert A.~Welch Foundation.
The research of J.~L. is supported by the Heisenberg Fellowship
of the DFG.
The collaboration of the two authors is additionally supported by
NATO, under grant CRG~931380.

\APPENDIX{A}{A\break
No-Renormalization Theorems for Locally supersymmetric EQFTs}
The key to the no-renormalization theorems is the locality of the
Wilsonian renormalization group:
Integrating out some high-momentum modes
of a quantum field (or all the modes, if the field is heavy) adds to the
Wilsonian action a spacetime-local term \ie, a $\int\!\!d^4x$ of a convergent
power series in quantum fields and their derivatives that has no
singularities at zero momentum.
In a manifestly supersymmetric EQFT which is both quantized and cut-off
in terms of explicit superfields,
the renormalization of the Wilsonian action is local in superspace, that is,
amounts to a full-superspace integral
$$
\Delta{\cal S}\ = \int\!d^8z\,{\bf E}\,\Delta{\cal L}
\eqn\Wilrenorm
$$
of a non-singular function $\Delta\cal L$ of quantum and background superfields
and their derivatives.

In the rigidly supersymmetric case, non-renormalization of the superpotential
is an immediate consequence of the fact that a  chiral-superspace
integral of a derivative-less $\Delta W(\modul,\matter)$ {\sl cannot}
be re-written as a full-superspace integral \Wilrenorm\ of a non-singular
$\Delta\cal L$.\refmark{\GGRS}
In the locally supersymmetric case, the situation is somewhat different
since the entire action~\SUGRACT\ has a form of a full-superspace integral
and there is no reason why the superspace-curvature superfields such as
$\cal R$ or ${\cal W}_{\alpha\beta\gamma}$ cannot appear in
$\Delta\cal L$ along with the other superfields of the theory.
However, the {\sl locality} of $\Delta\cal L$ restricts the manner of
appearance of
$\cal R$, \etc, in $\Delta\cal L$ to expressions that do not lead to any
singularities at zero curvature or zero derivatives of other fields.
In particular, $\Delta\cal L$ {\it cannot} contain a
$\Delta\widetilde W/2{\cal R}$
term with a derivative-less chiral $\Delta\widetilde W$ in the numerator
and the curvature $\cal R$ in the denominator,
and there are no other local non-singular full-superspace integrals
whose flat-superspace limit is equivalent to a superpotential.
Therefore, in locally supersymmetric EQFTs, just as in their
rigidly supersymmetric analogues, the
superpotential function $\widetilde W(\ccf,\modul,\matter)$
does not renormalize.

A similar argument reproduces
the no-renormalization theorem for the gauge couplings $f_\ind$:
Following ref.~[\SVa] (where the rigidly supersymmetric case is treated),
we quantize the gauge superfields in a background-covariant gauge
and require the UV cutoff of the theory to be background-gauge invariant
as well.
Under these conditions, not only the integral~\Wilrenorm\ renormalizing the
Wilsonian action should remain invariant under the background gauge
transformations,
but its integrand $\Delta\cal L$ itself should be background-gauge invariant.
Alas, no full-superspace integral of a {\sl local, non-singular,
gauge-invariant} $\Delta\cal L$ has a component-field expansion
containing $F_{mn}^2$.
Consequently, the supersymmetric gauge couplings $\tilde f_\ind(\ccf,\modul)$
appear to be renormalization-free.

As explained by Shifman and Vainshtein, this argument has a one-loop-hole:
In background gauges, the normalization of matter, ghost and quantum
vector superfields depends on the background.
Consequently,
integrating out the high-momentum modes of those superfields results in the
{\it anomalous} renormalization of the action for the
background gauge fields.
{}From the path-integral point of view, this anomaly is the effect of the
background-dependence of the measure for the modes being integrated out;
in terms of the Feynman diagrams, this means that the entire anomaly is
given by one-loop diagrams in which one of the background gauge vertices
is replaced by a quantum vertex.\refmark{\GGRS}
Hence, the Wilsonian supersymmetric gauge couplings
$\tilde f^W_\ind(\ccf,\modul)$ do renormalize, but only at the one-loop level.

\APPENDIX{B}{B\break
    Regularizing a Locally Supersymmetric EQFT\break
    While Preserving All of its Symmetries}
%
\chapterstyle={\Alphabetic}
\chapternumber=2
The research that lead to this article began as a collaboration between
Brian Warr and the two present authors.
The results presented in this Appendix should be credited to Brian
as much as to V.~K. and J.~L. combined;
alas, Brian's untimely death prevented him from further participation
in this project.
The two of us would like to use this opportunity to express our deep
sorrow at his passing and to acknowledge the work he has done.

\ldf\warr{B.~Warr \journal Annals of Phys. &183 (1988) p.~1 and p.~59.}
\ldf\PS{O.~Piguet and K.~Sibold,  {\it Renormalized Supersymmetry},
	Birkh\"auser, Boston, 1986.}
Many formal arguments about supersymmetric gauge theories presume the
existence of an ultraviolet regularization scheme that preserves
the manifest supersymmetry of the superfield formalism and at the
same time has manifest four-dimensional background gauge invariance
and a BRST symmetry protecting the quantum gauge invariance.
In refs.~[\warr],
Brian Warr proved that there indeed exists such a
regularization scheme.
\foot{For the discussion of alternative regularization schemes
    see refs.~[\PS,\GGRS].}
Specifically, he showed that for the non-chiral supersymmetric gauge theories,
\foot{A gauge theory, supersymmetric or otherwise, is called chiral or
    non-chiral according to whether the left-handed Weyl fermions together
    form a complex or a self-conjugate representation of the gauge group.
    For a supersymmetric gauge theory, this criterion has nothing to do with
    the chirality of the matter superfields $\matter^I$ --- they are always
    chiral --- and everything to do with their gauge quantum numbers.}
it is possible to combine a huge-but-finite set of
Pauli-Villars-like compensating superfields and a carefully chosen
pattern of fully-covariant-higher-derivative propagator-regulating
terms in the Lagrangian and thus regularize {\it all} divergencies
of the perturbation theory.
The manifest covariance of this regularization scheme is plain to see;
the difficult part of the Warr's proof was to show that it does
regularize all the perturbative divergences.
Regularizing all the divergences of a chiral gauge theory is more difficult,
and thus the chiral version of the Warr's regularization scheme (WRS)
is not quite manifestly gauge invariant;
nevertheless, this scheme is sufficient to prove all the Ward identities
of a chiral gauge symmetry that is free of the triangle
and linear (trace) anomalies.

\section{Rigidly Supersymmetric Warr's Scheme.}
Before we proceed with adapting the WRS to
locally supersymmetric EQFTs, let us review its rigidly supersymmetric version.
To be precise, we present a superfield version of the non-chiral WRS which
is manifestly invariant under the background gauge transformations and
regularizes all the divergences of a non-chiral gauge theory;
we shall briefly discuss the chiral case later in this section.
The formalism of this Appendix and of the following Appendix~C
is based upon ref.~[\GGRS], which also describes the supersymmetric
quantization procedure.

Our first step towards constructing the WRS is to soften the
UV behavior of the matter superfields' propagators.
For this purpose, we add to the bare action of the theory
covariant higher derivative operators that generically look like
$$
A^{(n)} \int\!\!d^4x d^4\Theta\,\matterb e^{2V} \dalamcov^n\matter ,
\eqn\rigidFCHDmat
$$
where $n$ is a positive integer, the normalization factor
$A^{(n)}$ is a real constant of the order $O(\Lambda^{-2n})$,
all indices are suppressed and
$$
\dalamcov\ =\ \coeff{1}{16}\overline{D}^2 e^{-2V} D^2 e^{+2V}
\eqn\rigidDAlam
$$
is the covariant D'Alambertian for the charged chiral superfields.
The name ``covariant D'Alambertian'' means that
$\dalamcov\matter$ is a chiral superfield that transforms under the gauge
transformations exactly like the $\matter$, but in the trivial gauge
background $V\equiv0$, $\dalamcov\matter$ reduces to simply
$\dalamord\matter\equiv\partial^\mu\partial_\mu\matter$.
The covariance of the operator $\dalamcov$ provides for the
gauge invariance of the regulators~\rigidFCHDmat\ while its D'Alambertian
nature tells us the effect of those regulators on the scalar propagators
of the theory.
For a massless chiral superfield, the regularized superspace propagator is
$$
{{1\over16} D^2 \overline{D}^2 \delta^{(4)}(\Theta) \over p^2 A(p^2) }\
\equiv\ {{1\over16}D^2 \overline{D}^2 \delta^{(4)}(\Theta) \over
    p^2 \left( 1+ A^{(1)} p^2 + A^{(2)} p^4 +\cdots\right) }
\eqn\Qprop
$$
while for a massive scalar supermultiplet, the four
$(\matter,\matterb)\mapsto(\matterb,\matter)$ propagators can be summarized
in a two-by-two matrix
$$
{1\over |M|^2+p^2 A^2(p^2)}\,\times \pmatrix{
    A(p^2)({1\over16} D^2 \overline{D}^2) &
    M({1\over4} D^2) \cr
    M^* ({1\over4}\overline{D}^2) &
    A(p^2)({1\over16} \overline{D}^2 D^2)\cr }
\delta^{(4)}(\Theta) .
\eqn\Qpropmass
$$
The ultraviolet limits of these propagators are controlled by the
regulator \rigidFCHDmat\ with the largest number of derivatives, \ie,
with the largest $n$.
Specifically, apart from the $D^2$ and the $\overline{D}^2$ factors, the
propagators \Qprop\ and \Qpropmass\ decrease with $|p|\gsim\Lambda$
like $(1/p^2)^{\nmax +1}$, in obvious notations;
we shall refer to the $\nmax $ as the {\it type}
of the matter superfield.

Similarly, we soften the UV behavior of the gauge superfields' propagators
by adding to the bare action of the theory covariant higher derivative
operators
$$
\coeff14 B^{(n)} \int\!\!d^4x d^2\Theta\, W^\alpha \dalamcov^n W_\alpha\quad
+\ \rm H.\,c.
\eqn\rigidFCHDgauge
$$
(gauge indices suppressed).
Again, the $\dalamcov$ here is the covariant D'Alambertian \rigidDAlam\
for the chiral, gauge-covariant superfields (such as $W^\alpha)$;
this assures that the terms \rigidFCHDgauge\ are both gauge-invariant and
supersymmetric.
The normalization factors $B^{(n)}$ are constants of the order
$O(\Lambda^{-2n})$; for simplicity, we assume real $B^{(n)}$.
The precise effect of the regularizing terms \rigidFCHDgauge\ on the gauge
superfield's propagator depend on the gauge-fixing scheme;
in the super-Feynman gauge, the propagator is
$$
{\delta^{(4)}(\Theta)\over
(\Re f)p^2 + B^{(1)} p^4 + B^{(2)} p^6 +\cdots }
\eqn\FGprop
$$
while in the `super-transverse' gauge favored by Warr the propagator is
somewhat more complicated.

The gauge-invariance of the higher-derivative
propagator-regularizing terms has its price, namely the higher-derivative
couplings of the gauge superfields to themselves and to the charged
scalar superfields;
these couplings tend to counteract the regularizing effect of the
propagator softening.
Careful counting of powers of all the momenta
involved in a generic superfield Feynman diagram (including the momenta
implicit in the $D$ operators) shows that all the
multi-loop diagrams are superficially convergent as long as
the gauge superfields are of a higher type than any scalar superfield,
\ie, $\nmax ({\rm gauge})>\nmax ({\rm any\ scalar})$.
\refmark{\warr}
(If there are Yukawa couplings between the matter superfields,
one also needs $\nmax ({\rm matter})\ge {1\over3}\nmax ({\rm gauge})$.)
Some of the one-loop diagrams however remain superficially divergent
in spite of the covariant-higher-derivative regulators and thus must
be regularized in some other manner, both for their own sake and to
eliminate possible sub-diagram divergencies of the multi-loop diagrams.

A divergent one-loop diagram (or sub-diagram) may have an arbitrary number
of external gauge legs but no external legs belonging to matter or ghost
superfields.
For non-chiral gauge theories, we may calculate the one-loop diagrams using
the second order formalism (see Appendix~C for details), which automatically
avoids any quadratic or linear divergences while the logarithmic divergences
consistently assemble into infinite renormalization of the gauge coupling.
To cancel these logarithmic divergences, we add to the theory an array of
Pauli-Villars superfields (PVs), which are heavy
(mass of the order $O(\Lambda)$)
scalar superfields with covariant higher-derivative Lagrangians;
generically, we need several kinds of PVs with different gauge quantum numbers,
different types and both Bose and Fermi statistics.
The complete spectrum of the PVs should satisfy the equations
$$
\displaylines{
\sum_{\cal P}^{\rm PVs} \pm T_\ind({\cal P})\left( 2\nmax({\cal P})+1\right)
\hfill\eqname\WarrEquations \cr
\hfill{}=\ T(G_\ind)\left(2\nmax({\rm gauge})+3\right)\
-\sum_{r} n_r({\rm matter}) T_\ind(r) \left(2\nmax({\rm matter})+1\right)
\qquad\cr }
$$
for all gauge group factors $G_\ind$;
in our notations, $\pm T_\ind({\cal P})$ refers to the statistics and
to the gauge quantum numbers of the Pauli-Villars superfield $\cal P$.
The details of the solution of all the eqs.~\WarrEquations\ are not germane
to the subject matter of this article; all we really need to know is that the
solution exists.

Regularizing all the one-loop divergences of a chiral gauge theory is more
difficult since the second order formalism does not work for the chiral
superfields in a complex representation of the gauge group.
The first order formalism is plagued by the quadratic and linear
divergences that must be pre-regularized before one can put together
Feynman diagrams with different arrangements of external legs.
In the chiral WRS, the pre-regularization is achieved with the
help of additional Pauli-Villars superfields of mass $M_{\rm pre}\gg\Lambda$.
The pre-regularizing PVs form complex representations of the gauge group;
their higher-derivative kinetic energy terms are gauge-covariant, but their
mass terms are not.
The gauge invariance of the effective theory is restored by adding a local
counter-term to the bare action of the theory; this is possible whenever
the theory is free from the triangle and linear (trace) anomalies.
The gauge invariance of the chiral WRS is thus not quite manifest;
nevertheless, thanks to $M_{\rm pre}\gg\Lambda$,
the gauge symmetry is completely restored at energies well above the ordinary
UV regularization scale $\Lambda$ and the exact form of the counter-term
is completely determined at the one-loop level of the theory.
Therefore, as far as the Ward identities --- and their corollaries --- are
concerned, the chiral WRS is as good as if it was
manifestly gauge invariant.

We close this overview by stressing that the WRS
regularizes the UV divergencies of Feynman diagrams and has nothing
to do with the convergence/divergence of the perturbative expansion of
the EQFT.
Moreover, the finite theory defined by the WRS is unitary
only in the limit of the UV cutoff $\Lambda$ being much bigger then
the momenta of any external particles involved in the S-matrix;
when the external momenta become comparable with the cutoff, the
unitarity breaks down.
In short, the WRS is not a supersymmetric substitute
for a lattice regularization, and we do not propose a
locally super\-symmetric substitute for a lattice either.
Instead, we are about to construct a manifestly supersymmetric and
gauge invariant regularization scheme for locally supersymmetric
EQFTs whose UV limits are perturbative (as long as the supergravity
itself is limited to the background).

\section{Warr's Scheme for Locally-Supersymmetric EQFTs.}
Adapting the WRS to locally supersymmetric EQFTs involves two tasks:
First, we modify the regularizing terms in the action to make them
manifestly invariant under the local SUSY, which in our formalism
comprises the general super-coordinate, local Lorentz and super-Weyl
transformations.
In addition, we would also like to have manifest background gauge
symmetry of the entire action; therefore, we will use the background-%
covariant version of the super-Feynman gauge.
The second task is to regularize the divergencies that involve non-trivial
supergravitational or moduli backgrounds; this work is still in progress,
so we shall present only the results which are relevant to the calculations
of the Appendix~C.

We begin with fully-covariant derivatives for the chiral superfields.
When acting upon spinless, Weyl invariant superfields, the
operators
$$
\ccf^{-2}\left( \coeff14 \overline{\cal D}^2 -2{\cal R}\right) \ccfb
\quad{\rm and}\quad
\ccfb^{-2}\left( \coeff14 {\cal D}^2 -2\overline{\cal R}\right) \ccf
\eqn\doublederivative
$$
are super-covariant as well as Lorentz and Weyl invariant
(albeit not gauge-co\-variant).
Hence, for a charged but spinless and Weyl invariant chiral superfield
such as $\matter$, the fully-covariant D'Alambertian can be written as
$$
\dalamFC\ =\ {1\over\ccf^2} \left( \coeff14 \overline{\cal D}^2
-2{\cal R}\right) {1\over\ccfb} e^{-2V}
\left( \coeff14 {\cal D}^2 -2\overline{\cal R}\right) \ccf e^{+2V} .
\eqn\localDAlam
$$
By {\sl full} covariance we mean that $\dalamFC\matter$ transforms
exactly like $\matter$ under all the local symmetries of the theory,
namely gauge, SUSY, Lorentz and super-Weyl.
Given this fully-covariant D'Alambertian, we can write the fully-symmetric
analogues of the matter-propagator regulating terms \rigidFCHDmat\ as
$$
\int\!\!d^8z\,{\bf E}\ccf\ccfb\, A^{(n)}_{\Ib J}\, \matterb^\Ib e^{2V}
\dalamFC^n\matter^J .
\eqn\localFCHDmat
$$
Generically, the matrices $A^{(n)}_{\Ib J}$ have the same structure as the
$Z_{\Ib J}$ matrix, but without any loss of regularizing power we may choose
to have $A^{(n)}_{\Ib J}=A^{(n)}\delta_{\Ib J}$ where the overall coefficient
$A^{(n)}=O(\Lambda^{-2n})$, $\Lambda$ being the scale parameter of the UV
cutoff.
In any case, the matrix elements $A^{(n)}_{\Ib J}$ must be $\ccf$ independent
and it is best to have them moduli-independent as well.

The fully-symmetric analogues of the gauge-propagator regulating operators
\rigidFCHDgauge\ can be written in a similar fashion, namely
$$
\coeff14 B^{(n)} \int\!\!d^8z\,{\cal E}\, {\cal W}^\alpha
\left(\ccf^{3/2}\dalamFC\ccf^{-3/2}\right)^n {\cal W}_\alpha
\quad +\ \rm H.\,c.,
\eqn\localFCHDgauge
$$
where $\ccf^{3/2}\dalamFC\ccf^{-3/2}$ is the fully-covariant D'Alambertian
for the gauge field stress tensor ${\cal W}^\alpha$ whose Weyl weight is 3.
Generically, one may have different coefficient $B^{(n)}_\ind$ for
the different irreducible factors $\ind$ of the gauge group, but without
any loss of regularizing power we may choose the same
$B^{(n)}=O(\Lambda^{-2n})$ for all~$\ind$.
In any case, all the $B^{(n)}_\ind$ must be $\ccf$ independent
and it is best to have them moduli-independent as well.

So far we have described the minimal covariantization of the higher
derivative regularizing operators with respect to the local SUSY.
To these minimal operators \localFCHDmat\ and \localFCHDgauge\
we may (and should) add non-minimal operators that involve covariant
superspace curvature tensors such as ${\cal W}^{\alpha\beta\gamma}$ or
${\cal R}-{1\over8}\overline{\cal D}^2\log\ccfb$.
These non-minimal higher derivative operators play an important role in
regularizing Feynman diagrams with gravitational external lines.

Now consider the background-covariant super-Feynman gauge for
locally supersymmetric gauge theories.
As in the rigid case,\refmark{\GGRS}
 we split the gauge superfield $V$ into the
quantum and the background gauge superfields according to
$$
\exp(2V)\ =\ \exp(\Omega^\dagger) \exp(2{\bf v}) \exp(\Omega) .
\eqn\gaugesplit
$$
The quantum gauge superfield $\bf v$ is hermitian; the background
gauge superfield $\Omega$ is neither hermitian nor chiral.
The locally-supersymmetric, background-covariant analogues of the
super-Landau gauge conditions can be written as
$$
\eqalign{
\left[\ccf^{\ww-2} \left( \coeff14 \overline{\cal D}^2 -2{\cal R}\right)
    \ccfb \left(e^{-\Omega}\right)^{(a)}_{(b)} \right]{\bf v}^{(b)}\ &
=\ {\bf G}^{(a)} ,\cr
\left[\ccfb^{\ww-2} \left( \coeff14 {\cal D}^2 -2\overline{\cal R}\right)
    \ccf \left(e^{\Omega^\dagger}\right)^{(a)}_{(b)}\right]{\bf v}^{(b)}\ &
=\ \overline{\bf G}^{(a)} ,\cr
}\eqn\LLandau
$$
where ${\bf G}^{(a)}$ and $\overline{\bf G}^{(a)}$ are fixed but arbitrary
chiral and anti-chiral superfields.
Eqs.~\LLandau\ fix the quantum gauge symmetry of $\bf v$, but they are
covariant with respect to the background gauge symmetry, under which
$\bf G$ and $\overline{\bf G}$
behave like ordinary adjoint multiplets of scalar superfields.
Generically, the Weyl weight of the superfields $\bf G$ and
$\overline{\bf G}$ is an arbitrary real parameter $\ww$ of the
gauge-fixing conditions \LLandau,
but in order to simplify the ghost structure of the EQFT (see below),
we prefer to use $\ww=3$.

The Faddeev-Popov ghost action corresponding to the gauge conditions
\LLandau\ follows from the general rules\refmark{\GGRS}:
$$
\eqalign{
{\cal S}_{\rm FP}\ &
= \int\!\!d^8z\, {\cal E}\,c' \left[\ccf^{\ww-2}
	\left( \coeff14 \overline{\cal D}^2 -2{\cal R}\right)
	\ccfb e^{-\Omega} \right]
    \left(\partder{V}{\xi}c +\partder{V}{\bar\xi}\bar c\right)\quad
    +\ \rm H.\,c. \cr
&= \int\!\!d^8z\, {\bf E}\,\left( \ccf^{\ww-2}\ccfb\,c' e^{-\Omega}\,
	+\,\ccf\ccfb^{\ww-2}\,\bar c' e^{+\Omega^\dagger}\right)\cr
&\omit\hfil$\displaystyle{
    \left[ {\bf v}\left( e^{+\Omega}c +e^{-\Omega^\dagger}\bar c\right)\,
	+\,{\bf v\over\tanh\bf v}
            \left( e^{+\Omega}c -e^{-\Omega^\dagger}\bar c\right)\right]
    }$\cr
}\eqn\FPghostsgen
$$
(gauge indices suppressed).
The ghosts $c$ and $\bar c$ have the same quantum numbers as the parameters
$\xi$ and $\bar\xi$ of infinitesimal quantum gauge transforms: They are
chiral/antichiral Weyl-invariant superfields in the adjoint representation
of the background gauge group.
The anti-ghosts $c'$ and $\bar c'$ are also chiral/antichiral superfields
in the adjoint representation of the background gauge group, but their
Weyl weights are $(3-\ww)$.
For $\ww=3$, all Faddeev-Popov ghosts are Weyl invariant and the ghost
action \FPghostsgen\ becomes
$$
{\cal S}_{\rm FP}\ = \int\!\!d^8z\, {\bf E}\ccf\ccfb\,
\left( \bar c' e^{\Omega^\dagger}e^{\Omega} c
+ \bar c e^{\Omega^\dagger}e^{\Omega} c'
+ c'c -\bar c'\bar c\
+\ {\cal L}_{\rm int}(c,c',\bar c,\bar c',{\bf v},\Omega,\Omega^\dagger)
\right) .
\eqn\FPghosts
$$
Apart from the overall factor ${\bf E}\ccf\ccfb$, this ghost Lagrangian
is identical to its rigid counterpart;
furthermore, the Faddeev-Popov ghosts are of type zero --- they do not have
any higher-derivative terms in their Lagrangian.

In Feynman gauges, one averages the path integral over the Landau gauge
parameters ${\bf G}^{(a)}$ and $\overline{\bf G}^{(a)}$
with a Gaussian weight;
this is equivalent to replacing the constraints \LLandau\ with a
gauge-fixing action term.
To simplify the gauge superfields' propagators, the gauge-fixing action
should contain background-covariant higher-derivative terms that exactly
parallel the higher-derivative terms in the gauge-invariant action.
Thus, in matrix notations,
$$
{\cal S}_{GF}\ = \int\!\!d^8z\,{\bf E} (\ccf\ccfb)^{-2}\;
\overline{\bf G}\, e^{\Omega^\dagger}e^{\Omega} \left[ (\Re f)\
+ \smash{\sum_{n=1}^{\nmax ({\rm gauge})}}
    \vphantom{\sum}B^{(n)}
    \bigl(-\ccf^3\dalamFC({\rm bg})\ccf^{-3}\bigr)^n
    \right] {\bf G} ,
\eqn\FGaveraging
$$
where $f$ is the matrix of the bare gauge couplings
$f_{(a)(b)}=f_\ind\delta_{(a)(b)}$ of the un-regularized EQFT;
in the manifestly super-Weyl invariant formalism of this Appendix,
$f$ does not depend on the $\ccf$.
The coefficients $B^{(n)}$ in eq.~\FGaveraging\
are exactly as in eq.~\localFCHDgauge\ and
the background-covariant $\dalamFC({\rm bg})$ is obtained from the
fully-covariant $\dalamFC$ via disregarding the quantum gauge superfield
$\bf v$, \ie, via replacing $e^{2V}$ with $e^{\Omega^\dagger}e^{\Omega}$;
the peculiar powers of $\ccf$ and $\ccfb$ in eq.~\FGaveraging\ correspond
to $\ww=3$.
To properly compensate for the gauge averaging, we should add to the
theory the Nielsen-Kallosh ghosts $b^{(a)}$ and $\bar b^{(a)}$,
whose action mirrors \FGaveraging:
$$
{\cal S}_{NK}\ = \int\!\!d^8z\,{\bf E} (\ccf\ccfb)^{-2}\;
\bar b\, e^{\Omega^\dagger}e^{\Omega} \left[
    (\Re f)\
    + \smash{\sum_{n=1}^{\nmax ({\rm gauge})}}
    \vphantom{\sum}B^{(n)}
    \bigl(-\ccf^3\dalamFC({\rm bg})\ccf^{-3}\bigr)^n
    \right] b \,;
\eqn\NKghost
$$
as usual, the Nielsen-Kallosh ghosts do not couple to the quantum gauge
superfields ${\bf v}^{(a)}$ or to any other quantum fields,
but they are sensitive
to $\Omega$, $\Omega^\dagger$ and other background fields.
Similar to the Faddeev-Popov ghosts and antighosts,
Nielsen-Kallosh ghosts are fermionic spinless (anti) chiral superfields
in the adjoint representation of the background gauge group.
Unlike the Faddeev-Popov ghosts, the Nielsen-Kallosh ghosts have non-zero
Weyl weight $\ww=3$ and type $\nmax (b,\bar b)=\nmax \rm(gauge)$.

Finally, consider the Pauli-Villars superfields $\cal P$ of a
locally super\-symmetric EQFT.
In the non-chiral case when $\matter^I$ together form a self-conjugate
representation of the gauge group, we need only the PVs that are
massive scalar superfields with fully covariant actions
$$
\eqalign{
\int\!\!d^8z\,{\bf E} (\ccf\ccfb)^{1-w}\; &
\overline{\cal P} \left[ \sum_{n=0}^\nmax  A^{(n)}_{\cal P}\,e^{2V}
        \bigl(-\ccf^\ww\dalamFC\ccf^{-\ww}\bigr)^n \right] {\cal P}\cr
&+\int\!\!d^8z\,{\cal E}\,\half\ccf^{3-2\ww} M_{\cal P} {\cal P}^2
    \quad +\ \rm H.\,c.,\cr
}\eqn\PVaction
$$
where $A^{(n)}_{\cal P}=O(\Lambda^{-2n})$, $M_{\cal P}=O(\Lambda)$
and neither $A^{(n)}_{\cal P}$ nor $M_{\cal P}$ depends on either $\ccf$
or the moduli of the EQFT.
Different Pauli-Villars superfields may have different
types $\nmax$ and different Weyl weights $\ww$ as well as different
statistics and gauge-group quantum numbers.
In addition, similar to the matter and the gauge superfields, the PVs
may have non-minimal higher-derivative couplings to the covariant
superspace curvature tensors.

As in the rigid case, the spectrum of the PVs must satisfy
equations \WarrEquations\ in order to cancel all the divergences that
do not involve external gravitational, $\ccf$ or moduli superfields.
Cancelation of the divergences that do involve superspace curvatures
or derivatives of the super-Weyl compensators imposes additional equations
similar to eq.~\WarrEquations
\foot{Actually, all multi-loop Feynman diagrams, with or without
    gravitational or $\ccf$ external lines, are superficially regularized
    by the minimal higher-derivative operators alone.
    It is the one-loop diagrams that impose constraints on the spectrum
    of the PVs.
    In this aspect, locally supersymmetric EQFTs  do not differ from
    their rigid analogues.
    };
it is these additional equations that
determine the Weyl weights of the PVs and the coefficients of the
non-minimal higher-derivative operators for both the PVs and the
matter and the gauge superfields.
On the other hand, any possible divergences involving the moduli fields
are completely regularized by the higher-derivative regulators and impose
no constraints on the PVs;
this follows from the moduli independence of the higher-derivative
coefficients $A^{(n)}$, $B^{(n)}$ and $A^{(n)}_{\cal P}$.

For locally-supersymmetric EQFTs with complex spectra of the matter
superfields $\matter^I$, we need additional, heavier PVs to pre-regularize
all the one-loop divergences of the first order formalism.
We have not worked out all the details of the locally-supersymmetric
pre-regularization; however,
by analogy with the rigid case, we believe that complete pre-regularization
of the one-loop divergences
can be achieved using PVs with fully-covariant kinetic and higher-derivative
actions but with masses that break gauge invariance.

\APPENDIX{C}{C\break
Gauge Couplings of a Locally Supersymmetric EQFT:\break
Explicit Superfield Calculations of the One Loop Effects.}
In Appendix~B, we described a regularization
scheme that is manifestly symmetric with respect to both
local SUSY (including manifest super-Weyl symmetry) and background
gauge symmetry.
In this Appendix, we use this scheme to explicitly calculate
to the one-loop order the differences between the bare, the Wilsonian
and the effective gauge couplings;
as promised, our results agree with
eqs.~\Wilsanom, \gWilson, \SV\ and \Katerm.
Whereas the super-Weyl anomaly responsible for the \K\ terms in
eqs.~\Wilsanom\ and \Katerm\ is exhausted at the one loop level,
the calculations of this Appendix effectively confirm those equations
to all orders of the perturbation theory.

The effective couplings are properties of the generating functional of the
EQFT; in a background-gauge covariant formalism, the effective gauge
couplings are given by the two-point Green's functions for the background
gauge fields;
the superspace analogue of eq.~\geffdef\ is
$$
{\cal A}\bigl(W^{(a)\alpha}(-p),W^{(a)}_\alpha(p)\bigr)\
=\ {-4\over \overline{D}^2}\, {1\over \{g_\ind(p)\}^2} .
\eqn\SUSYgeff
$$
The two-point Green's function \SUSYgeff\ is even and, at the one-loop level
of analysis, it does not distinguish whether
the matter superfields form a representation $r$ of the gauge group
or a complex conjugate representation $\bar r$.
Therefore, once we derive a general formula for the $\{g_\ind\}$ of an EQFT
with a non-chiral spectrum of the chiral matter superfields $\matter^I$,
the same formula should apply verbatim to the chiral EQFTs as well.
In light of this observation, we shall limit our explicit calculations
to the non chiral case and thus avoid the pre-regulation and all the
associated difficulties.

At the one-loop level of analysis, the entire generating functional for
all the background fields is the logarithm of the super-determinant of the
background-dependent matrix of second variational derivatives of the
EQFT's bare Lagrangian with respect to all the quantum superfields of
the theory.
Since we are interested only in the gauge couplings and their moduli
and $\ccf$ dependence, we presume a gravitationally trivial background
(\ie, flat vielbein $E^A_M$) while allowing arbitrary configurations of the
background gauge superfields.
Similarly, we presume the moduli superfields and the $\ccf$ to have
uniform values throughout the superspace, although we make no assumptions
about the actual values of the $\vev{\modul^i}$ and the $\vev\ccf$.
These assumptions effectively reduce a locally-supersymmetric EQFT
with moduli fields to a rigidly-supersymmetric EQFT with adjustable
couplings and regularizing terms.
For example, fully-covariant higher-derivative regulator terms
\localFCHDmat\ reduce to \rigidFCHDmat, except that the coefficients
$A^{(n)}$ become $\vev\ccf$-dependent $\tilde A^{(n)}=|\vev\ccf|^{2-2n}
A^{(n)}$.
Note however that the un-rescaled Minkowski metric of the effective
rigid theory is not the physical Einstein metric of the locally
supersymmetric theory; we shall return to this point later in this section.

In the background-covariant super-Feynman gauge, the quantum gauge
superfields ${\bf v}^{(a)}$ are full (off-shell) vector superfields
and, according to the standard lore,\refmark{\GGRS}
their loops do not contribute to the $\{g_\ind^{-2}\}^\ol$.
Ref.~[\GGRS] does not consider the higher-derivative terms
\localFCHDgauge, but the argument works even when such terms are present.
Indeed, let us take the second variational derivative
$\delta^2/\delta{\bf v}^2$ of the
entire bare Lagrangian of the EQFT, including the higher-derivative terms
\localFCHDgauge\ and the gauge-fixing terms \FGaveraging.
The result is a background-gauge covariant operator, which, after some
algebra, can be written as a polynomial of degree $\nmax ({\rm
gauge})+1$ in operator products $\eta_{ab}\nabla^a\nabla^b$,
$W^\alpha_{\rm bg}\nabla_\alpha$ and
${\overline{W}}^{\dot\alpha}_{\rm bg}\nabla_{\dot\alpha}$
($\nabla_{\dot\alpha}\equiv e^{+\Omega}
\overline{D}_{\dot\alpha}e^{-\Omega}$,
$\nabla_\alpha\equiv e^{-\Omega^\dagger}D_\alpha e^{+\Omega^\dagger}$
and $\nabla^a\equiv{1\over4}\sigma^a_{\alpha\dot\alpha}
\{\nabla^\alpha,\nabla^{\dot\alpha}\}$).
There also are terms involving higher covariant derivatives of
the background gauge fields, but all such terms are suppressed by
negative powers of the UV cutoff scale $\Lambda$ and have negligible
effects on Green's functions of gauge particles with momenta $p\ll\Lambda$.
The remaining terms have at most one spinorial
derivative $\nabla^\alpha$ or $\nabla^{\dot\alpha}$ acting upon the
quantum gauge superfield $\bf v$ for each power of a background gauge
connection or curvature;
consequently, any one-loop Feynman diagram with the $\bf v$ superfield
running around the loop has at most one spinorial derivative {\sl in the
loop} per external leg.
According to the superfield Feynman rules, a loop integral vanishes
unless the loop contains at least four spinorial derivatives;
therefore, in the background-covariant Feynman gauge,
the one-loop diagrams with the $\bf v$ loop do not
contribute to the two-point and three-point Green's functions for
the background gauge fields.

Besides the ${\bf v}^{(a)}$, all other quantum superfields of the EQFT ---
the matter superfields, the ghosts and the PVs --- are chiral
and their respective contributions to the $\{g_\ind^{-2}\}^\ol$
can be computed in similar ways.
In flat superspace, contribution of a generic chiral superfield $C$
to the one-loop generating functional of the background fields can be
formally written as
$$
\pm\Sdet\pmatrix{
    {\delta^2{\cal L}_+\over C^2} &
    {\delta^2{\cal L}_-\over C\overline{C}} \cr
    {\delta^2{\cal L}_+\over \overline{C}C} &
    {\delta^2{\cal L}_-\over \overline{C}^2} \cr
}\eqn\chisdet
$$
where the overall sign $\pm$ depends on the statistics of $C$ and
the ${\cal L}_+$ and the ${\cal L}_-$ are respectively the
chiral- and the antichiral-superspace Lagrangians, \ie,
$$
{\cal S}\ = \int\!\!d^4x\,d^2\Theta\,{\cal L}_+\
= \int\!\!d^4x\,d^2\bar\Theta\,{\cal L}_-\,.
\eqn\chilagdef
$$
For a chiral superfield $C$ of type $\nmax $
in a background of constant moduli superfields and $\ccf$, we have
$$
\eqalign{
{\delta^2{\cal L}_+\over C^2}\ &
=\ \widetilde M,\qquad
{\delta^2{\cal L}_-\over \overline{C}^2}\
=\ \widetilde M^*,\cr
{\delta^2{\cal L}_+\over \overline{C}{C}} &
=\ \coeff14 D^2 e^{2V} A(-\dalamcov),\cr
{\delta^2{\cal L}_-\over {C}\overline{C}} &
=\ \coeff14 \overline{D}^2 \left(e^{2V} A(-\dalamcov)\right)^\top\
    =\ A(-\dalamcov) \overline{D}^2 e^{-2V} ,\cr
}\eqn\chimatels
$$
where $\widetilde M=\vev\ccf^{3-2\ww}M$ is the un-normalized,
$\ccf$-dependent bare mass of $C$, $A$ is a polynomial of degree
$\nmax $ with $\ccf$-dependent coefficients $\tilde A^{(n)}=
|\vev\ccf|^{2-2n-2\ww}A^{(n)}$ ($n=0,1,\ldots,\nmax $)
and $V$ is the background gauge
superfield, $e^{2V}=e^{\Omega^\dagger}e^{\Omega}$;
the second equality in the last equation \chimatels\ is valid
for self-conjugate representations of the background gauge group.
Thus, for non-chiral gauge theories, we substitute eqs.~\chimatels\
into eq.~\chisdet, simplify the determinant and arrive at
$$
\pm\Sdet\left( |\widetilde M|^2\,
    -\,A(-\dalamcov)\dalamcov A(-\dalamcov) \right)\
=\ \pm\Sdet\left( H(-\dalamcov)\right) ,
\eqn\SecondOrder
$$
where $H$ is a polynomial defined according to
$H(x)=|\widetilde M|^2+xA^2(x)$.

Eq.~\SecondOrder\ defines fairly simple
second-order Feynman rules for the charged chiral superfields:
The propagator of the scalar superfield $C$ is
$\delta^{(4)}(\Theta)/H(k^2)$ while the vertices that come
with one spinorial derivative $D_\alpha$ {\sl in the loop}
per external gauge line are
$H'(k^2)W_{\rm bg}^\alpha D_\alpha$ and
$\half H''(k^2)\bigl(W_{\rm bg}^\alpha D_\alpha\bigr)^2$;
there are other kinds of vertices, but they carry fewer
$D_\alpha$'s and thus do not contribute to the two-point functions.
As described in ref.~[\GGRS],
{\sl one} of the external gauge lines connected to the loop
comes with an additional factor $\overline{D}^2$ in the loop
at the expense of a similar factor on the external line;
this is precisely the mechanism that gives rise to the two-point
functions of the form \SUSYgeff.

Topologically, there are two one-loop Feynman diagrams with two
external legs:
$$
\checkex 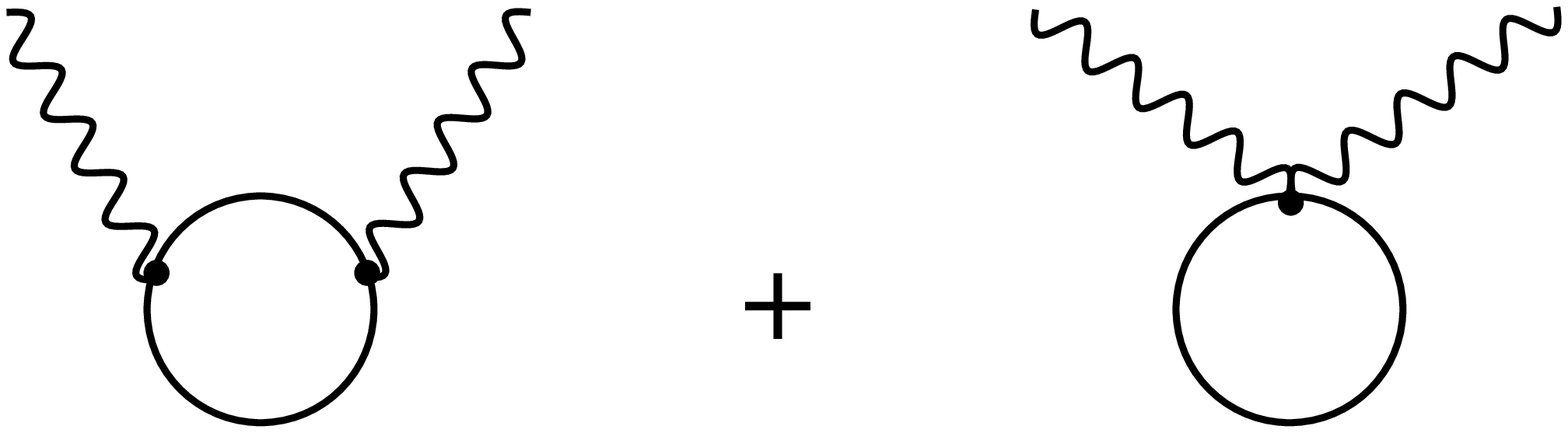
\iffigureexists \diagram TwoDiagrams.eps
\else \missbox{Missing Diagrams} \fi
\eqn\TwoDiagrams
$$
The $d^4\Theta$ integrals associated with the two diagrams are very
simple; generically, both diagrams contribute to the effective gauge
couplings and the respective contributions are
$$
\pm T_\ind \int\!{d^4 k\over (2\pi)^4}\,
{\bigl(H'(k^2)\bigr)^2\over H(k^2)\,H((k+p)^2)}\qquad
{\rm and}\qquad -(\pm T_\ind) \int\!{d^4 k\over (2\pi)^4}\,
{H''(k^2)\over H(k^2)} .
\eqn\LoopIntegrals
$$
In these formul\ae, $\pm T_\ind$ refers to the statistics and the
gauge quantum numbers of
the chiral superfield that runs around the loop,
$k$ is the loop momentum and $p$ is the momentum
of the external legs; we presume $p$ to be much smaller than the UV
cutoff scale $\Lambda$ while $A^{(n)}=O(\Lambda^{-2n})$.
Evaluating the Feynman integrals \LoopIntegrals\ shows that their
sum does not depend on the details of the polynomial $H(k^2)$ but only
on its limiting behaviors for $k^2\to\infty$ and for $k=O(|p|)\ll\Lambda$.
The result is particularly simple for a chiral superfield that is
either much heavier or much lighter than $p$:
for $|\widetilde M|^2\gg p^2$, the two integrals \LoopIntegrals\
total to
$$
{\pm T_\ind\over 16\pi^2}\left(
(2\nmax +1)(\log k^2_{\rm max}-1)\,
+\,2\log{\tilde A^{(\nmax )}\over |\widetilde M|}\right) ,
\eqn\HeavyLoop
$$
while for $\widetilde M=0$ or $|\widetilde M|^2\ll p^2$,
the result is
$$
{\pm T_\ind\over 16\pi^2}\left(
(2\nmax +1)(\log k^2_{\rm max}-1)\,
+\,2\log{\tilde A^{(\nmax )}\over \tilde A^{(0)}}\,
-\,\log p^2\,+3\right) .
\eqn\LightLoop
$$
The appearance of the $\log k^2_{\rm max}$ terms in these formul\ae\
reflects the logarithmic UV divergence of the individual superfields'
contributions.
However, once we sum over all the superfields that may run around the
loop, the divergences cancel out.
Indeed,  eqs.~\WarrEquations\ for the types, statistics and
gauge quantum numbers of the Pauli-Villars superfields are equivalent
to
$$
\sum_C \pm T_\ind(C)\bigl(2\nmax (C)+1\bigr)\ =\ 0,
\eqn\WarrsSum
$$
where the sum is over all the charged chiral superfields, including
the PVs, the physical matter superfields $\matter^I$ and the
ghosts $c$, $c'$ and $b$ of the gauge fixing.

\goodbreak
Now consider the finite terms in formul\ae~\HeavyLoop\ and \LightLoop.
For the matter superfields $\tilde A^{(\nmax)}=|\ccf|^{2-2\nmax}
A^{(\nmax)}$ where $A^{(\nmax)}$ is a constant of the order
$O(\Lambda^{-2\nmax})$ while
$\tilde A^{(0)}=|\ccf|^2\exp({-\kappa^2\over3}K)Z$.
Assuming all the matter superfields to be light compared to the
renormalization momentum $p$, we can write the total of their finite
contributions to the $\{g_\ind^{-2}(p)\}$ as
$$
\eqalign{
\sum_r^{\rm matter} {T_\ind(r)\over 16\pi^2}\Bigl( &
n_r\left( \coeff23\kappa^2K
	-4\nmax({\rm matter})\left[\log\Lambda+\log|\vev\ccf|\right]
	-\log p^2\right)\cr
&-2\log\det Z^{(r)} \Bigr), \cr
}\eqn\MatterContribs
$$
plus a constant that depends on the dimensionless
parameter $\Lambda^{2\nmax}A^{(\nmax)}$ of the WRS.

For the Faddeev-Popov ghosts $\nmax=0$ and $\tilde A^{(0)}=|\ccf|^2$;
for the Nielsen-Kallosh ghost $\nmax=\nmax({\rm gauge})$,
$\tilde A^{(\nmax)}=|\ccf|^{-4-2\nmax}B^{(\nmax)}$ where
$B^{(\nmax)}$ is a constant of the order $O(\Lambda^{-2\nmax})$
and $\tilde A^{(0)}=|\ccf|^{-4}Re f_\ind$.
All ghosts are massless; thus, the total of their finite
contributions to the $\{g_\ind^{-2}(p)\}$ is
$$
{ T(G_\ind)\over 16\pi^2}\left( 4\nmax({\rm gauge})
	\left[\log\Lambda+\log|\vev\ccf|\right]
	+3\log p^2+2\log\Re f_\ind\right)
\quad+\ \rm const.
\eqn\GhostContribs
$$
For the Pauli-Villars superfields,
$\tilde A^{(\nmax)}/|\widetilde M|=|\ccf|^{-1-2\nmax}\times\rm a$
constant of the order $O(\Lambda^{-1-2\nmax})$, regardless of the Weyl
type of a particular PVs.
Hence, the total finite contribution of the PVs to
the $\{g_\ind^{-2}(p)\}$ is
$$
-\sum_{\cal P} {\pm T_\ind({\cal P})\over 16\pi^2}
\left( 4\nmax({\cal P})+2\right) \left[\log\Lambda+\log|\ccf|\right]
\quad+\ \rm const.
\eqn\PVContribs
$$
Using eq.~\WarrEquations, we can re-express this sum over the
Pauli-Villars supermultiplets as a sum over the matter and ghost
supermultiplets.
Combining the result with eqs.~\MatterContribs\ and \GhostContribs,
we obtain
$$
\eqalign{
\{g_\ind^{-2}(p)\}^\ol &
    =\ \Re f_\ind\  +\ \rm const\cr
&+\sum_r^{\rm matter} {T_\ind(r)\over 8\pi^2}
    \left( n_r\left( \coeff23\kappa^2K +\log{\Lambda^2\over p^2}
	+2\log|\ccf| \right)\ -\ 2\log\det Z^{(r)} \right)\cr
&+\ {T(G_\ind)\over 16\pi^2} \left( -3\log{\Lambda^2\over p^2}
    -6\log|\ccf| +2\log\Re f_\ind \right) .
}\eqn\Penultimate
$$
Note that the ultraviolet types $\nmax$ of any of the charged superfields
cancel out of this formula;
indeed, only the constant (\ie, momentum-, moduli- and $\ccf$-independent)
term in eq.~\Penultimate\ is affected in any
way by any of the intricacies of the Warr's regularization scheme.
The actual value of this constant term depends on the scheme's
dimensionless parameters $\Lambda^{2\nmax}A^{(\nmax)}$,
$\Lambda^{2\nmax}B^{(\nmax)}$, $\Lambda^{2\nmax}A^{(\nmax)}_{\cal P}$
and $M_{\cal P}/\Lambda$;
these parameters are freely adjustable since the
regularizing properties of the Warr's cutoff do not depend
on their values.
By adjusting the cutoff's parameters it is easy to eliminate
the overall constant in eq.~\Penultimate;
henceforth, we shall assume that the cutoff's parameters are indeed
adjusted in this way.

Our next step is to re-organize eq.~\Penultimate\ by grouping together all
the momentum, $\ccf$ or \K\ dependent terms.
Using the $\ccf$-dependent \K\ function $\widetilde K$ instead of $K$,
we have
$$
\!\eqalign{
\{g_\ind^{-2}(p)\}^\ol
=\ \Re f_\ind\ &
+\ {3\cano_\ind\over 8\pi^2}\, \Re\log\ccf\
+\ {\cano_\ind\,\kappa^2\over 16\pi^2}\,\widetilde K\
+\ {\cren_\ind\over 16\pi^2}\left( \log{\Lambda^2\over p^2}\,
	-\,{\kappa^2\widetilde K\over3}\right)\cr
&+{T(G_\ind)\over 8\pi^2}\,\log\Re f_\ind\
-\sum_r^{\rm matter} {T_\ind(r)\over 8\pi^2}\,
	\log\det Z^{(r)} ,\cr
}\!\eqn\Ultimate
$$
where $\cren_\ind$ and $\cano_\ind$ are as in eqs.~\crendef\ and \canodef.
Formula \Ultimate\ relates the effective gauge
couplings of an EQFT to the bare Lagrangian couplings of the theory
regularized in a manifestly locally supersymmetric, background-gauge
invariant and super-Weyl invariant manner.
Before we proceed with further analysis of this formula, however,
we would like to turn our attention to the Wilsonian gauge couplings.

The price of the super-Weyl invariance of the UV regulators
we have used so far
is its $\ccf$ dependence, which obscures the $\ccf$ and moduli
dependence of the Wilsonian couplings of the theory.
As discussed in section 2.3, we identify the Wilsonian couplings
with the bare Lagrangian couplings of a theory whose UV regulator is
totally independent of the moduli and of the $\ccf$ while retaining
as many of the symmetries of the theory as possible under circumstances.
For the problem at hand, this means that the regulator should be locally
supersymmetric and background-gauge invariant but the super-Weyl
invariance has to be given up.
Consequently, the bare couplings of the theory may acquire non-classical
dependences on the Weyl compensators $\ccf$ and $\ccfb$;
in particular, we may have $\ccf$-dependent bare gauge couplings $f_\ind$.

It is very easy to render the Warr's regularization scheme
$\ccf$-independent at the expense of the super-Weyl invariance:
One simply erases all appearances of $\ccf$ and $\ccfb$ in eqs.\localDAlam\
through \PVaction.
The implications of this erasure for the way we calculate the effective
gauge couplings are also simple: Everything works exactly as before,
except for the use of the $\ccf$-independent $A^{(n)}$, and $M$ instead
of the $\tilde A^{(n)}$ and the $\widetilde M$.
Only the lowest-derivative coefficients $\exp(-\coeff13\kappa^2\widetilde K)
Z$ for the matter superfields $\matter^I$ remain $\ccf$-dependent
since they are part of the classical Lagrangian of the theory rather than
its UV regulator.
Therefore, repeating the steps that lead to eq.~\Ultimate\ almost
verbatim, we arrive at
$$
\eqalign{
\{g_\ind^{-2}(p)\}^\ol
=\ \Re f_\ind\ &
+\ {\cano_\ind\,\kappa^2\over 16\pi^2}\,\widetilde K\
+\ {\cren_\ind\over 16\pi^2}\left( \log{\Lambda^2\over p^2}\,
	-\,{\kappa^2\widetilde K\over3}\right)\cr
&+{T(G_\ind)\over 8\pi^2}\,\log\Re f_\ind\
-\sum_r^{\rm matter} {T_\ind(r)\over 8\pi^2}\,
	\log\det Z^{(r)} .\cr
}\eqn\WUltimate
$$
The only difference between eqs.~\Ultimate\ and \WUltimate\ is that
the latter lacks the $(3\cano_\ind/8\pi^2)\Re\log\ccf$ term, plus the
fact that the bare couplings involved in the two formul\ae\
differ from each other; this difference is the finite renormalization
effect of changing the UV regulators of the theory.
Identifying the bare gauge coupling in eq.~\WUltimate\
with the true Wilsonian gauge coupling $\tilde f^W_\ind(\modul,\ccf)$
and re-naming the $\ccf$-independent bare gauge coupling in eq.~\Ultimate\
`$f^W_\ind(\modul)$', we have
$$
\tilde f^W_\ind(\modul,\ccf)\
=\ f^W_\ind(\modul)\ +\ {3\cano_\ind\over8\pi^2}\,\log\ccf .
\eqn\WilsanomAgain\&\Wilsanom
$$
As described in section~2.3, neither $f^W_\ind$ nor $\tilde f^W_\ind$
renormalize beyond the one-loop level (\cf\ Appendix~A);
therefore, eq.~\WilsanomAgain\ holds without any modifications to all
orders of the perturbation theory.

Finally, let us consider the momentum dependence of
the effective gauge couplings.
In eqs.~\Ultimate\ and \WUltimate, the reference momentum appears in
combination
$$
\log{\Lambda^2\over p^2\equiv\eta^{mn}p_mp_n}\
-\ {\kappa^2\widetilde K\over3}\,,
\eqn\MomentumFormula
$$
where $\eta^{mn}=\rm diag(-,+,+,+)$ is the flat un-rescaled Minkowski metric;
this un-rescaled metric is implicit in the rigid-SUSY formalism we used
throughout this section.
On the other hand,
supergravitationally trivial background of a locally supersymmetric
EQFT means conventionally normalized flat vielbein $E^A_M$, which
for generic moduli VEVs $\vev{\modul^i}$ and generic
$\vev\ccf$ implies that the physical spacetime metric is
$$
g_{mn}\ =\ \eta_{mn}\,\exp(-\coeff13\kappa^2\widetilde K)
\eqn\scaledmetric
$$
\looseness=-1
(\cf\ eq.~\weylmetric)
rather than the un-rescaled $\eta_{mn}$.
Therefore, the \K\ term in eq.\ \MomentumFormula\ is the artifact
of using the wrong metric:
In terms of the physical metric $g_{mn}$,
$$
\log{\Lambda^2\over \eta^{mn}p_mp_n}\ -\ {\kappa^2\widetilde K\over3}\
=\ \log{\Lambda^2\over p_{\rm phys}^2\equiv g^{mn}p_mp_n}\,.
\eqn\RightMomentum
$$
(In the Wess-Zumino gauge \WZK\ $g^{mn}=\eta^{mn}$ and
eq.~\RightMomentum\ is a tautology, but in other super-Weyl gauges
one has to distinguish between the physical momentum square and the
$\eta^{mn}p_mp_n$.)
Thus, the physical meaning of eqs.~\Ultimate\ and \WUltimate\ is
$$
\displaylines{
\Re \tilde f^W_\ind\ +\ {\cano_\ind\,\kappa^2\over 16\pi^2}\,\widetilde K\
+\ {\cren_\ind\over8\pi^2}\log\Lambda\
=\ \Re f^W_\ind\ +\ {\cano_\ind\,\kappa^2\over 16\pi^2}\, K\
+\ {\cren_\ind\over8\pi^2}\log\Lambda\
=\ F_\ind\hfill\cr
{}\hfill =\ \{g_\ind^{-2}(p)\}^\ol\,
+\ {\cren_\ind\over 16\pi^2}\,\log p_{\rm phys}^2\
-\ {T(G_\ind)\over 8\pi^2}\,\log g_\ind^{-2}\
+\sum_r^{\rm matter} {T_\ind(r)\over 8\pi^2}\,\log\det Z^{(r)} ,\cr
\hfill\eqname\Confirmation \cr }
$$
which is precisely the one-loop approximation to the eqs.~\SV\ and \Katerm.

\APPENDIX{D}{D\break
Effective Axionic Couplings}
Classically, moduli-dependence of the $\theta_\ind(\modul,\modulb)$ ---
the coefficients of the CP-odd terms $\tr_a(F\tilde F)$ in the gauge-field
Lagrangian --- gives rise to the axionic couplings of the moduli scalars.
In a quantum theory, derivative couplings of the moduli to the
charged fermions also lead to the same effect:
The 3-point Green's function for two gauge bosons, $A^{(a)}_m(p_1)$
and $A^{(a)}_n(p_1)$,
and a modulus scalar $\modul^i$ (or $\modulb^\ib$), acquires a CP-odd component
$(1/16\pi^2)\epsilon^{mnkl}p_{1k} p_{2l}\,
    \{\theta_{\ind,{i\,({\rm or}\,\ib)}}\}$.
In supersymmetric theories, the effective axionic couplings
$\{\theta_{\ind,{i\,({\rm or}\,\ib)}}(\vev\modul,\vev\modulb\}$
are related to the moduli-dependence of the effective gauge
couplings\refmark{\DKLb,\VKa}:
$$
\eqalign{
\{\theta_{\ind,i}(\vev\modul,\vev\modulb)\}\ &
=\ {+8\pi^2i}\,\partder{\{g_\ind(\vev\modul,\vev\modulb)\}^{-2}}%
	{\vev{\modul^i}}\,,\cr
\{\theta_{\ind,\ib}(\vev\modul,\vev\modulb)\}\ &
=\ {-8\pi^2i}\,\partder{\{g_\ind(\vev\modul,\vev\modulb)\}^{-2}}%
	{\vev{\modulb^\ib}}\,.\cr
}\eqn\axionic
$$
The basis for these relations is SUSY on the level of Green's functions,
and as long as it is unbroken, it does not matter exactly
how the EQFT is quantized or whether SUSY is rigid or local.\refmark{\DKLb}

Combining eqs.~\axionic\ and \SV,
we obtain
$$
\eqalign{
\left( 1- {T(G_\ind)\over 8\pi^2}\{g_\ind\}^2\right)\cdot\{\theta_{\ind,i}\}\ &
=\ {+i} \partder{}{\vev{\modul^i}}\left[ 8\pi^2\, F_\ind\
    	-\sum_r T_\ind(r)\,\log\det\{ Z^{(r)}\} \right] ,\cr
\left( 1- {T(G_\ind)\over 8\pi^2}\{g_\ind\}^2\right)\cdot\{\theta_{\ind,\ib}\}\
&
=\ {-i} \partder{}{\vev{\modulb^\ib}}\left[ 8\pi^2\, F_\ind\
    	-\sum_r T_\ind(r)\,\log\det\{ Z^{(r)}\} \right] ;\cr
}\eqn\SVaxionic
$$
again, it does not matter whether SUSY is rigid or local
or whether it is manifest in the Wilsonian action and the cutoff.
Therefore, in order to completely determine the moduli dependence
of the renormalization group integrals $F_\ind$, all we have to do is
to calculate the effective axionic couplings, and we
can perform this calculation in any formalism we like.
(Similar computations are also performed in refs.~[\JLpascos--\CLO].)

At the one-loop level of analysis (which is sufficient to completely
determine $F_\ind$), the easiest way to compute an effective axionic
coupling is to use component fields and ordinary Feynman rules.
There are three contributions to $\{\theta_{\ind,i}\}$:
the tree-level term $\partial\theta^W_\ind/\partial\modul^i$;
the renormalization of that term due to loops of gauge bosons;
and, finally, the axial anomaly graphs involving charged fermions
$\Psi^I_\alpha$ and $\lambda^{(a)}_\alpha$.
Assuming all such fermions retained in an EQFT are massless, we have
$$
\{\theta_{\ind,i}\}^\ol\
=\ \partder{\theta^W_\ind}{\modul^i}\cdot\left(1+{T(G_\ind)\over
4\pi^2}\,g^2\right)\
-\ 2i\sum_r^{\rm all\ charged\atop fermions} T_\ind(r)\,\Tr\Gamma_i^{(r)} ,
\eqn\oneloopax
$$
where $\Gamma_i^{(r)}$ is the matrix of couplings of $\partial_m\modul^i$
to the `flavor' currents of charged fermions;
a similar formula holds for the axionic couplings of the anti-chiral
moduli $\modulb^\ib$.
To make our notations clear, let us write down the terms in the Wilsonian
Lagrangian of the EQFT that involve two charged fermionic fields,
one left-handed and one right-handed:
$$
\def\bs{\mathop{\raise 0.3ex\rlap{\kern 0.15em /}\nabla}
	\limits^{\longleftrightarrow}}
L_{\bar ff}\
=\ \coeff i2 Z_{\Ib J}\left(\overline{\Psi}^\Ib\bs\Psi^J\right)\
+\ \coeff i2 (\Re f)^{-1}_{(a)(b)}
	\left(\bar\lambda^{(a)}\bs\lambda^{(b)}\right) ,
\eqn\Lff
$$
where ${\raise 0.3ex\rlap{\kern 0.15em /}\nabla}\equiv\gamma^m\nabla_m$,
$\nabla_m$ being the covariant derivative containing gravitational,
gauge and flavor connections:
$$
\nabla_m\ =\ \partial_m\ +\ \half\omega_m^{ab}\sigma_{ab}\
-\ iA_m^{(a)}T_{(a)}\ +\ (\partial_m\modul^i)\Gamma_i\
+\ (\partial_m\modulb^\ib)\Gamma_\ib\ .
\eqn\Gammadef
$$
In this formula, $\Gamma_i$ and $\Gamma_\ib$ are matrices carrying `flavor'
indices, \ie, $(\Gamma_i\Psi)^J=\Gamma^J_{iK}\Psi^K$, \etc;
$\Gamma_i^{(r)}$ in eq.~\oneloopax\ is the restriction of the full $\Gamma_i$
matrix to fermions transforming like $r$ under the gauge group.

Formula \oneloopax\ presumes background gauge invariance of the quantum theory
but it does not need manifest SUSY.
Therefore, we can obtain the $\Gamma_i$ and $\Gamma_\ib$ matrices
of a locally supersymmetric EQFT by simply looking up the component-field
Lagrangian in any standard reference on SUGRA and identifying the appropriate
terms as belonging to~\Lff.
\foot{Actually, not all standard references can be used verbatim but only those
    following the ``American'' convention for the Weyl rescaling
    in which fermionic fields are rescaled but their phases remain unchanged;
    in the super-Weyl invariant language, this convention corresponds to
    $\arg\ccf=0$ and hence $\theta^W_\ind=-8\pi^2 \Im\tilde f^W_\ind
    = -8\pi^2 \Im f^W_\ind$.
    In the ``European'' phase convention,
    the relation between the properly defined Wilsonian $\theta$ angles
    and the imaginary parts of $f_\ind$ is not so straightforward.}
Using as our source ref.~[\WB], we have, in matrix notations,
$$
\eqalign{
\Gamma_i^{(\Psi)}\ &
=\ {1\over 2Z}\,\partder{Z}{\modul^i}\
	-\ {\kappa^2\over 4}\partder K{\modul^i}\,,\cr
\Gamma_\ib^{(\Psi)}\ &
=\ {-1\over 2Z}\,\partder{Z}{\modulb^\ib}\
	+\ {\kappa^2\over 4}\partder K{\modulb^\ib}\,,\cr
\Gamma_i^{(\lambda)}\ &
=\ {1\over 2(f+f^*)}\,\partder{f}{\modul^i}\
	+\ {\kappa^2\over 4}\partder K{\modul^i}\,,\cr
\Gamma_\ib^{(\lambda)}\ &
=\ {-1\over 2(f+f^*)}\,\partder{f^*}{\modulb^\ib}\
	-\ {\kappa^2\over 4}\partder K{\modulb^\ib} \cr
}\eqn\SUGRAgammas
$$
Rigidly supersymmetric EQFTs have similar $\Gamma_i$ matrices,
but without the K\"ahler terms.

At this point, all that remains to do is some straightforward algebra.
Substituting $\theta^W_\ind=-8\pi^2 \Im f^W_\ind$, $g_\ind^2=1/\Re f_\ind^W$
and the matrices~\SUGRAgammas\ into eq.~\oneloopax\ leads to
$$
\!\eqalign{
\!\{\theta_{\ind,i}\}^\ol &
= {\let\vcenter=\vtop \eqalign{
	i\left(4\pi^2+T(G_\ind)g_\ind^2\right) \partder{f_\ind^W}{\modul^i}\, &
	-\, \coeff i2 T(G_\ind)\left(g_\ind^2\partder{f_\ind^W}{\modul^i}
	    +\kappa^2\partder{K}{\modul^i}\right)\cr
	&+\, \coeff i2 \partder{}{\modul^i}\sum_r T_\ind(r)\left[
	    n_r\kappa^2 K -2\log\det Z^{(r)}\right]\cr }}\!\cr
&=\, i\left(8\pi^2+T(G_\ind)g_\ind^2\right) \partder{\Re f_\ind^W}{\modul^i}\cr
&\hphantom{\left(4\pi^2+T(G_\ind)g_\ind^2\right) \partder{}{\modul^i}}
    +\ i\partder{}{\modul^i}\left[ \cano_\ind\cdot\half\kappa^2 K\,
	-\sum_r T_\ind(r)\,\log\det Z^{(r)}\right]\cr }\!
\eqn\SUGRAax
$$
and ditto for the $\{\theta_{\ind,\ib}\}$.
Note how the coefficients of all the K\"ahler terms
assemble into precisely the $\cano_\ind$ defined in eq.~\canodef.
Now we substitute eqs.~\SUGRAax\ and their complex conjugates
into eqs.~\SVaxionic, which immediately gives us
$$
\displaylines{
F_\ind(\modul,\modulb)\ -\ \Re f_\ind^W(\modul)\
    -\ \cano_\ind\,{\kappa^2 K(\modul,\modulb)\over 16\pi^2}
    \hfill\eqname\Faxionic\cr
{}=\ \hbox{a moduli-independent constant}\ +\ O(g^2).\cr }
$$
The nature of the constant here being obvious from the Wilsonian
renormalization of $f_\ind^W$, eq.~\Faxionic\ is essentially the same as
eq.~\Katerm, except for the $O(g^2)$ terms, which are artifacts of the
one-loop approximation to the axionic couplings and should cancel out
of more accurate higher-order calculations.

We conclude that eq.~\Katerm\ is quite robust: Any formulation of
a locally supersymmetric EQFT in which $\theta_\ind^W(\modul,\modulb)=
-8\pi^2\Im f^W_\ind(\modul)$ for holomorphic $f_\ind^W$ would lead to the
same relation between $\Re f_\ind^W$ and the renormalization
group integrals $F_\ind$ for the running effective gauge couplings
$\{g_\ind(p)\}$.
In particular, the K\"ahler term eq.~\Katerm\ is an inalienable
feature of the local SUSY; the compensator formalism may be the
easiest way to see why this term should be present, but all other
ways should lead to the same destination.

\refout
\bye